\documentclass[11pt]{article}
\usepackage{color,amsmath, amsthm, amsfonts, amssymb, dsfont, graphicx, listings, graphics, epsfig, enumerate, bbm,dsfont,setspace,soul
}

\usepackage[pdftex]{hyperref}
\usepackage[margin=1in]{geometry}
\usepackage[fleqn,tbtags]{mathtools}
\usepackage[english]{babel}

\usepackage[numbers]{natbib}   

\usepackage{cleveref}

\addto\captionsenglish {}
\makeatletter \renewenvironment{proof}[1][\proofname] {\par\pushQED{\qed}\normalfont\topsep6\p@\@plus6\p@\relax\trivlist\item[\hskip\labelsep\bfseries#1\@addpunct{.}]\ignorespaces}{\popQED\endtrivlist\@endpefalse} \makeatother
\newtheorem{theorem}{Theorem}[section]
\newtheorem{corollary}[theorem]{Corollary}
\newtheorem{lemma}[theorem]{Lemma}
\newtheorem{proposition}[theorem]{Proposition}
\newtheorem{assumption}[theorem]{Assumption}
\newtheorem{definition}[theorem]{Definition}
\theoremstyle{definition}
\newtheorem{example}[theorem]{Example}
\newtheorem{remark}[theorem]{Remark}

\numberwithin{equation}{section}

\newcommand{\R}{\mathbb{R}}

\DeclareMathOperator{\cl}{cl}
\DeclareMathOperator{\conv}{conv}

\DeclareMathOperator{\dom}{dom}

\let\abs=\envert

\newcommand{\F}{\mathcal{F}}
\newcommand{\X}{\mathcal{X}}
\newcommand{\Y}{\mathcal{Y}}

\renewcommand{\P}{\mathbb{P}}
\newcommand{\E}{\mathbb{E}}
\newcommand{\M}{\mathcal{M}}

\newcommand{\Q}{\mathbb{Q}}
\renewcommand{\a}{\alpha}

\renewcommand{\S}{\mathbb{S}}
\newcommand{\N}{\mathbb{N}}
\renewcommand{\R}{\mathbb{R}}
\renewcommand{\H}{\mathcal{H}}

\newcommand{\of}[1]{\ensuremath{\left( #1 \right)}}
\newcommand{\cb}[1]{\ensuremath{ \left\{ #1 \right\} }}
\newcommand{\sqb}[1]{\ensuremath{ \left[ #1 \right] }}
\newcommand{\norm}[1]{\ensuremath{ \left\Vert #1 \right\Vert }}
\newcommand{\ip}[1]{\ensuremath{ \left\langle #1 \right\rangle }}

\def\prehp(#1,#2){\ensuremath{  #1 \cdot #2 }}

\begin{document}
	
	\title{Dual representations for quasiconvex compositions with applications to systemic risk measures}
	\author{\c{C}a\u{g}{\i}n Ararat\thanks{Bilkent University, Department of Industrial Engineering, Ankara, Turkey, cararat@bilkent.edu.tr.}
		\and
		M\"{u}cahit Ayg\"{u}n\thanks{University of Amsterdam, Department of Quantitative Economics, Amsterdam, \mbox{The Netherlands}, m.aygun@uva.nl.}
	}
	\date{\today}
	\maketitle

\begin{abstract}
	Motivated by the problem of finding dual representations for quasiconvex systemic risk measures in financial mathematics, we study quasiconvex compositions in an abstract infinite-dimensional setting. We calculate an explicit formula for the penalty function of the composition in terms of the penalty functions of the ingredient functions. The proof makes use of a nonstandard minimax inequality (rather than equality as in the standard case) that is available in the literature. In the second part of the paper, we apply our results in concrete probabilistic settings for systemic risk measures, in particular, in the context of Eisenberg-Noe clearing model. We also provide novel economic interpretations of the dual representations in these settings.\\
	{\bf Key words:} dual representation, quasiconvex function, penalty function, composition of functions, minimax inequality, systemic risk measure\\
	{\bf MSC Codes:} 46N10, 91G45, 46A20, 52A01	
\end{abstract}

\section{Introduction}

Starting with the 2008 financial crisis, measuring risk in interconnected financial systems has gained importance in the financial mathematics community. Functionals defined for this purpose are generally referred to as \emph{systemic risk measures}, shifting the focus of the research in risk measures from the univariate case to the multivariate case. The goal of this paper is to study quasiconvex systemic risk measures as quasiconvex compositions from a duality point of view.

In the original framework of \cite{artzner}, \emph{coherent risk measures} are defined as monotone, convex, translative, and positively homogeneous functionals defined on a space of real-valued random variables. These random variables can be used to model the uncertain future worth of investments, and a risk measure assigns to each random variable its minimum deterministic capital requirement. Among the properties of coherent risk measures, \emph{monotonicity} is a natural requirement which asserts that the risk of an investment with consistently higher future values should be lower. \emph{Convexity} is related to diversification; under this property, the risk of a mixture of two portfolios is not higher than the same type of mixture of the individual risks. \emph{Positive homogeneity} is a scaling property that is relaxed for defining \emph{convex risk measures} in \cite{follmer}. Finally, \emph{translativity} asserts that a deterministic increase in the value of a portfolio decreases its risk by the same amount. This is indeed the property that justifies the interpretation of risk measure as capital requirement.

One might question whether convexity provides the correct encoding of the impact of diversification on risk. A weaker alternative is \emph{quasiconvexity}, which bounds the risk of a mixture only by the maximum of the individual risks, hence the statement ``Diversification does not increase risk." is reflected properly. Under translativity, convexity is equivalent to quasiconvexity. Hence, the switch from convexity to quasiconvexity implies working with non-translative functionals in general. Indeed, the work \cite{kupper} proposes a minimalist framework for risk measures in which only monotonicity and quasiconvexity are taken for granted, such functionals are called \emph{quasiconvex risk measures}; see also \cite{fritelli}. For the use of quasiconvex risk measures in the context of financial optimization problems; see \cite{qcport,sigrid,elisa}.

The theory of risk measures outlined above is for univariate random variables. In more complex settings such as markets with transaction costs (\cite{hamelheyde, hamelheyderudloff}) and financial networks with interdependencies (\cite{chen, feinstein, biagini, ararat}), it becomes necessary to evaluate the risks of random vectors. In this paper, we are particularly interested in the latter situation where the participating financial institutions are subject to correlated sources of risk, typically affecting the future values of their assets. Hence, the resulting future values are naturally modeled as correlated random vectors, explaining the multivariate nature of the problem. At the same time, the institutions form a network through mutual obligations and the aforementioned uncertainty affects the ability of the institutions to meet these obligations. Hence, the aim of a \emph{systemic risk measure} is to quantify the overall risk associated to the financial network.

In the pioneering work \cite{chen}, a systemic risk measure $R$ is defined as the composition of a univariate risk measure $\rho$ with a so-called \emph{aggregation function} $\Lambda$, that is, $R=\rho\circ \Lambda$. The role of the aggregation function is to summarize the impact of the random shock vector $X$ on the economy (or society) as a scalar random quantity $\Lambda(X)$. The definition of $\Lambda$ is made precise by the structure of the network and the accompanying clearing mechanism. For instance, one can consider a clearing system in the Eisenberg-Noe framework (\cite{eisnoe}) and define the aggregation function as the total payment made to society as in \cite{ararat}, in which case $\Lambda$ is an increasing concave function. The output of $\Lambda$ is further given as input to a convex risk measure $\rho$ to calculate the value of $R(X)$. The resulting systemic risk measure $R$ is a monotone convex functional that is not translative in general. In \cite{ararat}, dual representations for \emph{convex} systemic risk measures are studied in detail. The mathematical machinery used in that work is the conjugation formula in \cite[Thm. 2.8.10]{zal} and \cite[Thm. 3]{bot} for convex compositions.

When $\rho$ is only assumed to be a quasiconvex risk measure, the resulting systemic risk measure $R$ is also quasiconvex. Providing dual representations for this case is the starting point of this paper. However, we will first study the problem in greater generality. We will explore the dual representation of a quasiconvex composition $f\circ g$, where the ingredients $f, g$ are defined on general preordered topological vector spaces.

In the literature, the study of $f\circ g$ from a duality point of view is not new in the convex case. For a single function, Fenchel-Moreau theorem provides a dual representation for a convex lower semicontinuous function in terms of its Legendre-Fenchel conjugate (\cite[Thm. 12.2]{Rocka}). Then, it is natural to ask how and when we can have a dual representation for the composition of convex functions. This question has been answered in the literature, for instance, in \cite[Thm. 2.8.10]{zal} and \cite[Thm. 3]{bot}, by using perturbation functions and convex duality arguments.

As a natural extension of the convex case from a theoretical point of view, we look for dual representations of $f\circ g$ when it is guaranteed to be quasiconvex. This is an open problem to the best of our knowledge. For a single function, the quasiconvex duality theory in \cite{penot} provides a suitable replacement of conjugate functions in convex duality. This is further explored in \cite{simone2} within an abstract framework, in \cite{fritelliSIAM} for vector-valued functions within a conditional setting, and also in \cite{simone, kupper, fritelli} within the context of risk measures. In line with \cite{kupper}, the dual functions for quasiconvex duality will be referred to as \emph{penalty functions} in the sequel.

In this paper, we provide a formula for the penalty function of $f\circ g$, roughly speaking, in terms of the penalty functions of $f$ and $g$. More precisely, apart from the more technical continuity conditions, we will assume that $f$ is an extended real-valued monotone, quasiconvex function. Since $g$ is a vector-valued function (in a possibly infinite-dimensional space), choosing the right notion of quasiconvexity requires extra care. To this end, we will use the notion of \emph{natural quasiconvexity}, which is introduced for vector-valued functions in \cite{tanaka} and for set-valued functions in \cite{kur}. When $g$ is a monotone, naturally quasiconcave function, the resulting composition $f\circ g$ is a monotone, quasiconvex function.

For the proof of our main duality theorem (\Cref{mainthm}), we need a nonstandard minimax result since the assumptions of the standard minimax theorem in \cite{sion} may not hold in our case. We are able to overcome this issue by using the minimax inequality in \cite{Liu} (see also \cite{liurelated1,liurelated2}), which works under weaker conditions. With additional arguments that use the properties of the involved functions, we are able to turn the inequality into an equality. Hence, the proof of the main theorem makes novel use of minimax theory.

After building the general theory, we go back to our motivating problem on systemic risk measures. Using a quasiconvex univariate risk measure $\rho$ and a concave aggregation function $\Lambda$, we are able to provide a dual representation for the systemic risk measure $R=\rho\circ\Lambda$ in a probabilistic framework. We also discuss the economic interpretations of the dual variables and penalty functions in terms of the underlying financial network. Thanks to our results on quasiconvex compositions, we are able to decompose the contributions of $\rho$ and $\Lambda$ to the penalty function as separate terms, which would not have been possible by an application of quasiconvex duality (\cite{kupper,penot}) directly on $R$.

The rest of this paper is organized as follows. In \Cref{sec:quasicon}, we review some basic notions and results about convex and quasiconvex functions. \Cref{sec:natquasi} is dedicated to some more technical notions for vector-valued functions: natural quasiconvexity, regular monotonicity, and lower demicontinuity. In \Cref{sec:comp}, we prove the main theorem on quasiconvex compositions together with some important special cases. This is followed by \Cref{sec:cone}, where we discuss the validity of a compactness assumption in concrete settings. In \Cref{sec:application}, we apply the theory to obtain dual representations for systemic risk measures. Among the various examples that we study, Eisenberg-Noe model is discussed separately as it has a more sophisticated aggregation function. \Cref{sec: conclusion} concludes the paper. Some proofs of the results are collected in Appendices~\ref{app: 2 and 3}, \ref{app:comp}, and \ref{app:sec 6}.

\section{Convex and quasiconvex functions}\label{sec:quasicon}
	
\subsection{Preliminaries}\label{sec:prelim}
	
We begin with some basic notations and definitions that are used throughout the paper. We denote by $\overline{\R}\coloneqq \R\cup \{+\infty,-\infty\}$ the extended real line. Given $a,b\in\overline{\R}$, we define $a\vee b \coloneqq \max\{a,b\}$, $a\wedge b\coloneqq \min\{a,b\}$. For each $n\in\N\coloneqq\{1,2,\ldots\}$, we denote by $\R^n$ the $n$-dimensional Euclidean space and by $\R^n_+$ the set of all $z=(z_1,\ldots,z_n)^{\mathsf{T}}\in\R^n$ with $z_i\geq 0$ for each $i\in\{1,\ldots,n\}$. For $w,z\in\R^n$, we define their Hadamard product by $w\cdot z\coloneqq (w_1z_1,\ldots,w_nz_n)^{\mathsf{T}}\in \R^n$.   We write $\R_+=\R^1_+$ and $\R_{++}=(0,+\infty)$.
	
Let $\mathcal{X}$ be a Hausdorff locally convex topological vector space. For a set $A\subseteq\X$, $\cl(A)$ and $\conv(A)$ denote the closure and convex hull of $A$, respectively. We denote by $\mathcal{X}^\ast$ the topological  dual space of $\X$, endowed with the weak$^\ast$  topology $\sigma(\mathcal{X}^\ast,\mathcal{X})$. The bilinear duality mapping on $\X^\ast\times\X$ is denoted by $\ip{\cdot,\cdot}$. For nonempty sets $A,B\subseteq\X$ and $\lambda\in\R$, we define the sum $A+B\coloneqq\{x+y\mid x\in A, y\in B\}$ and the product $\lambda A\coloneqq\{\lambda x\mid x\in A\}$ in the Minkowski sense. When $A=\{x\}$ for some $x\in\X$, we write $x+B\coloneqq \{x\}+B$. For a nonempty set $L\subseteq\R$, we define $L A\coloneqq\{\lambda x\mid \lambda\in L, x\in A\}$.
	
	Throughout this section, let $f\colon\X \rightarrow \overline{\R}$ be a function. Given $m\in\R$, the \emph{$m$-sublevel set} of $f$ is defined as
	\[
	S_f^m\coloneqq \cb{x\in \mathcal{X} \mid  f(x) \leq m}.
	\]
	A straightforward calculation yields that $f$ can be recovered from its sublevel sets via
	\begin{equation}\label{remf}
		f(x)=\inf\{m \in \R \mid  x\in S_f^m\},\quad x\in\X.
	\end{equation}

	The function $f$ is called \emph{positively homogeneous} if $f(\lambda x)=\lambda f(x)$ for every $\lambda >0$, $x\in \X$. It is called \emph{proper} if $f(x)>-\infty$ for every $x\in\X$ and $f(x)<+\infty$ for at least one $x\in\X$. The \emph{conjugate function} or the \emph{Legendre-Fenchel transform} $f^\ast\colon\mathcal{X}^\ast\rightarrow \overline{\mathbb{R}}$ of $f$ is defined by
	\[
	f^\ast(x^\ast)\coloneqq \sup_{x\in \mathcal{X}}\of{\ip{x^\ast,x}-f(x)},\quad x^\ast\in\X^\ast.
	\]
	As an important special case, we may take $f=I_A$ for some $A\subseteq \X$, where $I_A$ is the (convex analytic) \emph{indicator function} of $A$ defined by $I_A(x) \coloneqq 0$ if $x\in A$, and by $I_A(x)= +\infty$ if $x \in \X\setminus A$. Then, the conjugate function of $I_A$ is the \emph{support function} of $A$ given by
	\begin{equation}\label{eq:support}
		I_A^\ast (x^\ast) = \sup_{x\in A}\ip{x^\ast,x},\quad x^\ast\in\X^\ast.
	\end{equation}
	
	\begin{definition}
		(i) The function $f$ is called \emph{quasiconvex} if $f\of{\lambda x+(1-\lambda)y}\leq f(x)\vee f(y)$ for every $x,y \in \mathcal{X}$ and $\lambda\in [0,1]$. It is called \emph{quasiconcave} if $-f$ is quasiconvex.\\
		(ii) Let $x\in\X$. The function $f$ is called \emph{lower semicontinuous at $x$} if $f(x) \leq \liminf _{i\in I} f(x_i)$ whenever $(x_i)_{i\in I}$ is a net in $\X$ that converges to $x$. It is called \emph{lower semicontinuous} if it is lower semicontinuous at each $x \in \mathcal{X}$. It is called \emph{upper semicontinuous (at $x$)} if $-f$ is lower semicontinuous (at $x$).
	\end{definition}
	
	\begin{remark}\label{remclosed}
		It is well-known that $f$ is quasiconvex if and only if $S_f^m$ is convex for every $m\in\R$ (\cite[Sect.~2.1, p.~41]{zal}), and $f$ is lower semicontinuous if and only if $S_f^m$ is closed for every $m\in\R$ (\cite[Lem.~2.39]{aliprantis}). Moreover, every closed convex strict subset of $\X$ can be written as the intersection of all closed halfspaces that contain it (\cite[Cor.~5.83]{aliprantis}). Thus, when $f$ is 
        lower semicontinuous and quasiconvex, $S_f^m$ can be written as an intersection of closed halfspaces for each $m\in\R$.
	\end{remark}
	
	\subsection{The order structure}\label{sec:order}
	
	To be able to handle monotone functions, we introduce an order structure on $\X$. To that end, let $C\subseteq\X$ be a convex cone and define a relation $\leq_C$ on $\X$ by
	\begin{equation}\label{preorder}
		x\leq_C y \quad \Leftrightarrow\quad y-x\in C
	\end{equation}
	for each $x,y\in\X$. It follows that $\leq_C$ is a \emph{vector preorder}, that is,  $x \leq_C y$ implies $x+z \leq_C y+z $ and $\lambda x\leq_C \lambda y$ for every $x,y,z \in \mathcal{X}$ and $\lambda>0$.
	
	\begin{remark}
		By \cite[Sect.~8.1]{aliprantis}, every vector preorder $\preccurlyeq$ on $\X$ can be written as $\preccurlyeq=\leq_C$, where $C \coloneqq \{x\in\X\mid 0\preccurlyeq x\}$ is a convex cone. Hence, the assumption that $C$ is a convex cone is not a restriction on the vector preorder of interest.
	\end{remark}
	
	Every $x\in C$ is called a \emph{positive element} of $\X$. We define the \emph{polar cone} of $C$ by
	\[
	C^{\circ}\coloneqq \{x^\ast \in \X^\ast \mid  \forall x\in C\colon \ip{x^\ast,x}	\leq 0\},
	\]
	which is a closed convex cone in $\X^\ast$.	Then, we define the cone of \emph{strictly positive elements} of $\X$ by
	\begin{equation}\label{sharp}
		C^{\#}=\cb{x\in C\mid \forall x^\ast\in C^{\circ}\setminus\{0\}\colon \ip{x^\ast,x}< 0 }.
	\end{equation}
	Given $\pi\in C^{\#}$, we may scale the elements of $C^{\circ}$ and obtain the closed convex set
	\[
	C_{\pi}^{\circ}\coloneqq\{x^\ast\in C^{\circ}\mid \ip{x^\ast,\pi}=-1 \}.
	\]
    
	\begin{remark}
		When $\X$ is finite-dimensional, $C^{\#}$ coincides with the interior of $C$. In our infinite-dimensional setting, we work with $C^{\#}$ as the interior of $C$ can be empty for many important examples including Lebesgue spaces; see \cite[Ex. 2.12]{gluckweber}.
	\end{remark}
	
	The next lemma shows that $C^{\circ}$ can be recovered from the (much) smaller set $C_{\pi}^{\circ}$ if $\pi\in C^{\#}$. We omit its elementary proof for brevity.
	
	\begin{lemma}\label{projection}
		Assume that $C^{\#}\neq\emptyset$ and let $\pi \in C^{\#}$. Then, $C^{\circ}\setminus \{0\}=\R_{++} C_{\pi}^{\circ}$.
	\end{lemma}
	
	Thanks to the order structure provided by $\leq_C$, we may define the monotonicity of sets and functions. We say that a set $A\subseteq \mathcal{X}$ is \emph{monotone} if $A+C\subseteq A$. Similarly, we say that $f$ is a \emph{decreasing} 
	function if $x\leq _{C} y$ implies $f(x)\geq f(y)$ for every $x,y\in\X$; we say that $f$ is an \emph{increasing} function if $-f$ is decreasing.
	
	\begin{remark}\label{remmonotone}
		It is easy to check that $f$ is decreasing if and only if its sublevel sets are monotone. 
	\end{remark}
	
	\subsection{Dual representations}\label{sec:dualsingle}
	 
	Let $f\colon \X\to \overline{\R}$ be a function. When it is proper, lower semicontinuous, and convex, Fenchel-Moreau theorem (\cite[Thm.~2.3.4]{zal}) provides a dual representation for $f$ in terms of its conjugate function $f^\ast$:
	\[
	f(x)=\sup_{x^\ast\in\X^\ast}\of{\ip{x^\ast,x}-f^\ast(x)},\quad x\in\X.
	\]
	One immediate consequence of this theorem is that a set $A\subseteq\X$ and its closed convex hull have the same support function, that is,
	\begin{equation}\label{chull}
		I^\ast_A(x^\ast)=I^\ast_{\cl(\conv(A))}(x^\ast),\quad x^\ast\in\X^\ast.
	\end{equation}
	This observation will later be useful in proving \Cref{kkt}, which will then be used in the proof of \Cref{mainthm}, the main theorem of the paper.

	For monotone functions, the following refinement of Fenchel-Moreau theorem is possible. The proof is straightforward, hence omitted.
	
	\begin{proposition}\label{fenchel}
		Suppose that $f$ is proper, decreasing, convex, and lower semicontinuous. Then,
		\begin{equation}\label{eq:fenchel}
			f(x)=\sup_{x^\ast\in C^{\circ}} \of{\ip{x^\ast,x}-f^*(x^\ast)},\quad x\in\X.
		\end{equation}
	\end{proposition}
	
	For a quasiconvex function, a suitable generalization of conjugation is possible by the so-called penalty function, which is defined in terms of the support function of sublevel sets. The precise definition is given next.

	\begin{definition}\label{penaltydefn}
		The \emph{penalty function} $\alpha_f\colon\mathcal{X}^\ast \times \R\rightarrow \overline{\mathbb{R}}$ associated with $f$ is defined by 
		\begin{equation*}
			\alpha_f(x^\ast,m)\coloneqq I^\ast_{S_f^m}(x^\ast)=\sup_{x\in S_f^m} \ip{x^\ast,x}, \quad x^\ast\in \mathcal{X}^\ast, m\in \R.
		\end{equation*}	
	\end{definition}
	
	\begin{remark}\label{nondeca}
		It is clear that $\a_f$ is positively homogeneous in its first argument, i.e., $\a_f(\lambda x^\ast,m)=\lambda\a_f(x^\ast,m)$ for every $x^\ast\in\X^\ast$, $m\in\R$. Moreover, $\a_f$ is increasing in its second argument. Indeed, by taking $m_1,m_2\in\R$ with $m_1\leq m_2$, we have $S_f^{m_1}\subseteq S_f^{m_2}$ so that $\a_f(x^\ast,m_1)\leq \a_f(x^\ast,m_2)$ for every $x^\ast\in\X^\ast$.
	\end{remark}

	We continue with a remark that serves as a basis for dual representations. 
	
	\begin{remark}\label{separationlem}
    Suppose that $f$ is decreasing, lower semicontinuous, and quasiconvex. Let $m\in\R$. Then, $S_f^m$ is a monotone, closed, and convex set by Remarks~\ref{remclosed}, \ref{remmonotone}. Hence, as a consequence of Hahn-Banach theorem, for every $x\in\X$, we have
		\begin{equation*}
			x\in S_f^m \quad  \Leftrightarrow \quad \forall x^*\in C^{\circ}\setminus \{0\}\colon \ip{x^*,x} \leq \alpha_f\of{x^*,m}.
		\end{equation*}	
	\end{remark}
	
	When $f$ is lower semicontinuous and quasiconvex, its dual representation will be stated in terms of a special pseudoinverse of $\a_f$, as defined in the next definition.
	
	\begin{definition}\label{leftinv}
		(\cite[App.~B]{kupper}) Let $\a\colon \X^\ast \times \R\to\overline{\R}$ be a function which is increasing in its second argument. We define its \emph{left inverse} $\beta \colon \X^\ast \times \R\to\overline{\R}$ (with respect to the second argument) by
		\begin{equation}
	\beta(x^\ast,s)\coloneqq\sup \cb{m\in \R\mid \a(x^\ast,m)<s}= \inf \cb{m\in \R\mid \a(x^*,m)\geq s}
		\end{equation}
		for each $x^\ast\in \X^\ast$ and $s\in\R$. We denote by $\beta_f$ the left inverse of the penalty function $\alpha_f$ associated with $f$.
	\end{definition}

 The next lemma provides simple strong duality results that will be useful in later calculations.

\begin{lemma}\label{monlemma}
Let $\a\colon \X^\ast \times \R\to\overline{\R}$ be a function that is increasing in its second argument. Let $\beta$ denote its left inverse.\\
	(i) Let $A\subseteq \X^\ast$ be a nonempty set and $r\colon\X^\ast\rightarrow \overline{\R}$ a function. Then, we have
 \begin{equation}\label{eq:monlemma1}
		\inf\cb{m\in \R \mid \forall x^\ast \in A\colon r(x^\ast)\leq \alpha(x^\ast,m)}=\sup_{x^\ast\in A}\beta(x^\ast,r(x^\ast)).
	\end{equation}
	(ii) Let $B$ be a nonempty set and $r\colon \X^\ast\times B\rightarrow \overline{\R}$ a function. Then, for every $x^\ast\in \X^\ast$,
	\begin{equation*}
		\inf\cb{m\in \R \mid \forall s \in B\colon r(x^\ast,s)\leq \alpha(x^*,m)}=\sup_{s\in B}\beta(x^\ast,r(x^\ast,s)).
	\end{equation*}
\end{lemma}

\begin{proof}
Let us prove (i). Let $\overline{m}$ denote the infimum on the left of \eqref{eq:monlemma1}. By \Cref{leftinv}, 
\eqref{eq:monlemma1} is equivalent to
\begin{equation} \label{monlemmax}
	\overline{m}=\sup_{x^\ast\in A}\inf\cb{m\in \R \mid  r(x^\ast)\leq \alpha(x^*,m)} .
\end{equation}
The $\geq$ part is immediate by weak duality. For the $\leq$ part, to get a contradiction, assume that there exists $\tilde{m}\in \R$ such that
\begin{equation}\label{c1115}
	\overline{m}>\tilde{m}>\sup_{x^\ast \in A}\inf\cb{m\in \R \mid r(x^\ast)\leq \alpha(x^\ast,m)}.
\end{equation}
The first inequality in \eqref{c1115} implies that there exists $\tilde{x}^\ast \in A$ such that
$r(\tilde{x}^\ast)> \alpha(\tilde{x}^\ast,\tilde{m})$. The second inequality in \eqref{c1115} implies that $\tilde{m}>\inf\{m\in\R\mid r(\tilde{x}^\ast)\leq \a(\tilde{x}^\ast,m)\}$. Hence, by the monotonicity of $\a$, we must have $r(\tilde{x}^\ast)\leq \a(\tilde{x}^\ast,\tilde{m})$, a contradiction. Thus, \eqref{monlemmax} follows. The proof of (ii) is similar, hence omitted.
\end{proof}

	We state the dual representation theorem for lower semicontinuous quasiconvex functions, which is a part of \cite[Thm. 3]{kupper}. It is formulated in terms of the left inverse of the penalty function. We provide its short proof for completeness.
	
	\begin{theorem}\label{firstthm}
		Suppose that $f\colon \mathcal{X}\rightarrow \overline{\mathbb{R}}$ is a decreasing, lower semicontinuous and quasiconvex function. Then, $f$ has the dual representation 
		\begin{equation}\label{qconvexdualrep}
			f(x)=\sup_{x^\ast\in C^{\circ}\setminus \{0\}} \beta_f\of{x^\ast,\ip{ x^\ast,x}},\quad x\in \X.
		\end{equation}
	\end{theorem}
 
\begin{proof}
Let $x\in \X$. By \eqref{remf} and \Cref{separationlem}, we have
\[
f(x)=\inf\{m\in\R\mid x\in S_f^m\}=\inf\cb{m\in \R \mid \forall x^\ast \in C^{\circ}\setminus\{0\}\colon \ip{x^\ast,x} \leq \alpha_f(x^\ast,m) }.
\]
Since $\alpha_f$ is increasing by \Cref{nondeca}, we may apply \Cref{monlemma}(i), from which \eqref{qconvexdualrep} follows.
\end{proof}

	In \cite{kupper}, a decreasing quasiconvex function on $\X$ is called a \emph{risk measure} as a generalization of convex and coherent risk measures studied in the financial mathematics literature; see \cite[Ch. 4]{follmer}, for instance. Hence, \Cref{firstthm} provides a dual representation for a lower semicontinuous (quasiconvex) risk measure.
	
	In applications, it might be necessary to consider a function that is defined on a subset of the vector space $\X$. The next corollary is for this purpose, which is proved in \Cref{sec:app2}. To that end, let $\mathcal{K}\subseteq \X$ be a monotone convex set. Given a function $g\colon \mathcal{K}\to\overline{\R}$, we extend $g$ as a function $\bar{g}\colon\X\to\overline{\R}$ by setting $\bar{g}(x)\coloneqq g(x)$ for $x\in\mathcal{K}$ and $\bar{g}(x)\coloneqq+\infty$ for $x\in\X\setminus\mathcal{K}$. Then, the sublevel sets, penalty function, and algebraic properties (quasiconvexity, monotonicity, etc.) of $g$ are defined as those of $\bar{g}$.
	
	\begin{corollary}\label{cor:singleconv}
		Let $g\colon \mathcal{K}\to\overline{\R}$ be a quasiconvex, decreasing and lower semicontinuous (with respect to the relative topology) function. 
		Then, we have 
		\begin{equation}\label{eq:convexrest}
			g(x)=\sup_{x^\ast\in C^{\circ}\setminus \{0\}} \beta_g\of{x^\ast,\ip{ x^\ast,x}}, \quad x\in \mathcal{K}.
		\end{equation}
	\end{corollary}
	
	When $f$ is a proper lower semicontinuous convex function, two dual representations are possible: the one provided by Fenchel-Moreau theorem, and the one provided by \Cref{firstthm} since $f$ is also quasiconvex. To establish the link between the two representations, we calculate the left inverse of the penalty function in terms of the conjugate function in the next proposition. Its proof is given in \Cref{sec:app2}.
	
\begin{proposition}\label{convexcase}
		Assume that $C^\#\neq\emptyset$. Suppose that $f$ is proper, decreasing, convex, and lower semicontinuous.\\
  (i) For every $x^\ast\in\X^\ast\setminus\{0\}$, $m\in\R$ such that $\{x\in \X\mid f(x)<m\}\neq\emptyset$, we have
\[
			\alpha_f(x^\ast,m)= I^\ast_{\dom f}(x^\ast) \wedge \inf_{\lambda >0}\of{\lambda m+ \lambda f^\ast\of{\frac{x^*}{\lambda}}}.
		\]
  In particular, if $\dom f=\mathcal{X}$, then we have
  \[
			\alpha_f(x^\ast,m)= \inf_{\lambda >0}\of{\lambda m+ \lambda f^\ast\of{\frac{x^*}{\lambda}}}.
		\]
    (ii) For every $x^\ast\in\X^\ast\setminus\{0\}$, we have $\beta_f(x^\ast,s)=+\infty$ for every $s> I^\ast_{\dom f}(x^\ast)$ and
	   \begin{equation}\label{beta_f}
	   \beta_f(x^\ast,s)=\sup_{\gamma\geq 0}\of{\gamma s-f^\ast(\gamma x^\ast)}
		\end{equation}
  for every $s\leq I^\ast_{\dom f}(x^\ast)$. In particular, if $\dom f =\mathcal{X}$, then \eqref{beta_f} holds for every $s\in \R$.
	\end{proposition}
	
	\begin{remark}
		Under the assumptions of \Cref{convexcase}, we may rewrite the dual representation in \Cref{firstthm} using \Cref{convexcase} and the fact that $C^{\circ}$ is a cone, which gives
		\[
			f(x)
   = \sup_{x^\ast\in C^{\circ}\setminus \{0\}} \sup_{\gamma \geq 0}\of{\ip{\gamma x^\ast,x}-f^\ast(\gamma x^\ast) }
			=\sup_{x^*\in C^{\circ}} \of{ \ip{x^\ast,x}-f^*( x^\ast)}
		\]
		for each $x\in\dom f$ since $\ip{x^\ast,x}\leq I^{\ast}_{\dom f}(x^\ast)$. Hence, in the convex case, the representation in \Cref{firstthm} reproduces the standard Fenchel-Moreau-type representation in \Cref{fenchel}.
	\end{remark}
	
	\section{Naturally quasiconvex vector-valued functions}\label{sec:natquasi}
	
	Throughout this section, let $\X,\Y$ be Hausdorff locally convex topological vector spaces with vector preorders $\leq_C,\leq_D$, where $C\subseteq\X$ and $D\subseteq\Y$ are closed convex cones. We denote by $2^\Y$ the power set of $\Y$. Let $f\colon \Y\to\overline{\R}$ and $g\colon\X\to\Y$ be functions. Our goal is to provide a dual representation for a quasiconvex composition of the form $f\circ g$. While \Cref{sec:quasicon} provides the background for extended real-valued functions, we dedicate this section to vector-valued functions.
	
	We start by giving some generalized notions of convexity and monotonicity for vector-valued functions.
    
	\begin{definition}\label{Dconvex}
		Consider the following notions for $g\colon\X\to\Y$.\\
		(i) $g$ is called \emph{$D$-convex} if $g(\lambda x_1+(1-\lambda)x_2)\leq_D \lambda g(x_1)+(1-\lambda)g(x_2)$ for every $x_1,x_2\in\X$ and $\lambda\in (0,1)$.
		It is called \emph{$D$-concave} if $-g$ is $D$-convex.\\
		(ii) $g$ is called \emph{$D$-naturally quasiconvex} if, for every $x_1,x_2 \in \X$ and $\lambda \in [0,1]$, there exists $\mu \in [0,1]$ such that
		$g(\lambda x_1+ (1-\lambda) x_2)\leq_D \mu g(x_1)+(1-\mu)g(x_2)$.
		It is called \emph{$D$-naturally quasiconcave} if $-g$ is naturally $D$-quasiconvex.\\
		(iii) $g$ is called \emph{$D$-decreasing} if $x_1 \leq_C x_2$ implies $g(x_2)\leq_D g(x_1)$ for every $x_1,x_2\in \X$; it is called \emph{$D$-increasing} if $-g$ is $D$-decreasing.\\
		(iv) $g$ is called \emph{$D$-regularly decreasing} if it is $D$-decreasing and, for every $x_1,x_2 \in \X$, $x_1\leq_{C^{\#}} x_2$ implies $g(x_2)\leq_{D^{\#}} g(x_1)$; it is called \emph{$D$-regularly increasing} if $-g$ is $D$-regularly decreasing.
\end{definition}

From \Cref{Dconvex}, it is clear that $D$-convexity implies $D$-natural quasiconvexity. For real-valued functions with $D=\R_+$, $D$-natural quasiconvexity coincides with quasiconvexity; see the notes after \cite[Def.~2.1]{kur}.

For the main dual representation theorem (\Cref{mainthm}), we will need a notion of strict monotonicity for a vector-valued function and regular monotonicity is suitable for this purpose.  
Recall that $C^{\#}$ and $D^{\#}$ are the (convex) cones of strictly positive elements in $\X$ and $\Y$, respectively; see \eqref{sharp}. Although these cones are not closed in general, their induced preorders $\leq_{C^{\#}}$ and $\leq_{D^{\#}}$ are defined as in \eqref{preorder}. To be able to employ this definition, we work under the following assumption.

\begin{assumption}\label{asmp}
	The cones $C^{\#}$ and $D^{\#}$ are nonempty.
\end{assumption}

If $f\colon\Y\to\overline{\R}$ is decreasing and $g\colon\X\to\Y$ is $D$-increasing, then it is immediate that the composition $f\circ g\colon \X\to\overline{\R}$ is decreasing. Similarly, if $f$ is a decreasing convex function and $g$ is a $D$-concave function, then it can be checked that $f\circ g$ is a convex function. The following proposition provides an analogue of this observation for the quasiconvex case. We omit its simple proof.

\begin{proposition}\label{quasi2} Suppose that $f$ is quasiconvex and decreasing, and $g$ is $D$-naturally quasiconcave. Then, $f\circ g\colon\X\to \overline{\R}$ is quasiconvex.
\end{proposition}

We proceed with a continuity concept for $g$, which is defined through its set-valued extension $G\colon \X\to 2^\Y$ given by
\begin{equation}\label{eq:setext}
G(x)\coloneqq g(x)+D,\quad x\in\X.
\end{equation}
Given $M\subseteq\Y$, the sets
\[
G^L(M)\coloneqq \cb{x\in\X\mid G(x)\cap M\neq\emptyset},\quad G^U(M)\coloneqq \cb{x\in\X\mid G(x)\subseteq M}
\]
are called the \emph{lower inverse image} and \emph{upper inverse image} of $M$ under $G$, respectively. Note that $(G^{U}(M))^c=G^L(M^c)$ and $(G^L(M))^c=G^{U}(M^c)$.

\begin{definition}\label{ldc}
	(\cite[Def. 2.1]{ha}) The function $g$ is called $D$-lower demicontinuous if the lower inverse image $G^L(M)$ is open for every open halfspace $M\subseteq \Y$.
\end{definition}

When $\Y=\R$ and $D=\R_+$, note that \Cref{ldc} coincides with the usual notion of lower semicontinuity; see \Cref{remclosed}.

\begin{remark}\label{rem:ldc}
	Note that $g$ is $D$-lower demicontinuous if and only if the upper inverse image $G^{U}(M)$ is closed for every closed halfspace $M\subseteq\Y$. This follows from the observations that $M$ is a closed halfspace if and only if $M^c$ is an open halfspace, and that $G^{U}(M)= (G^L(M^c))^c$.
\end{remark}

\begin{remark}
    In general, the property in \Cref{ldc} is also referred to as the lower demicontinuity of a set-valued function $G\colon \X\to 2^{\Y}$. A stronger continuity concept for such $G\colon\X\to 2^{\Y}$ is lower hemicontinuity: $G$ is called \emph{lower hemicontinuous} if the lower inverse image $G^L(U)$ is open for every open set $U\subseteq \Y$; see \cite[Def.~17.2]{aliprantis}. In \cite[Ex.~17.39]{aliprantis}, it is shown that lower demicontinuity does not imply lower hemicontiuity in general. We work with the weaker notion of lower demicontinuity in this paper.
\end{remark}

For the function $g\colon \X\to\Y$, let us consider the \emph{scalarization} $y^{\ast}\circ g\colon\X\to\R$ defined by
\begin{equation}\label{eq:scalarization}
	y^{\ast}\circ g(x)\coloneqq \ip{y^\ast,g(x)},\quad x\in\X,
\end{equation}
for each $y^\ast\in D^{\circ}\setminus\{0\}$.	The next proposition provides useful characterizations of the convexity, quasiconvexity, monotonicity, and lower-demicontinuity of $g$ in terms of the analogous properties of the family of scalarizations; see \Cref{sec:app3} for the proof.

\begin{proposition}\label{star}
	We have the following equivalences for $g$ and its scalarizations.\\
		(i) $g$ is $D$-increasing if and only if $y^{\ast}\circ g$ is decreasing for every $y^\ast\in D^{\circ}\setminus\{0\}$.\\
		(ii) $g$ is $D$-concave if and only if $y^{\ast}\circ g$ is convex for every $y^\ast\in D^{\circ}\setminus\{0\}$.\\
		(iii) $g$ is $D$-naturally quasiconcave if and only if $y^{\ast}\circ g$ is quasiconvex for every $y^\ast \in D^{\circ}\setminus \{0\}$.\\
		(iv) $g$ is $D$-lower demicontinuous if and only if $y^{\ast}\circ g$ is lower semicontinuous for every $y^\ast\in D^{\circ}\setminus \{0\}$.
\end{proposition}

\begin{remark}
	The equivalent condition in \Cref{star}(ii) is sometimes called \emph{$\ast$-quasi-convexity}; see, e.g., \cite[Def. 2.1]{kur}. Conditions (ii), (iii) can be seen as modified versions of \cite[Prop. 2.2, Thm. 2.1]{kur}, which are stated in a set-valued setting there.
\end{remark}

Let $y^\ast\in D^{\circ}\setminus\{0\}$. In view of \Cref{star}, when $g$ is $D$-naturally quasiconcave, increasing and $D$-lower demicontinuous, the function $y^{\ast}\circ g$ is quasiconvex, decreasing and lower semicontinuous. In this case, we may apply \Cref{firstthm} for $y^{\ast}\circ g$ to get
\begin{equation}\label{dualh}
	y^{\ast}\circ g(x)=\sup_{x^\ast\in C^{\circ}\setminus \{0\}} \beta_{y^{\ast}\circ g}(x^\ast,\ip{x^\ast,x}), \quad x\in \X.
\end{equation}
The availability of \eqref{dualh} will be useful in \Cref{sec:comp} when obtaining dual representations for quasiconvex compositions.

\section{Quasiconvex compositions}\label{sec:comp}

In this section, we establish dual representations for quasiconvex compositions. We continue working in the framework of \Cref{sec:natquasi}, where we have locally convex topological vector spaces $\X,\Y$ with respective preorders $\leq_C,\leq_D$.

\subsection{The dual representation}\label{sec:dual}

Let us fix two functions $f\colon\Y\to\overline{\R}$, $g\colon\X\to\Y$. To motivate the discussion, we make the following simple observation: if $f$ is decreasing and quasiconvex, and $g$ is $D$-increasing and $D$-naturally quasiconcave, then $f\circ g$ is decreasing and quasiconvex by \Cref{quasi2}. Hence, in view of \Cref{firstthm}, a dual representation for $f\circ g$ is readily available once $f\circ g$ is guaranteed to be lower semicontinuous. This is achieved in the next proposition by suitable continuity assumptions on $f,g$.

\begin{proposition}\label{fognat}
	Suppose that $f$ is decreasing, lower semicontinuous, and quasiconvex; and that $g$ is $D$-increasing, $D$-lower demicontinuous, and $D$-naturally quasiconcave. Then, $f\circ g$ is a decreasing, lower semicontinuous, and quasiconvex function. Moreover, for every $x\in\X$, we have
	\begin{equation}\label{dualcomp}
		f\circ g(x)= \sup_{y^\ast\in D^{\circ}\setminus \{0\}} \beta_f \of{y^\ast,\sup_{x^\ast\in C^{\circ}\setminus \{0\}} \beta_{y^{\ast}\circ g}\of{x^\ast,\ip{x^\ast,x}}}.
	\end{equation}
\end{proposition}

\begin{proof}
By \Cref{quasi2}, the function $f \circ g$ is decreasing and quasiconvex. Let us show that it is also lower semicontinuous. To that end, let $m\in\R$. Note that
\begin{equation}\label{eq:GU}
S_{f\circ g}^m=\{x\in \X\mid g(x)\in S^m_f\}=\{x\in \X \mid G(x)\subseteq S^m_f\}=G^{U}(S_f^m),
\end{equation}
where $G$ is the set-valued extension of $g$ defined by \eqref{eq:setext}. Here, only the second equality needs proof. Since $f$ is decreasing, $S_f^m$ is monotone. Let $x\in \X$ with $g(x)\in S^m_f$ and let $d\in D$. Since $S_f^m$ is monotone, we have $g(x)+d\in S_f^m$. As this is true for every $d\in D$, we have $G(x)=g(x)+D\subseteq S_f^m$. Conversely, let $x\in \X$ with $G(x)\subseteq S^m_f$. Since $0\in D$, we have $g(x)\in g(x)+D=G(x)\subseteq S_f^m$. These observations verify the second equality in \eqref{eq:GU}.

By \Cref{remclosed}, we may write $S_f^m= \bigcap_{M\in \mathcal{M}}M$, where $\mathcal{M}$ is the collection of all closed halfspaces $M$ such that $S_f^m\subseteq M$. Therefore,
\[
G^{U}(S_f^m)=G^{U}\of{\bigcap_{M\in\mathcal{M}}M}=\bigcap_{M\in \mathcal{M}}G^{U}(M).
\]
Since $g$ is $D$-lower demicontinuous, $G^{U}(M)$ is closed for each $M\in\mathcal{M}$. By \eqref{eq:GU}, it follows that $S_{f\circ g}^m = G^U(S_f^m)$ is closed. Therefore, $f\circ g$ is lower semicontinuous by \Cref{rem:ldc}.

By \Cref{firstthm}, we obtain the dual representation
\[
		f\circ g(x)=\sup_{x^\ast\in C^{\circ}\setminus \{0\}} \beta_{f\circ g}\of{x^\ast,\ip{ x^\ast,x}},\quad x\in\X.
\]
To prove \eqref{dualcomp}, let $x\in\X$. By applying \Cref{firstthm} for $f$ at the point $g(x)$, we get
\[
f(g(x))=\sup_{y^\ast\in D^{\circ}\setminus \{0\}} \beta_f(y^\ast,\ip{ y^\ast,g(x)}).
\]
On the other hand, by \eqref{dualh}, we have
\[
\ip{y^\ast,g(x)}=y^{\ast}\circ g (x)=\sup_{x^\ast\in C^{\circ}\setminus \{0\}} \beta_{y^{\ast}\circ g}(x^\ast,\ip{x^\ast,x}),\quad y^\ast\in D^{\circ}\setminus\{0\}.
\]
Combining the last two observations gives \eqref{dualcomp}.
\end{proof}

The next theorem shows that, under the same assumptions, the inner supremum in the dual representation in \Cref{fognat} can be combined with the outer one.

\begin{theorem}\label{newthm}
	Suppose that $f$ is decreasing, lower semicontinuous, and quasiconvex; and that $g$ is $D$-increasing, $D$-lower demicontinuous, and $D$-naturally quasiconcave. Then, for every $x\in\X$, we have
	\begin{equation*}
		f\circ g(x)=\sup_{x^\ast\in C^{\circ}\setminus\{0\}} \sup_{y^\ast\in D^{\circ}\setminus\{0\}}\beta_f\Big(y^\ast,\beta_{y^{\ast}\circ g}\of{x^\ast,\ip{x^\ast,x}}\Big).
	\end{equation*}
\end{theorem}

\begin{proof}
Let $x\in\X$. By \eqref{remf}, \Cref{separationlem}, and \Cref{monlemma}(i), we have
\begin{align*}
f\circ g(x)&=\inf \{m\in \R\mid  g(x)\in S_f^m\}\notag \\
&=\inf \cb{m\in \R \mid\forall y^\ast \in D^{\circ}\setminus\{0\} \colon \ip{y^\ast,g(x)}\leq \alpha_f(y^\ast,m)}\notag \\
&=\sup_{y^\ast \in D^{\circ}\setminus\{0\}}\inf\cb{m\in \R \mid \ip{y^\ast,g(x)}\leq \alpha_f(y^\ast,m)}.
\end{align*}
Then, by using \eqref{dualh} and then applying \Cref{monlemma}(i), we obtain
\begin{align*}
f\circ g(x)
&=\sup_{y^\ast \in D^{\circ}\setminus\{0\}}\inf\cb{m\in \R \mid \ip{y^\ast,g(x)}\leq \alpha_f(y^\ast,m)}\\
&=\sup_{y^\ast \in D^{\circ}\setminus\{0\}}\inf\cb{m\in \R \mid \sup_{x^\ast \in C^{\circ}\setminus \{0\}} \beta_{y^\ast \circ g}(x^*,\ip{x^\ast,x})\leq \alpha_f(y^\ast,m)}\\
&=\sup_{y^\ast \in D^{\circ}\setminus\{0\}}\sup_{x^\ast \in C^{\circ}\setminus\{0\}}\inf\cb{m\in \R \mid \beta_{y^\ast \circ g}(x^*,\ip{x^\ast,x})\leq \alpha_f(y^\ast,m)}\\
&=\sup_{x^\ast\in C^{\circ}\setminus\{0\}} \sup_{y^\ast\in D^{\circ}\setminus\{0\}}\beta_f\Big(y^\ast,\beta_{y^\ast \circ g}\of{x^\ast,\ip{x^\ast,x}}\Big),
\end{align*}
which concludes the proof.
\end{proof} 

\subsection{The main theorem}\label{sec:main}

While Theorems~\ref{fognat},~\ref{newthm} provide dual representations for $f\circ g$, they do not provide formulae for the penalty function $\a_{f\circ g}$ as well as its left inverse $\beta_{f\circ g}$ in terms of the same type of functions for $f$ and $g$ (more precisely, the scalarizations of $g$). This problem will be addressed by \Cref{mainthm} and \Cref{cor1} below. It turns out that these results work under a mild compactness assumption on $D^\circ$ as we describe next.

\begin{definition}\label{defn:gen}
	A set $\bar{D}^{\circ}\subseteq D^{\circ}$ is called a \emph{cone generator} for $D^{\circ}$ if every $y^\ast\in D^{\circ}\setminus\{0\}$ can be written as $y^\ast=\lambda \bar{y}^\ast$ for some $\lambda >0$ and $\bar{y}^\ast\in \bar{D}^{\circ}$.
\end{definition}

	It is clear that if $\bar{D}^{\circ}$ is a cone generator for $D^{\circ}$, then $D^{\circ}$ is the conic hull of $\bar{D}^{\circ}$.

\begin{remark}\label{base}
	Suppose that $D^{\#}\neq\emptyset$ and let $\pi\in D^{\#}$. Then, $D^{\circ}_{\pi}$ is a closed convex cone generator for $D^{\circ}$ thanks to \Cref{projection}.
\end{remark}

In \Cref{sec:cone}, we will discuss the existence and compactness of cone generators for several examples that show up frequently in applications. For the theoretical development of this section, we work under the following assumption.

\begin{assumption}\label{asmp:cone}
	There exists a convex and compact cone generator $\bar{D}^{\circ}$ for $D^{\circ}$.
\end{assumption}

Now, we state the main theorem of the paper, which provides a formula for the penalty function of $f\circ g$. 
\begin{theorem}\label{mainthm}
	Suppose that Assumptions~\ref{asmp},~\ref{asmp:cone} hold. In addition, suppose that $f$ is decreasing, lower semicontinuous, and quasiconvex; and that $g$ is $D$-regularly increasing, $D$-lower demicontinuous, and $D$-naturally quasiconcave. Then, for every $x^\ast\in C^{\circ}\setminus\{0\}$ and $m\in\R$, we have
	\[
		\alpha_{f\circ g} (x^\ast,m)=\inf_{y^\ast\in D^{\circ}\setminus\{0\}} \alpha_{y^{\ast}\circ g}\of{x^\ast,\alpha_f(y^\ast,m)}=\inf_{y^\ast\in D_\pi^{\circ}\setminus\{0\}} \alpha_{y^{\ast}\circ g}\of{x^\ast,\alpha_f(y^\ast,m)}.
	\]
\end{theorem}

\begin{remark}
In \Cref{mainthm}, we do not require $\bar{D}^{\circ}$ to be the same as $D^{\circ}_{\pi}$.
\end{remark}

The proof of \Cref{mainthm} consists of several auxiliary results together with the use of a minimax inequality in \cite{Liu} for two functions. \Cref{asmp:cone} will be crucial in applying this inequality. The proofs of the auxiliary results will be provided in \Cref{app:comp}. We begin with some notations. Given $m\in\R$ and $y^\ast \in D^{\circ}$, let us define
\[
A^m_{y^\ast}\coloneqq \cb{x\in \X\mid \ip{y^\ast,g(x)}\leq \alpha_f(y^\ast,m)},\quad 
\tilde{A}^m_{y^\ast}\coloneqq \cb{x\in \X\mid \ip{y^\ast,g(x)} < \alpha_f(y^\ast,m)}.
\]
Clearly, $\tilde{A}^m_{y^\ast}\subseteq A^m_{y^\ast}$ and $A^m_{y^\ast}$ is the $\alpha_f(y^\ast,m)$-sublevel set of $y^{\ast}\circ g$; see \eqref{eq:scalarization}. Therefore, when $g$ is $D$-increasing, $D$-naturally quasiconcave, and $D$-lower demicontinuous, the set $A^m_{y^\ast}$ is closed, convex, and monotone set by \Cref{star}. We give the precise relationship between the sets $\tilde{A}^m_{y^\ast}$ and $A^m_{y^\ast}$ in the following proposition.

\begin{proposition}\label{chul}
Suppose that \Cref{asmp} holds. In addition, suppose that $g$ is $D$-regularly increasing, $D$-naturally quasiconcave, and $D$-lower demicontinuous. Let $m\in\R$ and $y^\ast\in D^{\circ}\setminus \{0\}$. Then,
\begin{equation}\label{eq:chull}
	A^m_{y^\ast}=\cl (\tilde{A}^m_{y^\ast})=\cl\conv (\tilde{A}^m_{y^\ast}).
\end{equation}
\end{proposition}

\begin{remark}\label{remcg}
Under \Cref{asmp:cone}, let $m\in\R$, $y^\ast\in D^{\circ}\setminus\{0\}$. We may write $y^\ast=\lambda \bar{y}^\ast$ for some $\lambda>0$ and $\bar{y}^\ast\in \bar{D}^{\circ}$. Then, it is easy to see that
$A^m_{y^\ast}=A^m_{\bar{y}^\ast}$.
\end{remark}

Next, under \Cref{asmp:cone}, given $m\in\R$ and $x^\ast\in C^{\circ}$, we define two auxiliary functions $K^m_{x^\ast}, \tilde{K}^m_{x^\ast}\colon \X\times \bar{D}^{\circ}\to\overline{\R}$ by
\begin{equation}\label{KK}
K^m_{x^\ast}(x,y^\ast)=\ip{x^\ast,x}-I_{A^m_{y^\ast}}(x),\qquad 
\tilde{K}^m_{x^\ast}(x,y^\ast)=\ip{x^\ast,x}-I_{\tilde{A}^m_{y^\ast}}(x),
\end{equation}
for each $(x,y^\ast)\in\X\times\bar{D}^\circ$. The next proposition shows the relationship between these two functions.

\begin{proposition}\label{kkt}
Let $m\in\R$, $x^\ast\in C^{\circ}$. Suppose that Assumptions~\ref{asmp},~\ref{asmp:cone} hold. In addition, suppose that $g$ is $D$-regularly increasing, $D$-naturally quasiconcave, and $D$-lower demicontinuous. Then, for each $y^\ast\in\bar{D}^{\circ}$, we have 
\[
\sup_{x\in \X}\tilde{K}^m_{x^\ast}(x,y^\ast)=\sup_{x\in \X}K^m_{x^\ast}(x,y^\ast).
\]
\end{proposition}

We will use a minimax theorem in the proof of \Cref{mainthm}. As a preparation, we check some properties of the functions defined in \eqref{KK}; these will be needed for the application of the minimax theorem.

\begin{proposition}\label{propk}
Let $m\in\R$, $x^\ast\in C^{\circ}$. Suppose that $g$ is $D$-naturally quasiconcave. Then, the following properties hold.\\
	(i) Suppose further that $g$ is $D$-lower demicontinuous. Then, $K^m_{x^\ast}$ is concave and upper semicontinuous in its first argument, and quasiconvex in its second argument.\\
	(ii) $\tilde{K}^m_{x^\ast}$ is concave in its first argument, and quasiconvex and lower semicontinuous in its second argument.
\end{proposition}

The next two propositions relate the functions given in \eqref{KK} to the main problem.

\begin{proposition}\label{alfamin}
Suppose that \Cref{asmp:cone} holds. In addition, suppose that $f$ is decreasing, lower semicontinuous, and quasiconvex; and that $g$ is $D$-naturally quasiconcave and $D$-lower demicontinuous. Then, for each $x^\ast\in C^{\circ}$, $m\in \R$,
\begin{equation*}
	\alpha_{f\circ g} (x^\ast,m)
	=\sup_{x\in \X} \inf_{y^\ast\in \bar{D}^{\circ}}K^m_{x^\ast}(x,y^\ast).
\end{equation*}
\end{proposition}
\begin{proposition}\label{infhg}
Suppose that \Cref{asmp:cone} holds. Let $x^\ast\in C^{\circ},m\in \R$. Then,
\[
	\inf_{y^\ast\in D^{\circ}\setminus \{0\}} \alpha_{y^{\ast}\circ g}(x^\ast,\alpha_f(y^\ast,m))= \inf_{y^\ast\in \bar{D}^{\circ}} \alpha_{y^{\ast}\circ g}(x^\ast,\alpha_f(y^\ast,m))=\inf_{y^\ast\in \bar{D}^{\circ}} \sup_{x\in \X} K^m_{x^\ast}(x,y^\ast).
\]
\end{proposition}

We will use a nonstandard minimax inequality in the proof of the main result. For completeness, we provide its statement from \cite{Liu}.

\begin{theorem}[Liu (1978) \cite{Liu}]\label{liuthm}
Let $\mathcal{U}, \mathcal{V}$ be nonempty convex sets of two topological vector spaces, and consider two functions $f,\tilde{f}\colon \mathcal{U}\times\mathcal{V}\to\overline{\R}$ satisfying the following conditions:\\
(i) $f$ is upper semicontinuous in its first argument and quasiconvex in its second argument.\\
(ii) $\tilde{f}$ is quasiconcave in its first argument and lower semicontinuous in its second argument.\\
(iii) $\tilde{f}(u,v)\leq f(u,v)$ for every $u\in \mathcal{U}$ and $v\in \mathcal{V}$.\\
(iv) $\mathcal{U}$ is compact.\\
	Then, we have
	\begin{equation*}
	\inf_{u\in \mathcal{U}}\sup_{v\in \mathcal{V}} \tilde{f}(u,v)\leq \sup_{v\in \mathcal{V}}\inf_{u\in \mathcal{U}}f(u,v).
	\end{equation*}
\end{theorem}

With the tools developed above, we are ready to prove the main theorem.

\begin{proof}[Proof of \Cref{mainthm}]
Let $x^\ast\in C^{\circ}\setminus\{0\}$ and $m\in\R$. For each $y^\ast\in \bar{D}^{\circ}$ and $x\in\X$, since $\tilde{A}^m_{y^\ast}\subseteq A^m_{y^\ast}$, we have $I_{\tilde{A}^m_{y^\ast}}(x)\geq I_{A^m_{y^\ast}}(x)$ so that 
\begin{equation}\label{ineqK}
\tilde{K}^m_{x^\ast}(x,y^\ast)\leq K^m_{x^\ast}(x,y^\ast).
\end{equation}
By \Cref{propk}, $K^m_{x^\ast}$ is upper semicontinuous and concave in its first variable, and quasiconvex in its second variable; $\tilde{K}^m_{x^*}$ is concave in its first variable, and quasiconvex and lower semicontinuous in its second variable. These properties, together with \eqref{ineqK}, and the convexity and compactness of $\bar{D}^{\circ}$ are sufficient to apply \Cref{liuthm} (see also \cite[Thm. 3.1]{liurelated2} and \cite[Cor. 11]{liurelated1}) to the functions $K^m_{x^\ast}, \tilde{K}^m_{x^\ast}$. Consequently, we obtain
\begin{equation}\label{thmmm}
\inf_{y^\ast\in \bar{D}^{\circ}}\sup_{x\in \X}\tilde{K}^m_{x^\ast}(x,y^\ast) \leq \sup_{x\in \X}\inf_{y^\ast\in \bar{D}^{\circ}} K^m_{x^\ast}(x,y^\ast).
\end{equation}
By \Cref{kkt}, we have
\[
\sup_{x\in \X} \tilde{K}^m_{x^\ast}(x,y^\ast)=\sup_{x\in \X} K^m_{x^\ast}(x,y^\ast).
\] 
Hence, \eqref{thmmm} yields
\[
\inf_{y^\ast\in \bar{D}^{\circ}} \sup_{x\in \X} K^m_{x^\ast}(x,y^\ast)\leq\sup_{x\in \X} \inf_{y^\ast\in \bar{D}^{\circ}} K^m_{x^\ast}(x,y^\ast).
\]
However, the reverse inequality already holds by weak duality. Therefore, we get 
\[
\inf_{y^\ast\in \bar{D}^{\circ}} \sup_{x\in \X} K^m_{x^\ast}(x,y^\ast)=\sup_{x\in \X} \inf_{y^\ast\in \bar{D}^{\circ}} K^m_{x^\ast}(x,y^\ast).
\]
Moreover, by Propositions \ref{alfamin}, \ref{infhg}, we have 
\begin{align*}
\alpha_{f\circ g} (x^\ast,m)&=\sup_{x\in \X} \inf_{y^\ast\in \bar{D}^{\circ}}K^m_{x^\ast}(x,y^\ast)=\inf_{y^\ast\in \bar{D}^{\circ}} \sup_{x\in \X} K^m_{x^\ast}(x,y^\ast)\\
&=\inf_{y^\ast\in D^{\circ}\setminus\{0\}} \alpha_{y^{\ast}\circ g}(x^\ast,\alpha_f(y^\ast,m))=\inf_{y^\ast\in \bar{D}^{\circ}} \alpha_{y^{\ast}\circ g}(x^\ast,\alpha_f(y^\ast,m)).
\end{align*}
Finally, by \Cref{base} and \Cref{infhg} applied to $D^{\circ}_{\pi}$, we have
\[
\alpha_{f\circ g} (x^\ast,m)=\underset{y^\ast\in D^{\circ}_{\pi}}{\inf} \alpha_{y^\ast\circ g}\of{x^\ast,\alpha_f\of{y^\ast,m}},
\]
which completes the proof.
\end{proof}

The next corollary complements \Cref{mainthm} by providing a formula for the left inverse of the penalty function of $f\circ g$, which is the actual function that shows up in the dual representation of $f\circ g$ in \Cref{fognat}.

\begin{corollary}\label{cor1}
In the setting of \Cref{mainthm}, for every $x^\ast\in C^{\circ}\setminus\{0\}$, $s\in\R$,
\[
	\beta_{f\circ g } (x^\ast,s)=\sup_{y^\ast\in D^{\circ}\setminus\{0\}}\beta_f\Big(y^\ast,\beta_{y^{\ast}\circ g}(x^\ast,s)\Big).
\]
\end{corollary}

\begin{proof}
Let $x^\ast\in C^{\circ}\setminus\{0\}$, $s\in\R$. By the definition of left inverse and \Cref{mainthm},
\begin{align*}
\beta_{f\circ g }(x^\ast,s)&=\inf\cb{m\in \R \mid \alpha_{f \circ g}(x^\ast,m)\geq s}\\
&= \inf\cb{m\in \R\mid \forall y^\ast\in D^{\circ}\setminus\{0\}\colon \alpha_{y^{\ast}\circ g}(x^\ast,\alpha_f(y^\ast,m))\geq s}.
\end{align*}

We claim that the following minimax equality holds:
\begin{align}
&\inf\cb{m\in \R\mid \forall y^\ast\in D^{\circ}\setminus\{0\}\colon \alpha_{y^{\ast}\circ g}(x^\ast,\alpha_f(y^\ast,m))\geq s}\notag\\
&=\sup_{y^\ast\in D^{\circ}\setminus\{0\}}\inf\cb{m\in \R\mid \alpha_{y^{\ast}\circ g}(x^\ast,\alpha_f(y^\ast,m))\geq s}.\label{eq:minmax}
\end{align}
The $\geq$ part of this inequality holds as a weak duality property. Next, we show the $\leq$ part. To get a contradiction, suppose that there exists $\bar{m}\in\R$ such that
\begin{align}\label{eq:mbar}
&\inf\cb{m\in \R\mid \forall y^\ast\in D^{\circ}\setminus\{0\}\colon \alpha_{y^{\ast}\circ g}(x^\ast,\alpha_f(y^\ast,m))\geq s}\notag \\
&>\bar{m}>\sup_{y^\ast\in D^{\circ}\setminus\{0\}}\inf\cb{m\in \R\mid \alpha_{y^{\ast}\circ g}(x^\ast,\alpha_f(y^\ast,m))\geq s}.
\end{align}
The first inequality in \eqref{eq:mbar} implies the existence of $\bar{y}^\ast\in D^{\circ}\setminus\{0\}$ satisfying
\begin{equation}\label{eqcor11}
\alpha_{\bar{y}^\ast \circ g}(x^\ast,\alpha_f(\bar{y}^\ast,\bar{m}))< s.
\end{equation}
On the other hand, the second inequality in \eqref{eq:mbar} implies that
\[
\bar{m}>\inf\{m\in \R\mid \alpha_{\bar{y}^\ast\circ g} (x^\ast, \alpha_f(\bar{y}^\ast,m)) \geq s\}.
\]
Hence, there exists $m_{\bar{y}^*}<\bar{m}$ such that
\begin{equation}\label{eqcor12}
\alpha_{\bar{y}^\ast \circ g}(x^\ast,\alpha_f(\bar{y}^\ast,m_{\bar{y}^\ast}))\geq s.
\end{equation}
Since $\alpha_f$ is increasing in the second argument by \Cref{nondeca}, we have
\[
\alpha_f(\bar{y}^\ast,\bar{m})\geq \alpha_f(\bar{y}^\ast,m_{\bar{y}^\ast}).
\]
Hence, by \eqref{eqcor12}, the monotonicity of $\alpha_{\bar{y}^\ast \circ g}$, and \eqref{eqcor11}, we obtain
\[
s \leq  \alpha_{\bar{y}^\ast \circ g}(x^\ast,\alpha_f(\bar{y}^\ast,m_{\bar{y}^\ast})) \leq \alpha_{\bar{y}^\ast \circ g}(x^\ast,\alpha_f(\bar{y}^\ast,\bar{m})) <s,
\]
which is a contradiction. Hence, \eqref{eq:minmax} follows so that 
\begin{equation}\label{eq:conc}
\beta_{f\circ g }(x^\ast,s) = \sup_{y^\ast\in D^{\circ}\setminus\{0\}}\inf\cb{m\in \R\mid \alpha_{y^{\ast}\circ g}(x^\ast,\alpha_f(y^\ast,m))\geq s}.
\end{equation}

Let $y^\ast\in D^{\circ}\setminus\{0\}$. We claim that 
\[
\inf\cb{m\in \R\mid \alpha_{y^{\ast}\circ g}(x^\ast,\alpha_f(y^\ast,m))\geq s}
=\inf \cb{m\in \R \mid \alpha_f(y^\ast,m)\geq \beta_{y^{\ast}\circ g}(x^\ast,s)}.
\]
For each $m\in\R$, by the definition of left inverse, 
\[
\alpha_{y^{\ast}\circ g}(x^\ast,\alpha_f(y^\ast,m))\geq s\quad\Rightarrow\quad  \alpha_f(y^\ast,m)\geq \beta_{y^{\ast}\circ g}(x^\ast,s).
\]
Hence, the $\geq$ part of the claim follows. Next, we prove that $\leq$ part. To get a contradiction, suppose that 
\[
\inf\cb{m\in \R\mid \alpha_{y^{\ast}\circ g}(x^\ast,\alpha_f(y^\ast,m))\geq s}
>\tilde{m}>\inf \cb{m\in \R \mid \alpha_f(y^\ast,m)\geq \beta_{y^{\ast}\circ g}(x^\ast,s)}
\]
for some $\tilde{m}\in\R$. By the first inequality, we have $
\alpha_{y^{\ast}\circ g}(x^\ast,\alpha_f(y^\ast,\tilde{m}))< s$;
by the second inequality together with the monotonicity of $\a_f$, we have $\alpha_f(y^\ast,\tilde{m})\geq \beta_{y^{\ast}\circ g}(x^\ast,s)$.
Hence, by the monotonicity of $\alpha_{y^{\ast}\circ g}$,
\[
s\leq \alpha_{y^{\ast}\circ g}(x^\ast,\beta_{y^{\ast}\circ g}(x^\ast,s))\leq \alpha_{y^{\ast}\circ g}(x^\ast,\alpha_f(y^\ast,\tilde{m}))< s,
\]
a contradiction. Therefore, the claim follows.

Finally, combining \eqref{eq:conc} with the preceding claim gives
\begin{align*}
\beta_{f\circ g }(x^\ast,s)& = \sup_{y^\ast\in D^{\circ}\setminus\{0\}}\inf \cb{m\in \R \mid \alpha_f(y^\ast,m)\geq \beta_{y^{\ast}\circ g}(x^\ast,s)}
\\&=\sup_{y^\ast\in D^{\circ}\setminus\{0\}}\beta_f(y^\ast,\beta_{y^{\ast}\circ g}(x^\ast,s)),
\end{align*}
which finishes the proof.
\end{proof}

\begin{remark}
Let $\mathcal{R}^{\max}$ denote the set of all functions $\beta\colon C^{\circ}\times \R \to [-\infty,\infty]$ that satisfy the following properties:\\
    (i) $\beta$ is increasing and left-continuous in its second argument;\\
    (ii) $\beta$ is jointly quasiconcave;\\
    (iii) $\beta (\lambda x^\ast,s)=\beta(x^\ast,\frac{s}{\lambda})$ for every $x^\ast \in C^\circ$, $s\in \R$, and $\lambda>0$;\\
    (iv) $\beta$ has a uniform asymptotic minimum, i.e., $\lim_{s\to -\infty} \beta(x^\ast,s)=\lim_{s\to -\infty} \beta(z^\ast,s)$ for every $x^\ast,z^\ast \in C^\circ$;\\
    (v) the right-continuous version $(x^\ast,s)\mapsto \beta^+(x^\ast,s)\coloneqq \inf_{s'>s}\beta(x^\ast,s')$ is upper semicontinuous in its first argument.\\
In \cite[Thm.~3]{kupper}, it is shown that $\beta_{f\circ g}$ is unique in $\mathcal{R}^{\max}$ for the dual representation of $f\circ g$ in the sense of \Cref{firstthm}, and $\beta_{f\circ g}$ is indeed the left inverse of the minimal penalty function, i.e., $\alpha_{f\circ g}$. Hence, \Cref{cor1} provides the formula for the calculation of this unique function $\beta_{f\circ g}$, whereas \Cref{mainthm} gives the formula for the associated penalty function in the sense of \Cref{penaltydefn}.
\end{remark}

\subsection{Two important special cases}\label{sec:special}

We consider special cases of the setting in \Cref{sec:main} where at least one of the functions in the composition is convex/concave. In these cases, we can obtain simplified formulae for the penalty function of the composition. The proofs will be given in \Cref{app:special}. As before, we work with two functions $f\colon\Y\to\overline{\R}$, $g\colon \X\to\Y$.

We first work on the case where both $f$ and $g$ satisfy a stronger convexity assumption so that $f\circ g$ becomes convex. As the next corollary shows, the reduced form of the dual representation is consistent with the ones available for convex compositions in the literature; see, for instance, \cite[Thm. 2.8.10]{zal} and \cite[Thm. 3]{bot}.

\begin{corollary}\label{bothconvex}
Suppose that $f\colon\Y\to \overline{\R}$ is convex, decreasing and lower semicontinuous; and that $g$ is $D$-increasing, $D$-lower demicontinuous, and $D$-concave. Then, for each $x\in\X$ such that $g(x)\in \dom f$, we have
\[
f\circ g(x)=\sup_{x^\ast\in C^{\circ}}\sup_{y^\ast\in D^{\circ}}\of{\ip{ x^\ast,x}-(y^{\ast}\circ g)^\ast(x^\ast)-f^\ast(y^\ast)}.
\]
\end{corollary}

Next, we work on the case where only one of the functions in the composition has a stronger convexity assumption. While \Cref{bothconvex} reproduces earlier results in the literature, the next result is novel to this work to the best of our knowledge. 

\begin{proposition}\label{gconvex}
Suppose that $f$ is decreasing, lower semicontinuous, and quasiconvex; and that $g$ is $D$-increasing, $D$-lower demicontinuous, and $D$-concave. Then, $f\circ g$ is a decreasing, lower semicontinuous, and quasiconvex function; moreover, the following dual representation holds for each $x\in\X$:
\begin{equation}\label{gconvexdual}
	f\circ g(x)=\sup_{x^\ast\in C^{\circ}\setminus\{0\}} \sup_{y^\ast\in D^{\circ}\setminus\{0\}}\beta_f\Big(y^\ast,\ip{x^\ast,x}-(y^{\ast}\circ g)^\ast(x^\ast)\Big).
\end{equation}
Suppose further that $g$ is also $D$-regularly increasing and Assumptions~\ref{asmp}, \ref{asmp:cone} hold. Then, we have the following:\\
	(i) Let $x^\ast\in C^{\circ}\setminus\{0\}$, $m\in\R$ with $\alpha_f(y^\ast,m)\in\R$ and $A^m_{y^\ast}\neq\emptyset$
 for every $y^*\in D^{\circ}\setminus\{0\}$. Then,
	\begin{equation*}
		\alpha_{f\circ g } (x^\ast,m)=\inf_{y^\ast\in D^{\circ}\setminus\{0\}}\big((y^{\ast}\circ g)^\ast(x^\ast)+\alpha_f(y^\ast,m)\big).
	\end{equation*}
	(ii) For every $x^\ast\in C^{\circ}\setminus\{0\}$ and $s\in\R$,
	\begin{equation*}
		\beta_{f\circ g } (x^\ast,s)=\sup_{y^\ast\in D^{\circ}\setminus\{0\}}\beta_f\big(y^\ast,-(y^{\ast}\circ g)^\ast(0) \vee  (s-(y^{\ast}\circ g)^\ast(x^\ast))\big).
	\end{equation*}
\end{proposition}

For a linear operator $T\colon\mathcal{X} \to \mathcal{Y}$ and its adjoint operator $T^\ast\colon \mathcal{Y}^\ast \to \mathcal{X}^\ast$, we have $\ip{y^\ast,Tx}=\ip{T^\ast y^\ast,x}$ for every $x\in \X$ and $y^\ast \in \mathcal{Y}^\ast$. In the following corollary, the dual representation will be given when the function $g$ is a linear operator.

\begin{corollary}\label{cor:linop}
    Let $T\colon \X \to \Y$ be a $D$-increasing linear operator with adjoint operator $T^\ast$. Then, the following dual representation holds for every $x\in\X$:
    \begin{equation*}
        f\circ T(x)= \sup_{y^\ast\in D^{\circ}\setminus\{0\}} \beta_f(y^\ast,\ip{T^\ast y^\ast,x})
    \end{equation*}
Furthermore, if $T$ is $D$-regularly increasing with $\alpha_f(y^\ast,m)\in\R$ and  $A_{y^\ast}^m\neq\emptyset$ for every $y^*\in D^{\circ}\setminus\{0\}$, then
    \begin{equation*}
        \alpha_{f\circ T}(x^\ast,m)=\inf_{\{y^\ast \in D^\circ \setminus \{0\}\colon T^\ast x^\ast = y^\ast\}}\alpha_f(y^\ast,m).
    \end{equation*}
\end{corollary}

\subsection{Quasiconvex composition on a convex set}\label{sec:quasi-con}

We turn our attention to the case where the composition is considered on a monotone convex set $\mathcal{K}\subseteq \X$ with $C\subseteq\mathcal{K}$, see \Cref{cor:singleconv}, the analogous result for a single function. 

We work with two functions $f\colon\Y\to\overline{\R}$ and $g\colon \mathcal{K}\to \Y$. The following results extend \Cref{mainthm} and \Cref{newthm}. Their proofs are given in \Cref{sec:appsec4}.

\begin{corollary}\label{convexthm}
Suppose that $f$ is decreasing, lower semicontinuous, and quasiconvex; and that $g$ is regularly increasing, $D$-lower demicontinuous (with respect to the relative topology), and $D$-naturally quasiconcave. Then, $f\circ g$ is a decreasing, lower semicontinuous, and quasiconvex function. Moreover, for each $x^\ast\in C^{\circ}\setminus\{0\}$ and $m\in\R$, we have
\[
\alpha_{f\circ g} (x^\ast,m)=\inf_{y^\ast\in D^{\circ}\setminus\{0\}} \alpha_{y^{\ast}\circ g}\of{x^\ast,\alpha_f(y^\ast,m)}.
\]
\end{corollary}

\begin{proposition}\label{convexcor}
Suppose that $f$ is decreasing, lower semicontinuous, and quasiconvex; and that $g$ is increasing, $D$-lower demicontinuous (with respect to the relative topology), and $D$-naturally quasiconcave. Then, we have
\begin{equation}\label{dualcompcon}
	f\circ g(x)= \sup_{x^\ast\in C^{\circ}\setminus \{0\}} \beta_{f\circ g}\of{x^\ast,\ip{ x^\ast,x}}, \quad x\in \mathcal{K},
\end{equation}
and
\begin{equation}\label{dualcompcon2}
	f\circ g(x)=\sup_{x^\ast\in C^{\circ}\setminus\{0\}} \sup_{y^\ast\in D^{\circ}\setminus\{0\}}\beta_f\Big(y^\ast,\beta_{y^{\ast}\circ g}\of{x^\ast,\ip{x^\ast,x}}\Big), \quad x\in \mathcal{K}.
\end{equation}
\end{proposition}	

For a more specific case, we have the following proposition.
\begin{proposition}\label{convexconvexcor}
Suppose that $f$ is decreasing, lower semicontinuous, and quasiconvex; and that $g$ is increasing, $D$-lower demicontinuous (with respect to the relative topology), and concave. Then,
\begin{equation}\label{gconvexdualxd}
	f\circ g(x)=\sup_{x^\ast\in C^{\circ}\setminus\{0\}} \sup_{y^\ast\in D^{\circ}\setminus\{0\}}\beta_f\Big(y^\ast,\ip{x^\ast,x}-(y^{\ast}\circ g)^\ast(x^\ast)\Big),\quad x\in\mathcal{K}.
\end{equation}
\end{proposition}

\section{Compact cone generators}\label{sec:cone}

In this section, we will discuss the existence of compact convex cone generators in some concrete spaces and show that \Cref{mainthm} is applicable in these examples.

As noted in \Cref{base}, $D^{\circ}_{\pi}$ is a closed convex generator but it is not always compact. However, we do not have to restrict ourselves to this generator and can search for other compact generators because after guaranteeing the existence of a compact convex cone generator $\bar{D}^{\circ}$, we can still work with $D^{\circ}_{\pi}$ thanks to the second equality in \Cref{mainthm}.

\subsection{Finite-dimensional spaces}

Let us take $\Y=\R^n$ with the Euclidean norm $\norm{\cdot}$. As a natural consequence, $\Y^\ast=\R^n$ with the same norm $\norm{\cdot}$. Let us choose a convex cone $D$ and denote the unit ball by $B=\cb{y\in \mathbb{R}^n: \norm{y}\leq 1 }$. We show the existence of a compact convex generator for $D^{\circ}$ so that we can use \Cref{mainthm} for the case $\Y=\R^n$.

\begin{proposition}\label{finitedim}
The set $\bar{D}^{\circ}\coloneqq D^{\circ} \cap B$ is a compact and convex cone generator for $D^+$.
\end{proposition}
\begin{proof}
Since $D^{\circ}$ and $B$ are closed and convex sets, their intersection is also closed and convex. Moreover, $B$ is compact since it is closed and bounded. By using this fact and that $\bar{D}^{\circ}$ is a closed subset of $B$, we conclude that $\bar{D}^{\circ}$ is compact. To show that $\bar{D}^{\circ}$ generates $D^{\circ}$, let us take $y^\ast \in D^{\circ}\setminus\{0\}$. We have $\frac{y^\ast}{\norm{y^\ast}}\in D^{\circ}$ since $D^{\circ}$ is a cone and $\norm{\frac{y^\ast}{\norm{y^\ast}}}=1$, which implies that $\frac{y^\ast}{\norm{y^\ast}}\in B$ and hence $\frac{y^\ast}{\norm{y^\ast}}\in \bar{D}^{\circ}$. We can write $y^\ast = \norm{y^\ast} \frac{y^\ast}{\norm{y^\ast}}$ where $\norm{y^\ast}>0$ and $\frac{y^\ast}{\norm{y^\ast}}\in \bar{D}^{\circ}$; hence, $\bar{D}^{\circ}$ is a cone generator for $D^{\circ}$.
\end{proof}

\subsection{Lebesgue spaces}\label{sec:lebesgue}

Let $(\Omega,\F,\P)$ be a probability space, and let $p\in [1,+\infty]$, $n\in\N$. We denote by $L^0(\R^n)$ the space of all $n$-dimensional random vectors that are identified up to $\P$-almost sure equality. We denote by $L^p(\R^n)$ the space of all $X\in L^0(\R^n)$ such that $\norm{X}_p<+\infty$, where $\norm{X}_p\coloneqq (\E[\norm{X}^p])^{1/p}$ for $p<+\infty$ and $\norm{X}_p\coloneqq\inf\{c>0\mid \P\{\norm{X}\leq c\}=1\} $ for $p=+\infty$. For $p\in \{0\}\cup[1,+\infty]$ and a set $A\subseteq\R^n$, we denote by $L^p(A)$ the set of all $X\in L^p(\R^n)$ such that $\P\{ X\in A\}=1$.

In this section, we fix $p\in [1,+\infty)$ and consider the case $\Y=L^p(\R^n)$, which is equipped with the norm $\norm{\cdot}_p$ and the induced topology.  
Then, $\Y^\ast=L^q(\R^n)$ with the norm $\norm{\cdot}_q$ and we consider it with the topology $\sigma(\Y^\ast,\Y)$, where $q\in (1,+\infty]$ is defined by $\frac{1}{p}+\frac{1}{q}=1$. Let $D\subseteq \Y$ be a closed convex cone and denote the unit ball in $L^q(\R^n)$ by $B^n_q=\{Y^\ast\in L^q(\R^n) \mid \norm{Y^\ast}_q\leq 1\}$. We show the existence of a compact convex cone generator for $D^{\circ}$ next.

\begin{proposition}\label{coneLp}
The set $\bar{D}^{\circ}\coloneqq D^{\circ} \cap B^n_q$ is a compact and convex cone generator for $D^{\circ}$.
\end{proposition}

\begin{proof}
Since $D^{\circ}$ and $B^n_q$ are closed convex sets, so is their intersection $\bar{D}^{\circ}$. Also, $B^n_q$ is (weakly) compact by Banach-Alaoglu Theorem (\cite[Thm. IV.21]{reed}). By using this fact and that $\bar{D}^{\circ}$ is a closed subset of $B_q$, we conclude that $\bar{D}^{\circ}$ is also compact. The proof of the claim that $\bar{D}^\circ$ is a cone generator for $D^{\circ}$ is similar to the proof of \Cref{finitedim}, hence omitted.
\end{proof}

\section{Applications to systemic risk measures}\label{sec:application}
	
In this section, we will explore the implications of the general theory developed in \Cref{sec:comp} on some quasiconvex risk measures for interconnected financial systems. Such risk measures are referred to as \emph{systemic risk measures}, which are of recent interest in the financial mathematics literature. We refer the reader to \cite{ararat,biagini,chen,feinstein} for detailed discussions on this subject.

Throughout this section, we fix a probability space $(\Omega,\F,\P)$. The proofs of the results in this section are given in \Cref{app:sec 6}.

\subsection{General results on quasiconvex systemic risk measures}\label{sec:systrisk}

We consider an interconnected financial system with $n\in\N$ institutions in a static setting. Due to their financial activities, the assets of the institutions are subject to uncertainty. Consequently, the future values of the assets of all institutions can be modeled as a random vector $X\in L^0(\R^n)$, which is sometimes called a \emph{random shock}. A systemic risk measure quantifies the overall risk of the system by taking into account the correlations between the components of the random shock as well as the underlying structure of the system. In line with \cite{chen} and \cite{biagini}, we study systemic risk measures of the form
\begin{equation}\label{systriskform}
R(X)=\rho(\tilde{\Lambda}\circ X),
\end{equation}
where $\tilde{\Lambda}\colon\R^n\to\R$ is an aggregation function and $\rho$ is a risk measure, see \Cref{defnriskagg} below for the precise descriptions of these terms. The aggregation function produces a univariate quantity $\tilde{\Lambda}\circ X\in L^0(\R)$ that summarizes the impact of the random shock on the economy (or society), which can be seen as an external entity of the system. The risk of this aggregate quantity is then evaluated through the univariate functional $\rho$ and the output $\rho(\tilde{\Lambda}\circ X)$ is the risk associated to the overall system when it faces random shock $X$.

To view the structure of $R$ in \eqref{systriskform} as a composition of two functions, we may simply define the functional version $\Lambda\colon L^0(\R^n)\to L^0(\R)$ of the aggregation function via $\Lambda(X)\coloneqq \tilde{\Lambda}\circ X$, that is,
\begin{equation}\label{lambdatilde}
\Lambda(X)(\omega)\coloneqq \tilde{\Lambda}(X(\omega)),\quad \omega\in\Omega.
\end{equation}
Then, \eqref{systriskform} can be rewritten as
\begin{equation}\label{systriskform2}
R=\rho\circ \Lambda.
\end{equation}

To obtain dual representations for systemic risk measures of the form \eqref{systriskform2}, we will consider random shocks that are sufficiently integrable. As in \Cref{sec:lebesgue}, we choose $\X=L^p(\R^n)$ and $\Y=L^p(\R)$, where $p\in [1,+\infty]$. These spaces are equipped with their norm topologies when $p<+\infty$ and with $\text{weak}^\ast$ topologies when $p=+\infty$. In all cases, we have $\X^\ast=L^q(\R^n)$ and $\Y^\ast=L^q(\R)$, with their weak topologies, where $q\in [1,+\infty]$ is determined by $\frac{1}{p}+\frac{1}{q}=1$. We denote by $\M^q_n(\P)$ the set of all vectors $\S=(\S_1,\ldots,\S_n)$, where $\S_i$ is a probability measure on $(\Omega,\F)$ that is absolutely continuous with respect to $\P$ and $\frac{d\S_i}{d\P}\in L^q(\R_+)$ for each $i\in\{1,\ldots,n\}$. For $X\in L^p(\R^n)$ and $\S\in\M_n^q(\P)$, we write $\E_{\S}[X]:=(\E_{\S_1}[X_1],\ldots,\E_{\S_n}[X_n])^{\mathsf{T}}$, where $\E_\Q$ denotes the expectation operator corresponding to a measure $\Q\in\M_1^q(\P)$. We take $C=L^p(\R^n_+)$ and $D=L^p(\R_+)$; hence, the dual cones are given by $C^{\circ}=L^q(\R^n_-)$ and $D^{\circ}=L^q(\R_-)$. With this choice of $D$, for convenience, we remove $D$ from the terminology; for instance, we simply call a function concave if it is $D$-concave.

The formal definitions of aggregation function and risk measure are given next.

\begin{definition}\label{defnriskagg}
(i) A function $\tilde{\Lambda}\colon \R^n\rightarrow \R$ is called an \emph{aggregation function} if it is increasing (with respect to $\R^n_+$ and $\R_+$).
(ii) A function $\rho\colon L^p(\R)\rightarrow \overline{\R}$ is called a \emph{quasiconvex risk measure} if it is quasiconvex and decreasing.
(iii) A function $R\colon L^p(\R^n)\rightarrow \overline{\R}$ is called a \emph{systemic risk measure} if it is of the form \eqref{systriskform}, where $\tilde{\Lambda}$ is an aggregation function and $\rho$ is a quasiconvex risk measure.
\end{definition}

In order for (iii) make sense in the above definition, we will impose the following assumption on the aggregation function.

\begin{assumption}\label{asmp:Lp}
    For an aggregation function $\tilde{\Lambda}$, its functional version $\Lambda$ defined by \eqref{lambdatilde} satisfies $\Lambda(X)\in L^p(\R)$ for every $X\in L^p(\R^n)$.
\end{assumption}

\begin{remark}
 A weaker version of the integrability condition in Assumption~\ref{asmp:Lp} reads as follows: $\Lambda(X)\in L^{p^\prime}(\R)$ for every $X\in L^p(\R^n)$, where $p^\prime\in [1,+\infty]$. Then, one can work with a risk measure $\rho$ defined on $L^{p^\prime}(\R)$ and obtain generalizations of the results presented in this section. To avoid cumbersome notation caused by working with two different exponents (and their conjugates), we will work under Assumption~\ref{asmp:Lp}, which is verified by all the examples we consider in Subsections~\ref{sec:ex} and~\ref{sec:eisnoe}.
\end{remark}

Consider a systemic risk measure $R=\rho\circ \Lambda$ as in \Cref{defnriskagg}. In view of \Cref{quasi2}, $R$ is quasiconvex whenever $\Lambda$ is naturally quasiconcave. We are particularly interested in the special case where $\Lambda$ is concave. As we will illustrate in \Cref{sec:ex}, such aggregation functions appear frequently in concrete examples. On the other hand, to ensure the lower demicontinuity of $\Lambda$, we need to impose sufficient regularity on $\tilde{\Lambda}$. This is done in the following lemma.

\begin{lemma}\label{continuitylemma}
	Let $\tilde{\Lambda}\colon \R^n\rightarrow \R$ be an aggregation function and define $\Lambda$ by \eqref{lambdatilde}.\\
	(i) If $\tilde{\Lambda}$ is concave and bounded from above, then $\Lambda$ is concave and lower demicontinuous. \\
	(ii) If $\tilde{\Lambda}$ is linear, then $\Lambda$ is linear and lower demicontinuous.\\
	(iii) If $\tilde{\Lambda}$ is regularly increasing (with respect to $\R^n_{+}$ and $\R_{+}$), then $\Lambda$ is regularly increasing.
\end{lemma}

	In the next proposition, we calculate the penalty function of a systemic risk measure when the aggregation function is concave and regularly increasing, and the univariate risk measure is quasiconvex and lower semicontinuous. It should be noted that, in \cite{ararat}, dual representations are provided for convex systemic risk measures, where $\rho$ is further assumed to be a convex (translative) risk measure. Hence, our results will extend these representations to the quasiconvex case. For convenience, we define the conjugate function $\tilde{\Phi}$ by
	\begin{equation}\label{conjtilde}
	\tilde{\Phi}(x^\ast)\coloneqq (-\tilde{\Lambda})^\ast(-x^\ast)=\sup_{x\in\R^n}\of{\Lambda(x)-(x^\ast)^{\mathsf{T}}x},\quad x^\ast\in\R^n,
	\end{equation}
	Similar to \eqref{lambdatilde}, we also define the functional version $\Phi$ of $\tilde{\Phi}$ by
	\begin{equation}\label{phi}
	\Phi(X^\ast)\coloneqq \tilde{\Phi}\circ X^\ast, \quad X^\ast \in L^q(\R^n).
	\end{equation}
	Moreover, for each $X^\ast\in L^q(\R^n)$, we introduce the set
	\begin{equation}\label{defnT}
	T_{X^\ast}\coloneqq \cb{Y^\ast\in L^q(\R_-)\mid \P\{X^\ast\neq 0, Y^\ast=0\}=0}.
	\end{equation}
	
	\begin{proposition}\label{propcom}
		Assume that $p\in [1,+\infty)$. Let $\tilde{\Lambda}\colon \R^n\rightarrow \R$ be a concave, regularly increasing aggregation function that is either bounded from above or linear. Let $\Lambda$ be defined by \eqref{lambdatilde}. Let $\rho$ be a lower semicontinuous quasiconvex risk measure. Let $X^\ast \in L^q(\R^n)$ and $m \in \R$ such that the strict sublevel set $\cb{X\in L^p(\R^n)\mid  \E\sqb{Y^\ast \Lambda (X)}<m}$ is nonempty for every $Y^\ast \in L^q(\R_-)\setminus\{0\}$. Then, 
		\[
		\alpha_{\rho\circ\Lambda}(X^*,m)=\inf_{Y^\ast\in T_{X^\ast}}\of{-\mathbb{E}\sqb{Y^*\Phi\of{\frac{X^*}{Y^*}}1_{\{Y^*<0\}}}+\alpha_{\rho}(Y^*,m)}.
		\]
	\end{proposition}
	
	Next, we aim to rewrite the formula in \Cref{propcom} in terms of probability measures. This reformulation will make it possible to provide economic interpretations of the dual representation in view of model uncertainty. Since $D_{1}^{\circ} = \{-\frac{d\mathbb{Q}}{d\mathbb{P}}\mid \Q\in \M^q_1(\mathbb{P})\}$ is a closed convex cone generator for $D^{\circ}=L^q(\R_-)$, we can write every $Y^\ast\in L^q(\R_-)\setminus\{0\}$ as $Y^\ast=-\lambda \frac{d\mathbb{Q}}{d\mathbb{P}}$ for some $\lambda > 0$ and $\Q\in \mathcal{M}^q_1(\P)$ by \Cref{base}. Similarly, every $X^\ast\in C^{\circ}=L^q(\R^n_-)$ can be written as $X^\ast=-w\cdot \frac{\mathbf{d\S}}{d\P}$, where $w\in \R^n_+$, $\S=(\S_1,\ldots,\S_n)\in \mathcal{M}_n^q(\P)$, and $\frac{\mathbf{d\S}}{d\P}\coloneqq(\frac{d\S_1}{d\P},\ldots, \frac{d\S_n}{d\P})^{\mathsf{T}}$. The interpretation of these dual variables is as follows. In the presence of model uncertainty, we consider $\Q$ as a probability measure that is assigned to an external entity, e.g., society, and, for each $i\in\{1,\ldots,n\}$, $\S_i$ is a probability measure that is assigned to internal entity $i$, e.g., a bank in the network, with corresponding weight $w_i$. Moreover, since we consider $X^\ast$ and $Y^\ast$ satisfying the condition $\P\{X^\ast\neq 0, Y^\ast=0\}=0$ in \Cref{propcom}, it follows from \cite[Lemma~6.3]{ararat} that $w_i\S_i$ is a finite measure that is absolutely continuous with respect to $\Q$, and we can write 
	\[
	\frac{w\cdot\frac{\mathbf{d\S}}{d\P}}{\frac{d\Q}{d\P}}=w\cdot\frac{\mathbf{d\S}}{d\Q},
	\]
	where all Radon-Nikodym derivatives are well-defined. Therefore, in probabilistic terms, the formula in \Cref{propcom} can be rewritten as
	\begin{equation}\label{eqprob}
	\alpha_{\rho\circ\Lambda}\of{-w\cdot \frac{\mathbf{d\S}}{d\P},m }=\inf_{\substack{\lambda>0 , \Q\in \M_1^q(\P)\colon \\ w_i\S_i\ll \Q\ \forall i }}\of{\E_{\Q}\sqb{\lambda \Phi \of{\frac{w}{\lambda}\cdot\frac{\mathbf{d\S}}{d\Q}}}+\lambda \alpha_{\rho}\of{-\frac{d\Q}{d\P},m}}.
	\end{equation}
	According to \eqref{eqprob}, the total penalty of choosing probability vector $\S$ and weight vector $w$ for the financial institutions is calculated by considering all possible choices of society's probability measure $\Q$ and an associated weight $\lambda$. As in the convex case studied in \cite{ararat}, $\Q$ is chosen from the absolute continuity interval defined via $w_i\S_i\ll\Q\ll\P$, $i\in\{1,\ldots,n\}$ using $w\cdot \S$. The infimum in \eqref{eqprob} can be seen as a directed distance from $w\cdot\S$ to $\P$ that is calculated through society's measure $\Q$. The first term inside the infimum is the \emph{multivariate divergence} of $w\cdot \S$ relative to $\Q$. The divergence function is determined by the structure of the network, see \Cref{sec:ex} and \Cref{sec:eisnoe} for concrete calculations. Moreover, this function is scaled by the weight $\lambda>0$ through $\lambda\Phi(\frac{\cdot}{\lambda})$, which is the conjugate function corresponding to $\lambda\Lambda(\cdot)$. In other words, society's weight $\lambda$ amplifies/shrinks the impact of the shock to society as a factor. The second term inside the infimum is the penalty of choosing $\Q$ with respect to the physical measure $\P$ in the presence of model uncertainty, which is quantified by the choice of the univariate risk measure $\rho$. Hence, the overall penalty is calculated as the least possible sum of these two distance terms. It is notable that the objective function of the penalty function has an additive structure in our quasiconvex framework, which generalizes the observations in \cite{ararat} for the convex case.
	
	As a continuation of \Cref{propcom}, we calculate the inverse of the penalty function in the next proposition.
	
	\begin{proposition}\label{dualfn}
	Assume that $p\in [1,+\infty)$. Let $\tilde{\Lambda}\colon \R^n\rightarrow \R$ be a concave, regularly increasing aggregation function. Let $\Lambda$ be defined by \eqref{lambdatilde}. Let $\rho$ be a lower semicontinuous quasiconvex risk measure.\\
	(i) Suppose that $\tilde{\Lambda}$ is bounded from above, that is, $\tilde{\Phi}(0)<+\infty$. Then, we have
			\begin{align}\nonumber 
			&\beta_{\rho\circ\Lambda}(X^\ast,s)
            \\&=\sup_{Y^\ast \in L^q(\R_-)\setminus \{0\}}\beta_{\rho}\of{Y^\ast,\Phi(0)\E[Y^\ast]}\vee 
			\sup_{Y^\ast \in T_{X^\ast}}\beta_{\rho}\of{Y^\ast, s+\E\sqb{Y^\ast\Phi\of{\frac{ X^\ast}{ Y^\ast}}1_{\{Y^\ast<0\}}}},\nonumber 
			\end{align}
			where $T_{X^\ast}$ is defined by \eqref{defnT}. In particular, when we transform the variables into the probabilistic setting, we get
			\begin{align}\nonumber 
			&\beta_{\rho\circ\Lambda}\of{-w\cdot \frac{\mathbf{d}\S}{d\P},s}
            \\& =\sup_{\Q\in\mathcal{M}^q_1(\P)}\beta_{\rho}\of{- \frac{d\Q}{d\P} ,-\Phi(0)}\vee \sup_{\substack{\Q\in \mathcal{M}_1^q(\P),\lambda >0\colon \\w_i\S_i\ll\Q\ \forall i}}\beta_{\rho }\of{-\frac{d\Q}{d\P},\frac{s}{\lambda}-\mathbb{E}_{\Q}\sqb{\Phi\of{\frac{w}{\lambda}\cdot\frac{ \mathbf{d}\S}{ d\Q}}}}.\nonumber 
			\end{align}
	(ii) Suppose that $\tilde{\Lambda}$ is linear and it is unbounded from above, that is, $\tilde{\Phi}(0)=+\infty$. Then, we have
			\begin{equation*}
			\beta_{\rho\circ\Lambda}(X^*,s)=\sup_{Y^\ast \in T_{X^\ast}}\beta_{\rho}\of{Y^\ast, s+\E\sqb{Y^\ast\Phi\of{\frac{ X^\ast}{ Y^\ast}}1_{\{Y^\ast<0\}}}},
			\end{equation*}
			and
			\begin{equation*}
			\beta_{\rho\circ\Lambda}\of{-w\cdot \frac{\mathbf{d\S}}{d\P},s}=\sup_{\substack{\Q\in \mathcal{M}_1^q(\P),\lambda >0\colon \\w_i\S_i\ll\Q\ \forall i}}\beta_{\rho }\of{-\frac{d\Q}{d\P}, \frac{s}{\lambda}-\mathbb{E}_{\Q}\sqb{\Phi\of{\frac{w}{\lambda}\cdot\frac{ \mathbf{d}\S}{d\Q}}}}.
			\end{equation*}
	\end{proposition}
	
	In the next proposition, we give a dual representation for quasiconvex systemic risk measures. Unlike Propositions \ref{propcom} and \ref{dualfn}, we allow for $p=+\infty$ here as we do not rely on the expression for the penalty function (hence not on the existence of a compact cone generator).
	
	\begin{proposition}\label{dualsystemic}
		Assume that $p\in [1,+\infty]$. Let $\tilde{\Lambda}\colon \R^n\rightarrow \R$ be a concave aggregation function that is either bounded from above or linear. Let $\Lambda$ be defined by \eqref{lambdatilde}. Let $\rho$ be a lower semicontinuous quasiconvex risk measure. Then, we have
		\begin{equation}\label{eq:dualsystemic}
		R(X)=\rho\circ\Lambda(X)=\sup_{\substack{w \in \mathbb{R}^n_+\setminus\{0\},\S\in \mathcal{M}_n^q(\P), \\ \Q \in \mathcal{M}^q_1(\P)\colon w_i\S_i\ll\Q\ \forall i}}\beta_{\rho }\of{-\frac{d\Q}{d\P},w^{\mathsf{T}}\E_{\S}\sqb{-X}-\mathbb{E}_{\Q}\sqb{\Phi\of{w\cdot\frac{ \mathbf{d\S}}{d\Q}}}}
		\end{equation}
		for every $X\in L^p(\R^n)$.
	\end{proposition}
	
	While the objective function of the penalty function has an additive structure in \Cref{propcom}, we see in \Cref{dualfn} that this might not be the case for its inverse. In other words, the inverse penalty function of $\rho$ and the divergence term including $\Phi$ might interact in a non-additive way. We will see such cases in \Cref{sec:ex}. Consequently, due to \Cref{dualsystemic}, the same structure also shows up in the final dual representation of the systemic risk measure. This is contrary to the convex framework of \cite{ararat}, where the penalty function directly appears in the dual representation of a convex systemic risk measure. Hence, our results shed light on a new feature of quasiconvex systemic risk measures that does not exist in convex systemic risk measures.

    We conclude this section by interpreting the dual representation in \eqref{eq:dualsystemic}, similar to the convex case considered in \cite{ararat}, in view of \emph{model uncertainty} and \emph{weight ambiguity}. To each institution $i\in \{1,\ldots,n\}$, we assign a probability measure $\S_i$ and a weight $w_i$. First, we calculate the weighted total expected loss of the institutions as $w^{\mathsf{T}}\E_{\S}\sqb{-X}$. Then, for each institution $i\in\{1,\ldots,n\}$, we calculate the weighted density $w_i\frac{d \S_i}{d \Q}$ as a measure of the discrepancy between $\S_i$ and society's probability measure $\Q$. Using the multivariate divergence function $\Phi$, we convert these weighted densities into a (directed) distance between the network and society. The weighted expected loss is adjusted by using this distance as a ``penalty'' term. Finally, $\beta_{\rho}$ uses this adjusted expected loss and the discrepancy between society's probability measure $\Q$ and the physical measure $\P$ to calculate an overall risk evaluation under $(w,\S,\Q)$. At the end, we report the most conservative risk evaluation over all choices of $(w,\S,\Q)$. We will discuss more concrete choices of $\Phi$ and $\beta_\rho$ in the next two sections.

   \begin{remark}\label{rem:nolambda}
   In the setting of \Cref{dualsystemic}, instead of exploiting the structure of $R=\rho\circ \Lambda$ as a quasiconvex composition, a more simplistic approach is to only use the dual representation of $\rho$ and apply it at $\Lambda(X)$ to obtain
   \[
    R(X)=\sup_{\Q\in \mathcal{M}_1^q(\P)}\beta_\rho\of{-\frac{d\Q}{d\P},\E^{\Q}[-\Lambda(X)]}
   \]
   for every $X\in L^p(\R^n)$. However, this representation does not have a useful interpretation in terms of the network since the ``systemic" nature of the problem hidden in $\Lambda$ is not dualized at all.
   \end{remark}
	
	\subsection{Examples}\label{sec:ex}
	
	In this section, we first recall some examples of quasiconvex risk measures and concave aggregation functions studied in the literature. Then, we will combine some choices of these two functions and illustrate the forms of the penalty functions and dual representations of the resulting systemic risk measures.
	
	We start by recalling two families of quasiconvex lower semicontinuous risk measures studied in \cite{kupper}. The first family consists of functionals of the form
	\[
	\rho(Y)=\ell^{-1}\of{\mathbb{E}[\ell\circ (-Y)]},\quad Y\in L^p(\R),
	\]
	where $p \in [1, +\infty]$, and $\ell\colon \R\to(-\infty,\infty]$ is a proper lower semicontinuous convex increasing function, called a \emph{loss function}. For simplicity, we assume that $\ell$ is differentiable. Such $\rho$ is called the \emph{certainty equivalent} associated to $\ell$. It is calculated in \cite{kupper} that
	\[
	\alpha_{\rho}\of{-\frac{d\Q}{d\P},m}=\E_{\Q} \sqb{h\circ \of{\theta \frac{d\Q}{d\P}}},\quad \Q \in \M^q_1(\P), m\in\R,
	\] 
	where $h$ is the right inverse of the derivative $\ell^\prime$, and $\theta=\theta(\Q,m)$ is the solution of the equation $\E[\ell\circ h\circ (\theta\frac{d\Q}{d\P})]=\ell^{+}(m)$ under some integrability and positivity conditions.
	
	Let us provide some concrete examples of the loss function $\ell$ and recall the penalty functions for the corresponding certainty equivalents, already calculated in \cite[Ex. 8]{kupper}.
	
	\begin{example}\label{risk1}
		(i) (Quadratic loss function) Let us take $p=2$, and $\ell(s)=s^2/2+s$ for $s\geq -1$, $\ell(s)=-\frac12$ for $s<-1$. Then, for each $\Q\in  \M_1^2(\P)$, we have $\alpha_{\rho}(-\frac{d\Q}{d\P},m)=-1$ for $m \geq -1$ and
			\[
			\alpha_{\rho}\of{-\frac{d\mathbb{Q}}{d\mathbb{P}},m}=(1+m)\norm{\frac{d\Q}{d\P}}_2-1,\quad m<-1,\qquad 
			\beta_{\rho}\of{-\frac{d\mathbb{Q}}{d\mathbb{P}},s}=\frac{s+1}{\norm{\frac{d\Q}{d\P}}_2}-1,\quad s<-1.
			\]
		(ii) (Logarithmic loss function) Let us take $p=1$ or $p=+\infty$, and $\ell(s)=-\ln(-s)$ for $s<0$, $\ell(s)=+\infty$ for $s\geq 0$. Then, for each $\Q\in\M^q_1(\P)$,
			\[
			\alpha_{\rho}\of{-\frac{d\mathbb{Q}}{d\mathbb{P}},m}=m e^{\mathbb{E}\sqb{\ln( \frac{d\mathbb{Q}}{d\mathbb{P}})}},\quad m<0,
			\qquad 
			\beta_{\rho}\of{-\frac{d\mathbb{Q}}{d\mathbb{P}},s}=se^{-\mathbb{E}\sqb{\ln( \frac{d\mathbb{Q}}{d\mathbb{P}})}},\quad s<0.
			\]
			(iii) (Power loss function) Let us take $p=1$ or $p=+\infty$, and fix some $\gamma\in (0,1)$. Take $\ell(s)=-\frac{(-s)^{1-\gamma}}{1-\gamma}$ for $s\leq 0$, $\ell(s)=\infty$ for $s>0$. Then, for each $\Q\in\M^q_1(\P)$,
			\[
			\alpha_{\rho}\of{-\frac{d\mathbb{Q}}{d\mathbb{P}},m}=\frac{m}{\norm{\frac{d\Q}{d\P}}_{\frac{\gamma-1}{\gamma}}},\quad m<0,
			\qquad 		\beta_{\rho}\of{-\frac{d\mathbb{Q}}{d\mathbb{P}},s}=s\norm{\frac{d\Q}{d\P}}_{\frac{\gamma -1}{\gamma}},\quad s<0.
			\]
			Here, for $Y^\ast\in L^1(\R)$, we use the notation $\norm{Y^\ast}_a \coloneqq (\E[\abs{Y^\ast}^a])^{\frac{1}{a}}$ for $a<1$ as well, although $\norm{\cdot}_a$ is not a norm in general.
	\end{example}
	
	We also revisit the \emph{economic index of riskiness} as another example of a quasiconvex risk measure. Based on a loss function $\ell$ as before, this risk measure is defined by 
	\[
	\rho(Y)=\frac{1}{\sup\{\lambda>0\mid \mathbb{E}[\ell\circ (-\lambda Y)]\leq c_0 \}} ,\quad Y\in L^p(\R),
	\]
	where $c_0\in \mathbb{R}$ is a fixed threshold for expected loss levels. To make this risk measure well-defined, $\ell$ is usually assumed to have the superlinear growth condition $\lim_{s\rightarrow \infty}\ell(s)/s=\infty$ and $p$ is chosen in accordance with $\ell$. Following the arguments in \cite{kupper}, it can be shown that
	\[
	\alpha_{\rho}\of{-\frac{d\mathbb{Q}}{d\mathbb{P}},m}=\mathbb{E}_{\mathbb{Q}} \sqb{mh\circ \of{m\theta \frac{d\mathbb{Q}}{d\mathbb{P}}}},\quad \Q \in \mathcal{M}^q_1(\mathbb{P}),m\in\R,
	\]
	where $\theta=\theta(\Q,m)$ is the solution of the equation $\mathbb{E}[\ell\circ h\circ (m\theta\frac{d\mathbb{Q}}{d\mathbb{P}})]=c_0$.
	
	The following example is the analogue of \Cref{risk1}(ii) for the economic index of riskiness; see \cite[Ex. 3, 9]{kupper} for more details.
	
	\begin{example}\label{risk2}
		Let us take $p=1$ and $c_0>0$, and consider $\ell(s)=-\ln(1-s)$ for $s<1$, $\ell(s)=+\infty$ for $s\geq 1$. Then, for each $\Q\in\M_1^\infty(\P)$, $m<0$, $s<0$, we have
\begin{align*}
		 \alpha_{\rho}\of{-\frac{d\mathbb{Q}}{d\mathbb{P}},m}&=m\of{1-\exp\of{\mathbb{E}\sqb{\ln\of{\frac{d\Q}{d\P}}}-c_0}},
\\
	\beta_{\rho}\of{-\frac{d\mathbb{Q}}{d\mathbb{P}},s}&=\frac{s}{1-\exp\of{\mathbb{E}\sqb{\ln \of{\frac{d\mathbb{Q}}{d\mathbb{P}}}}-c_0}},   
		\end{align*}
		where $\exp(x)=e^x$ for $x\in\R$.
	\end{example}
	
	Next, we recall some examples of concave aggregation functions from \cite[Sect. 4]{ararat}. In each example, we calculate the conjugate function $\tilde{\Phi}$ given by \eqref{conjtilde}. A more sophisticated aggregation function based on a clearing mechanism will be discussed separately in \Cref{sec:eisnoe}.
	
	\begin{example}\label{aggregation}
	(i) (Total profit-loss model) Let us take $\tilde{\Lambda}(x)=\sum_{i=1}^{n} x_i$ for each $x\in\R^n$. Then,
			\[
			\tilde{\Phi}(x^\ast)=
			\begin{cases}
			0& \text{ if } x^*=\mathbf{1},\\
			\infty & \text{ else}.
			\end{cases}
			\]
			The condition that $\Lambda(X)\in L^p(\R)$ for every $X\in L^p(\R^n)$ is satisfied for every $p\in [1,+\infty]$.\\
	(ii) (Total loss model) Let us take $\tilde{\Lambda}(x)=-\sum_{i=1}^{n} x_i^-$ for each $x\in\R^n$. Then,
			\[
			\tilde{\Phi}(x^\ast)=
			\begin{cases}
			0& \text{ if } x^*_i\in[0,1] \text{ for every } i \in \{1,\ldots,n\},\\
			\infty & \text{ else}.
			\end{cases}
			\]
			As in (i), for every choice of $p\in [1,+\infty]$, we have $\Lambda(X)\in L^p(\R)$ for every $X\in L^p(\R^n)$.\\
	(iii) (Exponential model) Let us take $\tilde{\Lambda}(x)=-\sum_{i=1}^{n} e^{-x_i-1}$ for each $x\in\R^n$. Then,
			\[
			\tilde{\Phi}(x^\ast)=\sum_{i=1}^{n} x_i^*\ln(x_i^*),
			\]
			where $\ln(0)\coloneqq-\infty$ and $0\ln(0)\coloneqq 0$ as conventions. The condition that $\Lambda(X)\in L^p(\R)$ for every $X\in L^p(\R^n)$ is satisfied only for $p=+\infty$. As a result, Propositions \ref{propcom} and \ref{dualfn} is not applicable. However, we can still use the dual representation in \Cref{dualsystemic}.\\
		Thanks to \Cref{continuitylemma}, each aggregation function $\tilde{\Lambda}$ above yields a lower demicontinuous concave functional version $\Lambda$ via \eqref{lambdatilde}. In (i) and (iii), the aggregation function is also regularly increasing.
	\end{example}
	
		By combining Examples \ref{risk1} and \ref{risk2} with \Cref{aggregation}, we will consider some examples of quasiconvex systemic risk measures and provide their penalty functions and dual representations in view of Propositions \ref{propcom} and \ref{dualsystemic}.
        
	\begin{example}
		(Total profit-loss model with economic index of riskiness)\\
        Take $\tilde{\Lambda}(x)=\sum_{i=1}^{n} x_i$ and $p\in[1,+\infty)$. By \eqref{eqprob}, we have
		\begin{equation*}\
		\alpha_{\rho\circ\Lambda}\of{-w\cdot \frac{\mathbf{d\S}}{d\P},m }=\inf_{\substack{\lambda>0 , \Q\in \M_1^q(\P)\colon \\w_i\S_i\ll \Q \ \forall i}}\of{\lambda \alpha_{\rho}\of{-\frac{d\Q}{d\P},m}+\mathbb{E}_{\Q}\sqb{\lambda \Phi\of{\frac{w}{\lambda}\cdot\frac{\mathbf{d}\S}{d\Q}}}}.
		\end{equation*}
		Thanks to the calculation in \Cref{aggregation}(i), it is enough to consider only the case where $\frac{w\cdot \mathbf{d\S}}{\lambda d\Q}=\mathbf{1}$ almost surely, that is, $w_1=\ldots=w_n=\lambda$ and $\S_1=\ldots=\S_n=\Q$. Therefore,
		\[
		\alpha_{\rho\circ\Lambda}\of{-w\cdot \frac{\mathbf{d\S}}{d\P},m}=
		\lambda \alpha_{\rho}(-\frac{d\Q}{d\P},m) 
		\]
		if $w\cdot \frac{\mathbf{d\S}}{ d\Q}=\lambda\mathbf{1}$ for some $\Q\in \M_1^q(\P)$, $\lambda >0$, and $\alpha_{\rho\circ\Lambda}\of{-w\cdot \frac{\mathbf{d\S}}{d\P},m}=+\infty$ otherwise. As a further special case, let us assume that $\rho$ is the economic index of riskiness in \Cref{risk2} corresponding to the logarithmic loss function with $p=1$. In this case, we obtain
		\[
		\alpha_{\rho\circ\Lambda} \of{-w\cdot \frac{\mathbf{d\S}}{d\P},m} = m \lambda \of{1-\exp\of{\E\sqb{\ln\of{\frac{d\Q}{d\P}}}-c_0}}
		\]
		if $w\cdot \frac{\mathbf{d\S}}{ d\Q}=\lambda\mathbf{1}$ for some  $\Q\in \M_1^{\infty}(\P)$ and $\lambda >0$, and $\alpha_{\rho\circ\Lambda} (-w\cdot \frac{\mathbf{d\S}}{d\P},m) =+\infty$ otherwise. 
	\end{example}
	
	\begin{example}
		(i) Let $\tilde{\Lambda}(x)=\sum_{i=1}^n x_i$ be the aggregation function in \Cref{aggregation}(i) and $p\in [1,+\infty]$. Then, by \Cref{dualsystemic} and \Cref{aggregation}, 
			\[\rho\circ\Lambda(X)=\sup_{\Q\in \mathcal{M}_1^q(\P)}\beta_{\rho}\of{-\frac{d\Q}{d\P},-\sum_{i=1}^{n}\mathbb{E}_{\Q}[X_i]}.
			\]
			In particular, if we take $\rho$ as the certainty equivalent corresponding to the power loss function (\Cref{risk1}(iii)) and $p=1$, then by \Cref{dualsystemic} and \Cref{risk1}, we get 
			\[
			\rho\circ\Lambda(X)=\sup_{\Q\in \mathcal{M}_1^\infty(\P)}-\norm{\frac{d\Q}{d\P}}_{\frac{\gamma-1}{\gamma}}\sum_{i=1}^{n}\mathbb{E}_{\Q}[X_i].
			\]
	(ii) Let us take the total loss model in \Cref{aggregation} and $p\in [1,+\infty]$. Then, we have the following dual representation by \Cref{dualsystemic}:
			\begin{equation}
			R(X)=\rho\circ\Lambda(X)=\sup_{\substack{w \in \mathbb{R}^n_+\setminus\{0\},\S\in \mathcal{M}_n^q(\P)\colon \\ w_i\frac{d\S_i}{d\P}\leq 1 \ \forall i,\\ \Q \in \mathcal{M}^q_1(\P)\colon w_i\S_i\ll\Q\ \forall i}}\beta_{\rho }\of{-\frac{d\Q}{d\P},-w^{\mathsf{T}}\E_{\S}\sqb{X }}.
			\end{equation}
			As a special case, let us take $p=2$ and consider the quadratic loss function in \Cref{risk1}(i), which gives
			\begin{equation}
			R(X)=\rho\circ\Lambda(X)=\sup_{\substack{w \in \mathbb{R}^n_+\setminus\{0\},\S\in \mathcal{M}_n^2(\P)\colon \\ w_i\frac{d\S_i}{d\P}\leq 1\ \forall i , w^{\mathsf{T}}\E_{\S}[X]<1,\\ \Q \in \mathcal{M}^2_1(\P)\colon w_i\S_i\ll\Q\ \forall i}}\frac{-w^{\mathsf{T}}\E_{\S}\sqb{X }+1}{\norm{\frac{d\Q}{d\P}}_2}-1.
			\end{equation}
		(iii) Let us suppose that $\rho$ is the certainty equivalent corresponding to the logarithmic loss function in \Cref{risk1}(ii) with $p=+\infty$. Then, by \Cref{dualsystemic} and \Cref{risk1}, we have
			\begin{equation*}
			\rho\circ\Lambda(X)=\sup_{\substack{w \in \mathbb{R}^n_+\setminus\{0\},\S\in \mathcal{M}_n^1(\P),\\ \Q \in \M_1^1(\P)\colon w_i\S_i\ll\Q\ \forall i}}-\frac{\mathbb{E}_{\Q}\sqb{\Phi\of{\frac{w\cdot \mathbf{d\S}}{d\Q}}}+w^{\mathsf{T}}\E_{\S}\sqb{X }}{e^{\mathbb{E}\sqb{\ln\of{\frac{d\Q}{d\P}}}}} .
			\end{equation*}
			In particular, let us assume that $\tilde{\Lambda}$ is the exponential aggregation function in \Cref{risk2}(iii). Then, \eqref{eqentropy} simplifies as
			\begin{equation}\label{eqentropy}
			\rho\circ\Lambda(X)=\sup_{\substack{w \in \mathbb{R}^n_+\setminus\{0\},\S\in \mathcal{M}_n^1(\P),\\ \Q \in \M_1^1(\P)\colon w_i\S_i\ll\Q\ \forall i}} \frac{w^{\mathsf{T}}\E_{\S}\sqb{-X}-\sum_{i=1}^n\H(w_i\S_i||\Q)}{e^{\mathbb{E}\sqb{\ln\of{\frac{d\Q}{d\P}}}}},
			\end{equation}
			where $\H(w_i\S_i||\Q)\coloneqq w_i\mathbb{E}_{\S_i}[\ln(\frac{w_id\S_i}{d\Q})]$ is the relative entropy of the finite measure $w_i\S_i$ with respect to society's probability measure $\Q$.
\end{example}
	
	We conclude this section by providing an economic interpretation of the dual representation in \eqref{eqentropy}. For given choices of the network's probability vector $\S$ and weight vector $w$, and society's probability $\Q$, the risk of the random shock $X$ is first calculated linearly as $w^{\mathsf{T}}\E_\S[-X]$. This linear evaluation is adjusted by the relative entropy term $\sum_{i=1}^n\H(w_i\S_i||\Q)$, which is a multivariate directed distance from $w\cdot \S$ to $\Q$. In the presence of model uncertainty for society, further adjustment of risk by the directed distance $e^{\E[\ln(\frac{d\Q}{d\P})]}$ from society's measure $\Q$ to the physical measure $\P$. The nonlinear interaction between the numerator and the denominator is due to the quasiconvex (but not convex) choice of $\rho$, as discussed in \Cref{sec:systrisk}. Finally, the systemic risk measure is calculated as the most conservative evaluation of the ratio over all choices of $w,\S,\Q$. Similar interpretations can be made for the other instances of systemic risk measures discussed above.
	
	\subsection{Eisenberg-Noe model}\label{sec:eisnoe}
	
	In some applications, random shocks might take values only in a certain subset of $\R^n$. In such cases, the aggregation function is naturally defined on this subset instead of the whole space. In this section, we will discuss the Eisenberg-Noe clearing model for which the aggregation function is of the form $\tilde{\Lambda}\colon\R^n_+\to\R$. Before describing this model in detail, as a preparation, we first state slightly different versions of Propositions \ref{propcom} and \ref{dualsystemic} for a generic aggregation function $\tilde{\Lambda}\colon\R^n_+\to\R$. Accordingly, we modify the definition of $\tilde{\Phi}$ in \eqref{conjtilde} as
	\[
	\tilde{\Phi}(x^\ast)=\sup_{x\in \R^n_+}(\Lambda(x)-(x^\ast)^{\mathsf{T}}x), \quad x^\ast\in \R^n,
	\]
	and we define the functional version $\Phi$ by \eqref{phi} as before.
	
	\begin{proposition}\label{propeisnoe1}
		Assume that $p\in [1,+\infty)$. Let $\tilde{\Lambda}\colon\R^n_+\to\R$ be a concave, regularly increasing and increasing function that is bounded from above. Let $\Lambda$ be defined by \eqref{lambdatilde} and suppose that $\Lambda(X)\in L^p(\R)$ for every $X\in L^p(\R^n_+)$. Let $\rho$ be a lower semicontinuous quasiconvex risk measure. Let $X^\ast\in L^q(\R^n_-)$ and $m\in\R$ such that the strict sublevel set $\cb{X\in L^p(\R^n_+)\mid \E\sqb{Y^\ast \Lambda (X)}<m}$ is nonempty for every $Y^\ast \in L^q(\R_-)\setminus\{0\}$. Then,
		\begin{align*}
		\alpha_{\rho \circ \Lambda}(X^\ast,m)=0\wedge \inf_{ Y^\ast\in L^q(\R_{--})}\of{-\E\sqb{ Y^*\Phi\of{\frac{X^*}{ Y^*}}}+\alpha_{\rho}\of{Y^\ast,m}}.
		\end{align*}
	\end{proposition}
	
	\begin{proposition}\label{propeisnoe3}
		Assume that $p\in[1,+\infty]$. Let $\tilde{\Lambda}\colon \R^n_+\to\R$ be a concave increasing function that is either bounded from above or linear. Let $\Lambda$ be defined by \eqref{lambdatilde} and suppose that $\Lambda(X)\in L^p(\R)$ for every $X\in L^p(\R^n_+)$. Let $\rho\colon L^p(\R)\to\overline{\R}$ be a lower semicontinuous quasiconvex risk measure. Then, for every $X\in L^p(\R^n_+)$,
		\begin{align*}
		\rho\circ \Lambda (X)=\sup_{\substack{X^*\in L^q(\R^n_-)\setminus \{0\},\\ Y^\ast \in L^q(\R_{--})}}\beta_{\rho }\of{Y^*,\E\sqb{(X^*)^{\mathsf{T}}X+Y^*\Phi\of{\frac{X^*}{Y^*}}}}.
		\end{align*}
	\end{proposition}
	
	As in \Cref{sec:systrisk}, we may switch to probability measures by writing $X^*=-w\cdot \frac{\mathbf{d\S}}{d\P}$ and $Y^*=-\lambda \frac{d\Q}{d\P}$, where $w\in \R^n_+\setminus \{0\}$, $\lambda>0$, $\Q\in \M_1^q(\P)$, and $\S \in \M_n^q(\P)$. Again, by \cite[Lem. 6.3]{ararat}, we have $w_i\S_i\ll \Q$ if $Y^\ast\in L^q(\R_{--})$. Hence, the representation in \Cref{propeisnoe3} can be rewritten as
	\begin{equation}\label{convexeq}
	\rho\circ\Lambda(X)=\sup_{\substack{w \in \mathbb{R}^n_+\setminus\{0\},\S\in \mathcal{M}^q_n(\P),\\ \Q \in \mathcal{M}_1^q(\P)\colon w_i\S_i\ll\Q\ \forall i}}\beta_{\rho }\of{-\frac{d\Q}{d\P},-\mathbb{E}_{\Q}\sqb{\Phi\of{w\cdot \frac{\mathbf{d}\S}{d\Q}}}-w^{\mathsf{T}}\E_{\S}\sqb{X}}.
	\end{equation}
	
	Next, we review the clearing model in \cite{eisnoe}, which takes into account the liabilities between the members of the financial network, hence the structure of the network. In this model, financial institutions are considered as the nodes of a graph, and their liabilities are considered as the corresponding arcs. More precisely, let $\mathcal{N}=\{0,1,\ldots,n\}$ denote the nodes, where nodes $1,\ldots,n$ typically represent the banks and node $0$ represents society. For each $i,j\in\mathcal{N}$, let $\ell_{ij}\geq 0$ denote the nominal liability of member $i$ to member $j$. Naturally, we assume no self-liabilities, that is, $\ell_{ii}=0$ for each $i\in\mathcal{N}$; and society has no liabilities to banks, that is, $\ell_{0i}=0$ for every $i\in \mathcal{N}$. We also assume that every bank has nonzero liability to society, that is, $\ell_{i0}>0$ for every $i\in\mathcal{N}\setminus\{0\}$. Then, the relative liability of member $i$ to member $j$ is defined by
	\[
	a_{ij}\coloneqq \frac{\ell_{ij}}{\bar{p}_i},
	\]
	where $\bar{p}_i\coloneqq \sum_{j=0}^{n} \ell_{ij}$ is the total liability of member $i$. Finally, let $x\in\R^n_+$ denote a possible realization of the uncertain value of the assets of the banks. A clearing payment vector $p(x)\in\R^n$ is defined as a solution of the following fixed point problem:
	\[
	p_i(x)=\min \cb{\bar{p}_i,\sum_{j=1}^{n}a_{ji}p_j(x)}\text{ for }i\in\mathcal{N}\setminus\{0\}.
	\]
	In words, at clearing, each bank either pays in full what it owes or it partially meets its obligations by paying what it receives from other banks. Obviously, every clearing payment vector $p=p(x)$ is a feasible solution for the following linear programming problem.
	\begin{align}\label{LP}
	\text{maximize}\quad &\sum_{i=1}^{n}a_{i0}p_i \\
	\text{subject to}\quad & p_i \leq x_i+\sum_{j=1}^{n}a_{ji}p_j\quad \forall i\in\{1,\ldots,n\},\notag \\
	& p_i\in [0,\bar{p}_i]\quad \forall i\in\{1,\ldots,n\}.\notag
	\end{align}
	It is shown in \cite[Lem. 4]{eisnoe} that every optimal solution of this problem is a clearing payment vector for the system. In addition, it is shown in \cite{eisnoe} that, for every $x \in \R^n_+$, the above linear programming problem is feasible, and hence it has an optimal solution; let us denote the optimal value by $\tilde{\Lambda}(x)$. It should be noted that $\tilde{\Lambda}(x)\in \R_+$ since $a_{i0}>0$ by definition and $p_i\in [0,\bar{p}_i]$.   $\tilde{\Lambda}$ calculates the effect of the realized values of the assets on society. Therefore, $\tilde{\Lambda}$ can be considered as an aggregation function. Let us take $D=L^p(\R_+)$ and $D^{\circ}=L^q(\R_-)$. Then, $\tilde{\Lambda}$ is concave and increasing as it is stated in \cite[Sect. 4.4]{ararat}; it is also bounded by $\sum_{i=1}^{n}a_{i0}\bar{p}_i$. Hence, the assumptions of \Cref{continuitylemma} are satisfied.
	
	Let us calculate the conjugate function $\tilde{\Phi}$: for every $x^\ast\in\R^n_+$, by \eqref{LP}, we have
	\begin{align*}
	\tilde{\Phi}(x^\ast)&=\sup_{x\in \R^n_+}\of{-x^{\mathsf{T}}x^\ast+\tilde{\Lambda}(x)}
	=\sup_{0\leq p\leq \bar{p}}\of{\sum_{i=1}^{n}a_{i0}p_i-\inf_{ \substack{ x\geq0 \\x\geq p-A^{\mathsf{T}}p}}\sum_{i=1}^n x_i^\ast x_i}\\
	&=\sup_{0\leq p\leq \bar{p}}\sum_{i=1}^{n}\Big(a_{i0}p_i- x_i^\ast\Big( p_i-\sum_{j=1}^{n}a_{ji}p_j\Big)^+\Big).
	\end{align*}
	Then, by \Cref{propeisnoe3}, we have
	\begin{align*}
	&\rho\circ \Lambda(X)
	=\sup_{\substack{X^\ast\in L^q(\R^n_-)\setminus \{0\},\\ Y^*\in L^q(\R_{--})}}\beta_{\rho }\of{Y^*,\E\sqb{X^{\mathsf{T}}X^*+Y^\ast \Phi \of{\frac{X^\ast}{Y^\ast}}}}.
	\end{align*}
	We can pass to the probabilistic setting by using \eqref{convexeq} as follows:
	\begin{align*}
	\rho\circ\Lambda(X)=\sup_{\substack{w \in \mathbb{R}^n_+\setminus\{0\},\S\in \mathcal{M}_n^q (\P),\\ \Q \in \mathcal{M}_1^q (\P)\colon w_i\S_i\ll\Q\ \forall i}}\beta_{\rho }\of{-\frac{d\Q}{d\P},-\mathbb{E}_{\Q}\sqb{\Phi\of{w\cdot \frac{\mathbf{d}\S}{d\Q}}}-w^{\mathsf{T}}\E_{\S}\sqb{X}}.
	\end{align*}
	As a special case, let us assume that $\rho$ is the certainty equivalent associated to the logarithmic loss function (see \Cref{risk1}(ii)) for the case $p=1$. Then, the dual representation simplifies as
	\begin{equation}\label{EisNoedual}
	\rho\circ \Lambda (X)=\sup_{\substack{w \in \R^n_+\setminus\{0\},\S\in \mathcal{M}_n^\infty (\P),\\ \Q \in \mathcal{M}^\infty_1(\P)\colon w_i\S_i\ll\Q\ \forall i}}
	\frac{w^{\mathsf{T}}\E_{\S}\sqb{-X}-\mathbb{E}_{\Q}\sqb{\Phi\of{w\cdot \frac{\mathbf{d}\S}{d\Q}}}}{e^{\mathbb{E}\sqb{\ln\of{\frac{d\mathbb{Q}}{d\mathbb{P}}}}}}.
	\end{equation}
	
	The economic interpretation of \eqref{EisNoedual} is similar to the one at the end of \Cref{sec:ex}. Different from the examples in \Cref{sec:ex}, the multivariate divergence term here is specific to the Eisenberg-Noe model. Hence, we focus on the interpretation of this term. With the help of \cite[Thm. 14.60]{Rockawets}, we can calculate the divergence term more explicitly as
	\begin{align*}
	\E_{\Q}\sqb{\Phi\of{w\cdot\frac{\mathbf{d}\S}{d\Q}}} &= \E_{\Q}\sqb{\sup_{0\leq p\leq \bar{p}}\sum_{i=1}^{n}\Big(a_{i0}p_i- w_i\frac{d\S_i}{d\Q}\Big( p_i-\sum_{j=1}^{n}a_{ji}p_j\Big)^+\Big)}\\
	&= \sup_{P\in L^1(\Q,[0,\bar{p}])}\of{\E_{\Q}\sqb{\sum_{i=1}^{n} a_{i0}P_i}- \sum_{i=1}^n w_i\E_{\S_i}\Bigg[\Big( P_i-\sum_{j=1}^{n}a_{ji}P_j\Big)^+\Bigg]},
	\end{align*}
	where $L^1(\Q,[0,\bar{p}])$ denotes the space of random vectors of the probability space $(\Omega,\F,\Q)$ that take values in the rectangle $[0,\bar{p}]$. Hence, under the supremum, we consider a scenario-dependent payment vector $P$. The term $\sum_{i=1}^{n} a_{i0}P_i$ represents the total payment received by society. Therefore, we calculate its expectation with respect to $\Q$, that is, with respect to society's own perspective. Let us fix a bank $i\in\{1,\ldots,n\}$. Then, $(P_i-\sum_{j=1}^{n}a_{ji}P_j)^+$ is the net equity of bank $i$; we calculate its expectation with respect to $\S_i$, that is, with respect to the bank's own perspective. Hence, the weighted sum $\sum_{i=1}^n w_i\E_{\S_i}[( P_i -\sum_{j=1}^{n}a_{ji}P_j)^+]$ can be seen as the expected net equity from the perspective of the overall network (besides society). Then, the difference $\E_\Q[\sum_{i=1}^n a_{i0}P_i]- \sum_{i=1}^n w_i\E_{\S_i}[( P_i-\sum_{j=1}^{n}a_{ji}P_j)^+]$ is a measure of the mismatch between society's expectation and the network's overall expectation for the payments. Finally, the multivariate divergence term, as a directed distance from $w\cdot\S$ to $\Q$, is calculated as the largest possible value of this mismatch over all choices of the random payment vector $P$.

\section{Conclusion}\label{sec: conclusion}

In the first part of the paper, we provide dual representation theorems for quasiconvex compositions in locally convex topological vector spaces. The provided formula for the penalty function of the composition is expressed in terms of the penalty functions of the ingredient functions. 
In the second part, we use these general results to obtain dual representations for a systemic risk measure that is the composition of a quasiconvex risk measure and an aggregation function. Such systemic risk measures were introduced in \cite{chen} in the coherent case and they simply quantify the risk of the random aggregate output of the network under a stress scenario. More capital-sensitive systemic risk measures that rely on a direct capital injection mechanism were studied later in \cite{feinstein,biagini,ararat} in the convex case using set-valued risk measures and their scalarizations. Studying the quasiconvex counterpart of these sensitive systemic risk measures is an interesting direction that we leave for future research.

\appendix

\section{Proofs of some results in Section~\ref{sec:quasicon} and Section~\ref{sec:natquasi}}\label{app: 2 and 3}
\subsection{Proof of some results in Section~\ref{sec:quasicon}}\label{sec:app2}

\begin{proof}[Proof of \Cref{cor:singleconv}]
Let us define a function $\tilde{g}\colon\X\to\R$ by
\[
\tilde{g}(x)\coloneqq \inf\cb{m\in\R\mid x\in \cl(S_g^m)},\quad x\in\X.
\]
Note that $S_{\tilde{g}}^m=\cl(S_g^m)$ for each $m\in\R$. Let $m\in\R$. Since $g$ is quasiconvex, it follows that $S^m_{\tilde{g}}$ is closed and convex. To show that it is also monotone, let $x\in S^m_{\tilde{g}}=\cl(S_g^m)$, $c\in C$. Let $U\subseteq\X$ be a neighborhood of $x+c$. Since $\X$ is a topological vector space, $\of{U-c}$ is an open set; hence, it is a neighborhood of $x$. Therefore, $\of{U-c} \cap S_g^m \neq \emptyset$. Let $z\in (U-c) \cap S_g^m$ so that $z+c \in U$. On the other hand, since $g$ is decreasing, $S_g^m$ is monotone, which yields that $z+c\in S_g^m$. It follows that $U\cap S_g^m\neq \emptyset$. Since $U$ is an arbitrary neighborhood of $x+c$, we conclude that $x+c\in \cl(S_g^m)=S_{\tilde{g}}^m$. Hence, $S_{\tilde{g}}^m$ is monotone. By Remarks \ref{remclosed}, \ref{remmonotone}, it follows that $\tilde{g}$ is decreasing, lower semicontinuous, and quasiconvex. Then, by \Cref{firstthm}, we get 
\begin{equation}\label{eqeqeq}
	\tilde{g}(x)=\sup_{x^\ast\in C^{\circ}\setminus \{0\}} \beta_{\tilde{g}}\of{x^\ast,\ip{ x^\ast,x}}, \quad x\in \X.	
\end{equation}
By definition, $S_{\tilde{g}}^m$ is the closed convex hull of $S_g^m$ for each $m\in\R$. Hence, \eqref{chull} yields
\begin{equation}\label{eqalpha}
	\alpha_{\tilde{g}}(x^\ast,m)=\sup_{y\in S_{\tilde{g}}^m}\ip{x^\ast,y}=\sup_{y\in S_g^m}\ip{x^\ast,y}=\alpha_g(x^\ast,m),\quad x^\ast\in\X^\ast,m\in\R.	
\end{equation}
For $x\in \mathcal{K}$, by \eqref{remf}, we have
\begin{equation}\label{eqg}
	\tilde{g}(x)=\inf\cb{m\in \R \mid x\in S_{\tilde{g}}^m}=\inf\cb{m\in \R \mid x\in S_{\tilde{g}}^m\cap \mathcal{K}}.
\end{equation}
We claim that $S_{\tilde{g}}^m\cap \mathcal{K}=S_g^m$. Indeed, it is clear that $S_{\tilde{g}}^m\cap \mathcal{K}=\cl(S_g^m)\cap\mathcal{K}\supseteq S_g^m$. On the other hand, since $g$ is lower semicontinuous with respect to the relative topology, we have $S_g^m=A\cap \mathcal{K}$ for some closed set $A\subseteq\X$. Since $S_g^m\subseteq A$, we have $\cl(S_g^m)\subseteq A$. It follows that $\cl(S_g^m)\cap \mathcal{K}\subseteq A\cap \mathcal{K}=S_g^m$. Hence, the claim follows. Then, \eqref{eqg} yields
$\tilde{g}(x)=\inf\{m\in \R \mid x\in S_g^m\}=g(x)$.
Combining this with \eqref{eqeqeq}, \eqref{eqalpha}, we get \eqref{eq:convexrest}.
\end{proof}

\begin{proof}[Proof of Proposition \ref{convexcase}]
Let $x^\ast\in\X^\ast\setminus\{0\}$, $m\in\R$ be such that $\{x\in \X\mid f(x)<m\}\neq\emptyset$. Note that $\alpha_f(x^\ast,m)=\sup_{x\in S_f^m}\ip{x^\ast,x}$ can be seen as the optimal value of the following convex optimization problem:
\[
\text{maximize } \ip{x^\ast,x}\text{ subject to }f(x)\leq m,\; x\in\X.
\]
By supposition, Slater's condition holds, that is, there exists $x_0\in\X$ such that $f(x_0)<m$. Hence, we have strong duality for this problem, that is,
\[
\alpha_f(x^\ast,m)=\inf_{\lambda\geq 0}\sup_{x\in \dom f}\of{\ip{x^\ast,x}-\lambda(f(x)-m)}.
\]
When $\lambda=0$,  $\sup_{x\in \dom f}\of{\ip{x^\ast,x}-\lambda(f(x)-m)}=I^\ast_{\dom f}(x^\ast)$. Moreover, we may evaluate the infimum over $\lambda>0$ separately as
\begin{align*}
 &\inf_{\lambda > 0}\sup_{x\in \dom f}\of{ \ip{x^\ast,x}-\lambda(f(x)-m)}\\
	&=\inf_{\lambda > 0}\of{\lambda m+\sup_{x\in \dom f}\of{ \ip{x^\ast,x}-\lambda f(x)}}=\inf_{\lambda >0}\of{\lambda m+ \lambda f^\ast\of{\frac{x^*}{\lambda}}}.
\end{align*}
Therefore, (i) follows. To prove (ii), first note that $\alpha_f(x^\ast,m)=-\infty$ for $m<\inf_{x\in \X}f(x)$ by definition. Hence,
\[
\beta_f(x^\ast,s)=\inf\cb{m \geq \inf_{x\in\X}f(x)\mid\alpha_f(x^\ast,m)\geq s }=\inf\cb{m \in F\mid\alpha_f(x^\ast,m)\geq s },
\]
where $F\coloneqq (\inf_{x\in\X}f(x),+\infty)$. Moreover,
the strict sublevel set $\cb{x\in \X\mid f(x)<m}$ is nonempty for $m\in F$. Then, by (i), we get
\[
\beta_f(x^\ast,s)
=\inf\cb{m\in F \mid I^\ast_{\dom f}(x^\ast) \wedge \inf_{\lambda >0}\of{\lambda m+ \lambda f^\ast\of{\frac{x^*}{\lambda}}}\geq s}.
\]
In particular, for each $s>I^{\ast}_{\dom f}(x^\ast)$, we immediately have $\beta_f(x^\ast,s)=+\infty$. Let $s\leq I^\ast_{\dom f}(x^\ast)$. Then, we have 
\begin{align*}
\beta_f(x^\ast,s)
&=\inf\cb{m\in F \mid \inf_{\lambda >0}\of{\lambda m+ \lambda f^\ast\of{\frac{x^*}{\lambda}}}\geq s}\\
&=\inf\cb{m\in F \mid \forall \lambda >0\colon \lambda m+ \lambda f^\ast\of{\frac{x^*}{\lambda}}\geq s}\\
&=\inf\cb{m\in F \mid \forall \lambda >0\colon m\geq \frac{s}{\lambda}-f^\ast\of{\frac{x^*}{\lambda}}}\\
&=\inf\cb{m>\inf_{x\in\X}f(x) \mid m\geq \sup_{\lambda>0}\of{\frac{s}{\lambda}-f^\ast\of{\frac{x^*}{\lambda}}}}\\
&=\inf_{x\in\X}f(x)\vee \sup_{\gamma>0}\of{\gamma s-f^\ast(\gamma x^\ast)}\\
&=-f^\ast(0)\vee \sup_{\gamma>0}\of{\gamma s-f^\ast(\gamma x^\ast)}=\sup_{\gamma\geq 0}\of{\gamma s-f^\ast(\gamma x^\ast)},
\end{align*}
which completes the proof.
\end{proof}

\subsection{Proofs of some results in Section~\ref{sec:natquasi}}\label{sec:app3}

\begin{proof}[Proof of \Cref{star}]
We prove (i) first. Let $x_1,x_2\in \X$ such that $x_1\leq _C x_2$. Note that we have $g(x_1)\leq_D g(x_2)$ if and only if $y^\ast\circ g(x_1)\geq y^\ast\circ g(x_2)$. Hence, $g$ is $D$-increasing if and only if $y^\ast\circ g$ is decreasing for every $y^\ast\in D^\circ\setminus\{0\}$.

Condition (ii) follows from \cite[Lem.~3.1]{kuroiwa2016}.

Next, we prove (iii) similar to the set-valued case in \cite[Thm.~2.1]{kur}. Assume that $g$ is $D$-naturally quasiconcave. Let $y^\ast \in D^{\circ}\setminus\{0\}$ and consider $y^{\ast}\circ g$. Let $x_1,x_2 \in \X$ and $\lambda \in [0,1]$. Since $g$ is $D$-naturally quasiconcave, there exists $\mu \in [0,1]$ such that $g(\lambda x_1+ (1-\lambda) x_2)\geq_D \mu g(x_1)+(1-\mu)g(x_2)$. Hence,
\[
	\ip{y^\ast,g(\lambda x_1+(1-\lambda)x_2)}
	\leq \ip{y^\ast, \mu g(x_1)+(1-\mu)g(x_2)}
	\leq \ip{y^\ast,g(x_1)}\vee \ip{y^\ast,g(x_2)}.
\]
Therefore, $y^{\ast}\circ g$ is quasiconvex.

Conversely, assume that $y^{\ast}\circ g$ is quasiconvex for each $y^\ast\in D^{\circ}\setminus\{0\}$. To get a contradiction, suppose that $g$ is not $D$-naturally quasiconcave. Hence, there exist $x_1,x_2\in \mathcal{X}$, $\lambda\in [0,1]$ such that
\[
\of{g(\lambda x_1+(1-\lambda) x_2)-\conv\of{\cb{g(x_1),g(x_2)}}}\cap D=\emptyset.
\]
Since the set $D$ is closed and convex, and the (shifted) line segment $g(\lambda x_1+(1-\lambda) x_2)-\conv(\{g(x_1),g(x_2)\})$ is compact and convex, by Hahn-Banach strong separation theorem, there exists $y_0^\ast\in \Y^\ast\setminus\{0\}$ with
\begin{equation}\label{sepresult}
	\sup_{d \in D} \ip{y_0^\ast,d} < \inf_{y\in g(\lambda x_1+(1-\lambda) x_2)-\conv(\{g(x_1),g(x_2)\})} \ip{y_0^\ast,y}
\end{equation}
Since $D$ is a cone, $\sup_{d\in D}\ip{y^\ast,d}$ is either $0$ or $+\infty$. However, the term on the right of \eqref{sepresult} is finite. Hence, we must have $\sup_{d\in D}\ip{y_0^\ast,d}=0$ so that $y_0^\ast\in D^{\circ}$. Using this information in \eqref{sepresult} implies
$
\ip{y_0^\ast,\mu g(x_1)+(1-\mu)g(x_2)}<\ip{y_0^\ast,g(\lambda x_1+(1-\lambda)x_2)}
$
for every $\mu\in[0,1]$. It follows that 
\[
\ip{y_0^\ast,g(x_1)} \vee \ip{y_0^\ast,g(x_2)}< \ip{y_0^\ast, g(\lambda x_1+(1-\lambda)x_2)},
\]
which contradicts the quasiconvexity of $y_0^{\ast}\circ g$. Hence, $g$ is $D$-naturally quasiconvex.

Finally, we prove (iv). Let $m\in\R$ and $y^\ast\in D^{\circ}\setminus\{0\}$. We claim that $S_{y^\ast\circ g}^m=G^U(S^m_{y^\ast})$. First, let $x\in S_{y^\ast\circ g}^m$ and take $d\in D$. Hence, $\ip{y^\ast,g(x)}\leq m$ and $\ip{y^*,d}\leq 0$. Combining these two inequalities yields $\ip{y^\ast,g(x)+d}\leq m$, that is, $g(x)+d\in S^m_{y^\ast}$. Since $d\in D$ is arbitrary, we have $g(x)+D\subseteq S^m_{y^\ast}$, i.e., $x\in G^U(S^m_{y^\ast})$. Conversely, let $x\in G^U(S^m_{y^\ast})$. In particular, $g(x)\in S^m_{y^\ast}$, i.e., $\ip{y^\ast,g(x)}\leq m$. Hence, $x\in S^m_{y^\ast\circ g}$, which completes the proof of the claim. By this claim and \Cref{rem:ldc}, (iv) follows.
\end{proof}

\section{Proofs of some results in Section~\ref{sec:comp}}\label{app:comp}
\subsection{Proofs of some results in Subsection~\ref{sec:main}}\label{sec:appsec4}

\begin{proof}[Proof of \Cref{chul}]
If $A^m_{y^\ast}=\emptyset$, then the result is obvious. Let us assume that $A^m_{y^\ast}\neq\emptyset$ and prove that $A^m_{y^\ast}=\cl(\tilde{A}^m_{y^\ast})$. Since $\tilde{A}^m_{y^\ast}\subseteq A^m_{y^\ast}$ and $A^m_{y^\ast}$ is closed, we have $\cl(\tilde{A}^m_{y^\ast})\subseteq A^m_{y^\ast}$. Next, let $x\in A^m_{y^\ast}$ and fix $c\in C^{\#}$, $\lambda >0$. Since $C^{\#}$ is a cone, we have $\lambda c \in C^{\#}$. Since $g$ is $D$-regularly increasing, we have $g(x+\lambda c)-g(x)\in D^{\#}$. In particular, since $y^\ast\in D^{\circ}\setminus\{0\}$, we have $\ip{y^\ast,g(x+\lambda c)-g(x) }<0$. Therefore,
\begin{align*}
	\ip{y^\ast,g(x+\lambda c)}&=\ip{y^\ast,g(x)}+\ip{y^\ast,g(x+\lambda c)-g(x)}\\
	&\leq \alpha_f(y^\ast,m)+\ip{y^\ast,g(x+\lambda c)-g(x)} < \alpha_f(y^\ast,m).
\end{align*}
Hence, $x+\lambda c \in \tilde{A}^m_{y^\ast}$. The net $(x+\lambda c)_{\lambda >0}$ in $\tilde{A}^m_{y^\ast}$ converges to $x$ as $\lambda \rightarrow 0$, which implies that $x\in\cl(\tilde{A}^m_{y^\ast})$. Hence, $A^m_{y^\ast}\subseteq \cl (\tilde{A}^m_{y^\ast})$ as well. Finally, since $A^m_{y^\ast}$ is convex, we have
$A_{y^\ast}^m=\conv(\cl(\tilde{A}_{y^\ast}^m))\subseteq\cl(\conv \tilde{A}_{y^\ast}^m)\subseteq A_{y^\ast}^m$. This shows that $A^m_{y^\ast}=\cl\conv(\tilde{A}_{y^\ast}^m)$.
\end{proof}

\begin{proof}[Proof of \Cref{kkt}]
Let $y^\ast\in\bar{D}^{\circ}$. By definition, we have
\begin{equation}\label{eq:Keq}
	\sup_{x\in \X} \tilde{K}^m_{x^\ast}(x,y^\ast)=\sup_{x\in \X} \big(\ip{x^\ast,x}-I_{\tilde{A}^m_{y^\ast}}(x)\big)=I^\ast_{\tilde{A}^m_{y^\ast}}(x^\ast).
\end{equation}
By \eqref{chull} and \Cref{chul}, we have
$I^\ast_{\tilde{A}^m_{y^\ast}}(x^\ast)=\sup_{x\in A^m_{y^\ast}}\ip{x^\ast,x}$.
Similarly,
\[
\sup_{x\in \X} K^m_{x^\ast}(x,y^\ast)=\sup_{x\in \X} \big(\ip{x^\ast,x}-I_{A^m_{y^\ast}}(x)\big)= \sup_{x\in A^m_{y^\ast}}\ip{x^\ast,x}.
\] 
Combining these gives the desired result.
\end{proof}

\begin{proof}[Proof of \Cref{propk}]
To prove (i), let $y^\ast\in \bar{D}^{\circ}$. Since $A^m_{y^\ast}$ is a closed convex set, $I_{A^m_{y^\ast}}$ is a lower semicontinuous convex function. Hence, $x\mapsto K^m_{x^\ast}(x,y^\ast)$ is an upper semicontinuous concave function.

Next, let $x\in \X$. We claim that $y^\ast\mapsto I_{A^m_{y^\ast}}(x)$ is a quasiconvex function. Indeed, let $y^\ast_1, y^\ast_2 \in \bar{D}^{\circ}$, $\lambda \in [0,1]$ and define $y^\ast\coloneqq\lambda y^\ast_1+(1-\lambda)y^\ast_2$. Since $\bar{D}^{\circ}$ is convex, $y^\ast \in \bar{D}^{\circ}$. If $x\in A^m_{y_1^*}$ or $x\in A^m_{y_2^*}$, then
$
I_{A^m_{y_1^\ast}}(x) \wedge I_{A^m_{y_2^\ast}}(x)=0 \leq I_{A^m_{y^\ast}}(x)
$
by the definition of indicator function. On the other hand, suppose that $x\notin A^m_{y_1^\ast}$ and $x\notin A^m_{y_2^\ast}$. Then, $\ip{y_1^\ast,g(x)}> \alpha_f(y_1^\ast,m)$ and $\ip{y_2^\ast,g(x)}> \alpha_f(y_2^\ast,m)$. Hence,
\begin{align*}
	\ip{y^\ast,g(x)}&>\lambda \alpha_f(y_1^\ast,m)+(1-\lambda)\alpha_f(y_2^\ast,m)\\
	&= \lambda \sup_{y\in S_f^m}\ip{ y_1^\ast,y}+(1-\lambda) \sup_{y\in S_f^m}\ip{y_2^\ast,y}\geq \sup_{y\in S_f^m}\ip{y^\ast,y}=\alpha_f(y^\ast,m).
\end{align*}
Therefore, $x\notin A^m_{y^\ast}$ so that 
$
I_{A^m_{y_1^\ast}}(x) \wedge I_{A^m_{y_2^\ast}}(x) \leq +\infty= I_{A^m_{y^\ast}}(x).
$
It follows that $y^\ast\mapsto I_{A^m_{y^*}}(x)$ is quasiconvex, hence so is $y^\ast\mapsto K^m_{x^\ast}(x,y^\ast)$.

To prove (ii), let $y^\ast\in \bar{D}^{\circ}$. We claim that $\tilde{A}^m_{y^\ast}$ is a convex set. Indeed, let $x_1, x_2 \in  \tilde{A}^m_{y^\ast}$ and $\lambda \in [0,1]$. Since $y^{\ast}\circ g$ is quasiconvex, we have
\[
y^{\ast}\circ g(\lambda x_1 + (1-\lambda)x_2)\leq y^{\ast}\circ g(x_1) \vee y^{\ast}\circ g(x_2)< \alpha_f(y^\ast,m),
\]
which implies that $\lambda x_1 + (1-\lambda)x_2 \in \tilde{A}^m_{y^*}$. Hence, the claim follows. Therefore, $I_{\tilde{A}^m_{y^\ast}}$ is a convex function and $x\mapsto \tilde{K}^m_{x^\ast}(x,y^\ast)$ is a concave function.

Let $x\in \X$. We show that $y^\ast\mapsto I_{\tilde{A}_{y^\ast}^m}(x)$ is quasiconvex. Let $y^\ast_1, y^\ast_2 \in \bar{D}^{\circ}$, $\lambda \in [0,1]$ and define $y^\ast\coloneqq \lambda y^\ast_1+(1-\lambda)y^\ast_2$. Since $\bar{D}^{\circ}$ is convex, $y^\ast\in \bar{D}^{\circ}$. If $x\in\tilde{A}^m_{y_1^\ast}$ or $x\in \tilde{A}^m_{y_2^\ast}$, then
$
I_{\tilde{A}^m_{y_1^\ast}}(x)\wedge I_{\tilde{A}^m_{y_2^\ast}}(x)=0 \leq I_{\tilde{A}^m_{y^\ast}}(x)$. Suppose that $x\notin \tilde{A}^m_{y_1^\ast}$ and $x\notin \tilde{A}^m_{y_2^\ast}$. Then, $\ip{y_1^\ast,g(x)}\geq \alpha_f(y_1^\ast,m)$ and $\ip{y_2^\ast,g(x)}\geq \alpha_f(y_2^\ast,m)$ so that
\begin{align*}
	\ip{y^\ast,g(x)}
	&\geq \lambda \alpha_f(y_1^\ast,m)+(1-\lambda)\alpha_f(y_2^\ast,m)\\
	&= \lambda \sup_{y\in S_f^m}\ip{ y_1^\ast,y}+(1-\lambda) \sup_{y\in S_f^m}\ip{y_2^\ast,y}\geq \sup_{y\in S_f^m}\ip{y^\ast,y}=\alpha_f(y^\ast,m),
\end{align*}
which implies that $x\notin \tilde{A}^m_{y^\ast}$. Hence, $I_{\tilde{A}^m_{y_1^\ast}}(x) \wedge I_{\tilde{A}^m_{y_2^\ast}}(x) \leq +\infty= I_{\tilde{A}^m_{y^\ast}}(x)$,
which completes the proof of quasiconvexity. It follows that $y^\ast\mapsto \tilde{K}^m_{x^\ast}(x,y^\ast)$ is quasiconvex.

Finally, to prove lower semicontinuity, let us define the set
\[
E^m_x\coloneqq\cb{y^\ast\in \bar{D}^{\circ}\mid \ip{y^\ast,g(x)}< \alpha_f(y^\ast,m)}=\Big\{y^\ast\in \bar{D}^{\circ}\mid 0< \sup_{y\in S_f^m}\ip{y^\ast,y-g(x)}\Big\}.
\]
Since the supremum of a family of continuous affine functions is lower semicontinuous, it follows that $E^m_x$ is open. On the other hand,	for each $y^\ast \in \bar{D}^{\circ}$, it is clear that $y^\ast\in E^m_x$ if and only if $x\in \tilde{A}^m_{y^\ast}$, that is, $I_{\tilde{A}^m_{y^\ast}}(x)=I_{E^m_x}(y^\ast)$. Hence, we indeed have 
\begin{equation}\label{eq:tildeK}
	\tilde{K}^m_{x^\ast}(x,y^\ast)=\ip{x^\ast,x}-I_{\tilde{A}^m_{y^\ast}}(x)=\ip{x^\ast,x}-I_{E^m_x}(y^\ast).
\end{equation}
Since $E^m_x$ is open, $I_{E^m_x}$ is upper semicontinuous. Then, by \eqref{eq:tildeK}, $y^\ast\mapsto \tilde{K}^m_{x^\ast}(x,y^*)$ is lower semicontinuous.
\end{proof}

\begin{proof}[Proof of \Cref{alfamin}]
Let $x^\ast\in C^{\circ}$, $m\in\R$. Since $f$ is decreasing, lower semicontinuous, and quasiconvex, by Remarks \ref{separationlem}, \ref{remcg}, we have
\begin{align*}
	\alpha_{f\circ g} (x^\ast,m)&= \sup_{x\in S_{f\circ g}^m} \ip{ x^\ast,x}= \sup \{\ip{ x^\ast,x}\mid g(x)\in S_f^m,\ x\in\X\} \\
	&=\sup_{x\in\X} \cb{\ip{x^\ast,x}\mid\forall y^\ast\in D^{\circ}\setminus \{0\}\colon \ip{y^\ast,g(x)}\leq \alpha_f(y^\ast,m)}\\
	&=\sup_{x\in\X} \cb{\ip{x^\ast,x}\mid\forall y^\ast\in \bar{D}^{\circ}\colon \ip{y^\ast,g(x)}\leq \alpha_f(y^\ast,m)}=\sup_{x\in B^m}\ip{x^\ast,x},
\end{align*}
where $B^m\coloneqq \bigcap_{y^\ast\in \bar{D}^{\circ}}A^m_{y^\ast}$.
Moreover,
\[
\sup_{x\in B^m}\ip{x^\ast,x}=\sup_{x\in\X}\of{\ip{x^\ast,x}-I_{B^m}(x)}=\sup_{x\in\X}\inf_{y^\ast\in \bar{D}^{\circ}}(\ip{x^\ast,x}-I_{A^m_{y^\ast}}(x)).
\]
Recalling the definition of $K^m_{x^\ast}$ in \eqref{KK}, the result follows.
\end{proof}

\begin{proof}[Proof of \Cref{infhg}]
Let $\bar{y}^\ast\in \bar{D}^{\circ}$. Clearly, we have
\[
\sup_{x\in \X} K^m_{x^\ast}(x,\bar{y}^\ast)=\sup_{x\in \X} (\ip{x^\ast,x}-I_{A^m_{\bar{y}^\ast}}(x)) =\sup_{x\in A^m_{\bar{y}^\ast}}\ip{x^\ast,x}.
\]
Hence,
\begin{align*}
	\inf_{\bar{y}^\ast\in \bar{D}^{\circ}\setminus\{0\}} \alpha_{\bar{y}^{\ast}\circ g}(x^\ast,\alpha_f(\bar{y}^\ast,m))
	&=\inf_{\bar{y}^\ast\in D^{\circ}\setminus\{0\}} \sup_{x\in\X}\cb{\ip{x^\ast,x}\mid\ip{\bar{y}^\ast,g(x)}\leq \alpha_f(\bar{y}^\ast,m)}\\
	&=\inf_{\bar{y}^\ast\in D^{\circ}\setminus\{0\}} \sup_{x\in A^m_{\bar{y}^\ast}} \ip{x^\ast,x}=\inf_{\bar{y}^\ast\in D^{\circ}\setminus\{0\}} \sup_{x\in \X} K^m_{x^\ast}(x,\bar{y}^\ast),
\end{align*} 
which proves the second equality in the proposition. On the other hand, given $y^\ast\in D^{\circ}\setminus\{0\}$, we may write $y^\ast=\lambda \bar{y}^\ast$ for some $\lambda >0$ and $\bar{y}^\ast\in\bar{D}^{\circ}$. Then, by \Cref{remcg},
\[
\alpha_{y^{\ast}\circ g}(x^\ast,\alpha_f(y^\ast,m))=\sup_{x\in A^m_{y^\ast}}\ip{x^\ast,x}=\sup_{x\in A^m_{\bar{y}^\ast}}\ip{x^\ast,x} = \alpha_{\bar{y}^{\ast}\circ g}(x^\ast,\alpha_{f}(\bar{y}^\ast,m)).
\]
Hence, the first equality in the proposition follows as well.
\end{proof}

\subsection{Proofs of the results in Subsection~\ref{sec:special}}\label{app:special}

\begin{proof}[Proof of \Cref{bothconvex}]
Let $x\in\X$ be such that $g(x)\in\dom f$. By \Cref{newthm}, we have
\[
f\circ g(x)=\sup_{x^\ast\in C^{\circ}\setminus\{0\}} \sup_{y^\ast\in D^{\circ}\setminus\{0\}} \beta_f\Big(y^\ast,\beta_{y^\ast \circ g}\of{x^\ast,\ip{x^\ast,x}}\Big).
\]
Since $\dom y^\ast \circ g=\mathcal{X}$, by applying \Cref{convexcase}(ii) to $y^\ast \circ g$, we get
\begin{align}
f\circ g(x)
&=\sup_{x^\ast\in C^{\circ}\setminus\{0\}} \sup_{y^\ast\in D^{\circ}\setminus\{0\}} \beta_f\Big(y^\ast,\sup_{\gamma \geq 0} \of{\ip{\gamma x^\ast,x}-(y^\ast \circ g)^\ast(\gamma x^\ast)}\Big)
\notag  \\ 
&=\sup_{x^\ast\in C^{\circ}\setminus\{0\}} \sup_{y^\ast\in D^{\circ}\setminus\{0\}} \sup_{\gamma \geq 0}\beta_f\Big(y^\ast, \ip{\gamma x^\ast,x}-(y^\ast \circ g)^\ast(\gamma x^\ast)\Big)
\notag  \\ 
&=\sup_{\tilde{x}^\ast\in C^{\circ}} \sup_{y^\ast\in D^{\circ}\setminus\{0\}}\beta_f\Big(y^\ast, \ip{\tilde{x}^\ast,x}-(y^\ast \circ g)^\ast(\tilde{x}^\ast)\Big), \label{beta_f_before}
\end{align}
where the last equality comes from the change-of-variables $\gamma x^\ast=\tilde{x}^\ast$ since $C^{\circ}$ is a cone. For each $\tilde{x}^\ast\in C^\circ$ and $y^\ast\in D^\circ\setminus\{0\}$, using $g(x)\in \dom f$, we have 
\begin{align*}
    \ip{\tilde{x}^\ast,x}-(y^\ast \circ g)^\ast(\tilde{x}^\ast)\leq \sup_{x^\ast \in C^{\circ}}(\ip{x^\ast,x}-(y^\ast\circ g)^\ast(x^\ast))=y^\ast \circ g(x)
    \leq I^\ast_{\dom f}(y^\ast),
\end{align*}
where we use Propositions~\ref{fenchel} and \ref{star} for the equality. Therefore, we can apply \Cref{convexcase}(ii) to $f$ in \eqref{beta_f_before} and obtain 
\begin{align}
f\circ g(x)
&=\sup_{x^\ast\in C^{\circ}} \sup_{y^\ast\in D^{\circ}\setminus\{0\}}\sup_{\gamma \geq0}\Big(\gamma\of{\ip{x^\ast,x}-(y^\ast \circ g)^\ast( x^\ast)}-f^\ast(\gamma y^\ast)\Big)
\notag  \\
&=\sup_{y^\ast\in D^{\circ}\setminus\{0\}}\sup_{\gamma \geq 0}\of{-f^\ast(\gamma y^\ast)+\sup_{x^\ast\in C^{\circ}} \gamma\of{\ip{x^\ast,x}-(y^\ast \circ g)^\ast( x^\ast)}}. \label{gamma_SUP}
\end{align}
Let $y^\ast\in D^\circ\setminus\{0\}$. For each $\gamma> 0$, since $\{\gamma x^\ast\mid x^\ast\in C^\circ\}=C^\circ$, we have
\begin{align}
   \sup_{x^\ast\in C^{\circ}} \gamma\of{\ip{x^\ast,x}-(y^\ast \circ g)^\ast( x^\ast)} &= \sup_{x^\ast \in C^{\circ}}\of{\ip{\gamma x^\ast,x}-\sup_{z\in \mathcal{X}}\of{\ip{\gamma x^\ast,z}-\ip{\gamma y^\ast,g(z)}}} \nonumber
   \\&=\sup_{\tilde{x}^\ast\in C^{\circ}}\of{\ip{\tilde{x}^\ast,x}-\sup_{z\in \mathcal{X}}\of{\ip{\tilde{x}^\ast,z}-\ip{\gamma y^\ast,g(z)}}} \nonumber
   \\&=\sup_{\tilde{x}^\ast\in C^{\circ}} \of{\ip{\tilde{x}^\ast,x}-((\gamma y^\ast)\circ g)^\ast(\tilde{x}^\ast)}.\label{gamma_RHS}
\end{align}
On the other hand, when $\gamma=0$, we have $(\gamma y^\ast\circ g)^\ast(\tilde{x}^\ast)=0$ for $x^\ast=0$ and $(\gamma y^\ast\circ g)^\ast(\tilde{x}^\ast)=+\infty$ for every $\tilde{x}^\ast\in C^\circ\setminus\{0\}$. Then, the supremum in \eqref{gamma_RHS} yields $0$ in this case. Hence, we have
\[
 \sup_{x^\ast\in C^{\circ}} \gamma\of{\ip{x^\ast,x}-(y^\ast \circ g)^\ast( x^\ast)} =\sup_{\tilde{x}^\ast\in C^{\circ}} \of{\ip{\tilde{x}^\ast,x}-(\gamma y^\ast\circ g)^\ast(\tilde{x}^\ast)}
\]
for every $y^\ast\in D^\ast\setminus\{0\}$ and $\gamma\geq0$. Then, by \eqref{gamma_SUP}, we obtain
\begin{align*}
f\circ g(x)
&=\sup_{y^\ast\in D^{\circ}\setminus\{0\}}\sup_{\gamma \geq 0} \of{-f^\ast(\gamma y^\ast)+\sup_{x^\ast\in C^{\circ}} \of{\ip{x^\ast,x}-(\gamma y^\ast \circ g)^\ast( x^\ast)}}
\\&=\sup_{\tilde{y}^\ast\in D^{\circ}} \sup_{x^\ast\in C^{\circ}}\of{-f^\ast( \tilde{y}^\ast)+ \ip{x^\ast,x}-(\tilde{y}^\ast \circ g)^\ast( x^\ast)},
\end{align*}
where the last equality is by the change-of-variables $\gamma y^\ast=\tilde{y}^\ast$ since $D^{\circ}$ is a cone.
\end{proof}

\begin{proof}[Proof of \Cref{gconvex}]
Note that $x\mapsto \ip{y^\ast,g(x)}$ is convex, lower semicontinuous by \Cref{star}. Let $x\in\X$. By \Cref{monlemma}(i)
and Fenchel-Moreau theorem,
\begin{align*}
f\circ g(x)&=\sup_{y^\ast \in D^{\circ}\setminus\{0\}}\inf\cb{m\in \R \mid \ip{y^\ast,g(x)}\leq \alpha_f(y^\ast,m)}
\\&=\sup_{y^\ast \in D^{\circ}\setminus\{0\}}\inf\Big\{m\in \R \mid \sup_{x^\ast \in C^{\circ}\setminus \{0\}}\of{\ip{x^*,x}-(y^\ast \circ g)^\ast(x^*)}\leq \alpha_f(y^\ast,m)
\Big\}
\\&=\sup_{y^\ast \in D^{\circ}\setminus\{0\}}\sup_{x^\ast \in C^{\circ}\setminus \{0\}}\inf\cb{m\in \R \mid \ip{x^*,x}-(y^\ast \circ g)^\ast(x^*)\leq \alpha_f(y^\ast,m)}
\\&=\sup_{x^\ast\in C^{\circ}\setminus\{0\}} \sup_{y^\ast\in D^{\circ}\setminus\{0\}}\beta_f\Big(y^\ast,\ip{x^\ast,x}-(y^\ast \circ g)^\ast(x^\ast)\Big),
\end{align*}
where the third equality is by \Cref{monlemma}(ii). Hence, \eqref{gconvexdual} follows.

From now on, we assume that $g$ is $D$-regularly increasing and Assumptions~\ref{asmp},~\ref{asmp:cone} hold. To prove (i), let $x^\ast\in C^{\circ}\setminus\{0\}$, $m\in\R$ with $\alpha_f(y^\ast,m)\in\R$ and $A^m_{y^\ast} \neq \emptyset$ for each $y^\ast\in D^{\circ}\setminus\{0\}$. By \Cref{mainthm},
\[
\alpha_{f\circ g } (x^\ast,m)=\inf_{y^\ast\in D^{\circ}\setminus\{0\}} \alpha_{y^\ast \circ g}(x^\ast,\alpha_f(y^\ast,m)).
\]
Also,  take $x\in S^{\alpha_f(y^\ast,m)}_{y^\ast \circ g}$ and let $c\in C^{\#}$. Then, there exists $d \in D^{\#}$ such that $g(x+c)=g(x)+d$ since $g$ is regularly increasing. Therefore, by using the definition of $D^{\#}$, we get
\[
\ip{y^\ast,g(x+c)}=\ip{y^\ast,g(x)+d}=\ip{y^\ast,g(x)}+\ip{y^\ast,d}< \ip{y^\ast,g(x)}\leq {\alpha_f(y^\ast,m)},
\]
which gives that $\{x\in \X\mid y^\ast \circ g(x)<{\alpha_f(y^\ast,m)}\}\neq\emptyset$. Note that $\dom y^\ast \circ g = \mathcal{X}$. Hence, by \Cref{convexcase}(i) applied to $y^\ast\circ g$, we have 
\[
\alpha_{f\circ g } (x^\ast,m)=\inf_{y^\ast\in D^{\circ}\setminus\{0\}}\inf_{\gamma >0}\of{\gamma (y^\ast \circ g)^\ast\of{\frac{x^\ast}{\gamma}}+\gamma\alpha_f(y^\ast,m)}.
\]
Then, by \cite[Thm. 2.3.1]{zal} on the elementary rules of conjugation, we have
\[
\alpha_{f\circ g } (x^\ast,m)=\inf_{y^\ast\in D^{\circ}\setminus\{0\}}\inf_{\gamma >0}\of{(\gamma y^\ast \circ g)^\ast(-x^\ast)+\gamma\alpha_f(y^\ast,m)}.
\]
By the positive homogeneity of $y^\ast \mapsto \alpha_f(y^\ast,m)$ and that of $y^\ast\mapsto y^\ast \circ g(x)$ for each $x\in\X$, we get
\[
\alpha_{f\circ g } (x^\ast,m)=\inf_{y^\ast\in D^{\circ}\setminus\{0\}}\inf_{\gamma >0}\of{(\gamma y^\ast \circ g)^\ast(x^\ast)+\alpha_f(\gamma y^\ast,m)}.
\]
Finally, since $D^{\circ}$ is a cone, we can make a change of variables and obtain (i).	

We prove (ii) next. By \Cref{cor1}, \Cref{convexcase}(ii) applied to $y^\ast \circ g$, and the definition of left inverse, we have
\begin{align*}
\beta_{f\circ g } (x^\ast,s)&=\sup_{y^\ast\in D^{\circ}\setminus\{0\}}\beta_f\Big(y^\ast,\beta_{y^\ast \circ g}(x^\ast,s)\Big)\\
&=\sup_{y^\ast\in D^{\circ}\setminus\{0\}}\beta_f\Big(y^*,\sup_{\gamma \geq 0} \big(\gamma s-(y^\ast \circ g)^\ast(\gamma x^\ast)\big)\Big)\\
&=\sup_{y^\ast\in D^{\circ}\setminus\{0\}}\inf\cb{m\in \R \mid \sup_{\gamma \geq 0} \big(\gamma s-(y^\ast \circ g)^\ast(\gamma x^\ast)\big)\leq \alpha_f(y^*,m)}\\
&=\sup_{y^\ast\in D^{\circ}\setminus\{0\}} \sup_{\gamma \geq 0} \inf\cb{m\in \R \mid \gamma s-(y^\ast \circ g)^\ast(\gamma x^\ast)\leq \alpha_f(y^*,m)},
\end{align*}
where the last equality comes from \Cref{monlemma}(ii). By the conjugation formula, for $\gamma >0$,
\begin{align*}
(y^\ast \circ g)^\ast(\gamma x^\ast)&=\sup_{x\in \X}\of{\ip{\gamma x^\ast,x}-\ip{y^\ast,g(x)}}\\
&=\gamma\sup_{x\in \X}\of{\ip{x^\ast,x}-\ip{\frac{y^\ast}{\gamma},g(x)}}
=\gamma \of{\frac{y^\ast}{\gamma} \circ g}^\ast(x^\ast).
\end{align*}
For $\gamma=0$, we have
\begin{align*}
&\inf\cb{m\in \R \mid \gamma s-(y^\ast \circ g)^\ast(\gamma x^\ast)\leq \alpha_f(y^*,m)}
\\&=\inf\cb{m\in \R \mid -(y^\ast \circ g)^\ast(0)\leq \alpha_f(y^*,m)}=\beta_f(y^\ast,-(y^\ast \circ g)^\ast(0))\eqqcolon c_{f,g}(y^\ast).
\end{align*}
Therefore, by using the previous two equations and the positive homogeneity of $\alpha_f$, we get
\begin{align*}
&\beta_{f\circ g } (x^\ast,s)=\sup_{y^\ast\in D^{\circ}\setminus\{0\}} \sup_{\gamma \geq 0}\inf\cb{m\in \R \mid \gamma s-(y^\ast \circ g)^\ast(\gamma x^\ast)\leq \alpha_f(y^*,m)}\\
&=\sup_{y^\ast\in D^{\circ}\setminus\{0\}}\of{c_{f,g}(y^\ast)\vee \sup_{\gamma >0}\inf\cb{m\in \R \mid \gamma s-\gamma \of{{\frac{y^\ast}{\gamma}\circ g}}^\ast(x^\ast)\leq \alpha_f(y^\ast,m)}}\\
&=\sup_{y^\ast\in D^{\circ}\setminus\{0\}}\of{c_{f,g}(y^\ast)\vee \sup_{\gamma >0}\inf\cb{m\in \R \mid s- \of{{\frac{y^\ast}{\gamma}}\circ g}^\ast(x^\ast)\leq \alpha_f\of{\frac{y^\ast}{\gamma},m}}}\\
&=\sup_{y^\ast\in D^{\circ}\setminus\{0\}}c_{f,g}(y^\ast)\vee \sup_{\substack{y^\ast\in D^{\circ}\setminus\{0\},\\ \gamma>0}}\inf\cb{m\in \R \mid  s- \of{{\frac{y^\ast}{\gamma}}\circ g}^\ast(x^*)\leq \alpha_f\of{\frac{y^\ast}{\gamma},m}}.
\end{align*}
Hence,
\begin{align*}
&\beta_{f\circ g } (x^\ast,s)\\
&=\sup_{y^\ast\in D^{\circ}\setminus\{0\}}c_{f,g}(y^\ast)\vee \sup_{y^\ast\in D^{\circ}\setminus\{0\}}\inf\cb{m\in \R \mid  s- (y^\ast \circ g)^\ast(x^*)\leq \alpha_f(y^\ast,m)}\\
&=\sup_{y^\ast\in D^{\circ}\setminus\{0\}}c_{f,g}(y^\ast)\vee \sup_{y^\ast\in D^{\circ}\setminus\{0\}}\beta_f\Big(y^\ast,s-(y^\ast \circ g)^\ast(x^\ast)\Big)\\
&=\sup_{y^\ast\in D^{\circ}\setminus\{0\}}\Big(c_{f,g}(y^\ast)\vee \beta_f\big(y^\ast,s-(y^\ast \circ g)^\ast(x^\ast)\big)\Big).
\end{align*}
By the monotonicity of $\beta_f$, we can also write the last line as
\[
\sup_{y^\ast\in D^{\circ}\setminus\{0\}}\beta_f\big(y^\ast,-(y^\ast \circ g)^\ast(0) \vee  (s-(y^\ast \circ g)^\ast(x^\ast))\big),
\]
which completes the proof.
\end{proof}

\begin{proof}[Proof of \Cref{cor:linop}]
        The corollary is a direct consequence of \Cref{gconvex} and the fact that $(\ip{y^\ast,Ax})^\ast (x^\ast)=0$ if $x^\ast=A^\ast y^\ast$ and $\infty$ otherwise. 
\end{proof}

\subsection{Proofs of the results in Subsection~\ref{sec:quasi-con}}\label{app:quasi-con}

Finally, we outline the proofs of the results in \Cref{sec:quasi-con}. Recall that we work with a monotone convex set $\mathcal{K}\subseteq \X$ with $C\subseteq\mathcal{K}$, and we consider two functions $f\colon\Y\to\overline{\R}$ and $g\colon \mathcal{K}\to \Y$. Let $x^\ast\in C^{\circ}$ and $m\in\R$. Similar to the constructions for the case $\mathcal{K}=\Y$ above, we define the sets
\[
\mathbb{A}^m_{y^*}\coloneqq \cb{x\in \mathcal{K}\mid \ip{y^\ast,g(x)}\leq \alpha_f(y^\ast,m)},\qquad  \tilde{\mathbb{A}}^m_{y^\ast}\coloneqq \cb{x\in \mathcal{K}\mid  \ip{y^\ast,g(x)}< \alpha_f(y^\ast,m)}
\]
for each $y^\ast \in D^{\circ}$, and the functions $\mathbb{K}^m_{x^\ast}, \mathbb{K}^m_{x^\ast}\colon\mathcal{K}\times \bar{D}^+\to \overline{\R}$ by 
\[
\mathbb{K}^m_{x^\ast}(x,y^\ast)\coloneqq \ip{x^\ast,x}-I_{\mathbb{A}^m_{y^\ast}}(x),\qquad \tilde{\mathbb{K}}_{x^\ast}(x,y^\ast) \coloneqq \ip{x^\ast,x}-I_{\tilde{\mathbb{A}}^m_{y^\ast}}(x).
\]
After giving these definitions, by using similar arguments, we can adapt Propositions \ref{chul}, \ref{kkt}, \ref{propk}, \ref{alfamin} and \ref{infhg}, and \Cref{remcg} for the following corollary.

\begin{proof}[Proof of Corollary \ref{convexthm}]
The proof follows the same reasoning as the proof of \Cref{mainthm}.
\end{proof}

\begin{proof}[Proof of Proposition \ref{convexcor}]
The proof of \eqref{dualcompcon} follows the same arguments as the proof of \Cref{fognat}. Here, we use \Cref{cor:singleconv} instead of \Cref{firstthm}. The proof of \eqref{dualcompcon2} follows by the same arguments as in \Cref{newthm}. 
\end{proof}
\begin{proof}[Proof of \Cref{convexconvexcor}]
The proof of \Cref{gconvex} is valid for this result.
\end{proof}

\section{Proofs of some results in Section~\ref{sec:application}}\label{app:sec 6}

\begin{proof}[Proof of \Cref{continuitylemma}]
To prove that $\Lambda$ is lower demicontinuous, by \Cref{rem:ldc}, we need to prove that $\Lambda^{U}(M)=\{X\in L^p(\R^n)\mid \Lambda(X) + L^p(\R_+)\subseteq M\}$ is closed for every closed halfspace $M=\{Y\in L^p(\R)\mid \E\sqb{Y^\ast Y}\geq 0\}$, where $Y^\ast\in L^q(\R)$.

We first claim that if $\Lambda(X)+L^p_+(\R)\subseteq M=\cb{Y\in L^p(\R)\mid \E\sqb{Y^*Y}\geq 0}$ for some $X\in L^p(\R^n)$, then $Y^\ast\in L^q(\R_+)$. To see this,	note that $\E\sqb{Y^\ast(\Lambda(X)+d)}\geq 0$ if and only if $\E[Y^\ast d]\geq -\E[Y^\ast \Lambda(X)]$ for every $d \in L^p_+(\R)$. Assume that $\E[Y^\ast d]<0$ for some $d \in L^p(\R_+)$. Since $L^p(\R_+)$ is a cone, for every $\lambda>0$, we have $\lambda d\in L^p(\R_+)$. Also, $\lambda \E[Y^\ast d]\to -\infty$ as $\lambda\to 0$. However, $\lambda \E[Y^\ast d]$ is bounded by $-\E[Y^\ast \Lambda(X)]$, hence we get a contradiction. Therefore, $\E[Y^\ast d]\geq0$ for all $d \in L^p(\R_+)$, which implies that $Y^\ast \in L^q(\R_+)$. This completes the proof of the claim.

In view of the claim, let us take $M=\{Y\in L^p(\R)\mid \E\sqb{Y^\ast Y}\geq 0\}$ for some $Y^\ast\in L^q(\R_+)$. We aim to show that $\{X\in L^p(\R^n)\mid \Lambda(X)+L^p(\R_+)\subseteq M \}$ is closed. Note that
\[
\cb{X\in L^p(\R^n)\mid \Lambda(X)+L^p(\R_+)\subseteq M }= \cb{X\in L^p(\R^n)\mid \E\sqb{Y^*\Lambda(X)}\geq 0 }.
\] 
Let us first consider case (i), where $\tilde{\Lambda}$ is concave and bounded from above. Thanks to concavity, the set $\{X\in L^p(\R^n)\mid\E[Y^*\tilde{\Lambda}(X)]\geq0\}$ is convex.

Suppose that $p<+\infty$. Take a sequence $(X^k)_{k\in \N}$ in $\{X\in L^p(\R^n)\mid \E\sqb{Y^*\Lambda(X)}\geq 0 \}$ that converges to some $\tilde{X}\in L^p(\R^n)$ strongly. Hence, there exists a subsequence $(X^{k_\ell})_{\ell\in\N}$ that converges to $\tilde{X}$ almost surely. By the continuity of $\tilde{\Lambda}$, and then reverse Fatou's lemma, we get
\begin{align}
\E[Y^\ast \Lambda(\tilde{X})]=\E[Y^\ast \tilde{\Lambda}\circ \tilde{X}]&= \E\sqb{Y^\ast \lim_{\ell\rightarrow\infty} \tilde{\Lambda}\circ X^{k_\ell} } \nonumber \\
& \geq \limsup_{\ell\rightarrow\infty} \E[Y^\ast \tilde{\Lambda}\circ X^{k_\ell}] = \limsup_{\ell\rightarrow\infty} \E[Y^\ast \Lambda(X^{k_\ell})]\geq 0.\label{revFat}
\end{align}
Hence, $\tilde{X}\in \{X\in L^p(\R^n)\mid \E\sqb{Y^*\Lambda(X)}\geq 0 \}$ and this set is closed. Note that we can use reverse Fatou's lemma in the above calculation since $\tilde{\Lambda}$ is bounded from above so that $(Y^\ast \Lambda (X^{k_\ell}))_{\ell\in\N}$ is bounded from above.

Suppose that $p=+\infty$. To prove $\text{weak}^\ast$ closedness, let $r>0$. By Krein-\v{S}mulian theorem, it is enough to prove that $\{X\in L^\infty(\R^n)\mid \E\sqb{Y^*\Lambda(X)}\geq 0, \norm{X}_\infty\leq r \}$ is closed in $L^1(\R^n)$. Let $(X^k)_{k\in\N}$ be a sequence in this set that converges to some $\tilde{X}\in L^1(\R^n)$ strongly in $L^1(\R^n)$. Hence, we may find a subsequence $(X^{k_\ell})_{\ell\in\N}$ that converges to $\tilde{X}$ almost surely. Repeating the argument in \eqref{revFat}, we see that $\E[Y^\ast\Lambda(\tilde{X})]\geq 0$. On the other hand, we have $\Vert X^{k_\ell}\Vert \leq r$ for all $\ell\in\N$ with probability one. Hence, $\Vert\tilde{X}\Vert\leq r$ with probability one so that $\Vert\tilde{X}\Vert_\infty\leq r$. It follows that $\tilde{X}\in\{X\in L^\infty(\R^n)\mid \E\sqb{Y^*\Lambda(X)}\geq 0, \norm{X}_\infty\leq r \}$, proving the closedness of this set in $L^1(\R^n)$.

Next we consider case (ii), where $\tilde{\Lambda}$ and hence $\Lambda$ are linear. In particular, there exists $a\in\R^n$ such that $\tilde{\Lambda}(x)=a^{\mathsf{T}}x$ for every $x\in\R^n$. Suppose that $p<+\infty$. Let us take a net $(X^k)_{k\in I}$ in $\{X\in L^p(\R^n)\mid \E\sqb{Y^*\Lambda(X)}\geq 0 \}$ that converges to some $\tilde{X}\in L^p(\R^n)$ weakly, where $I$ is an arbitrary index set. By linearity and weak convergence, we have
\[
\E[Y^*\Lambda(\tilde{X})]=\E[Y^*\tilde{\Lambda}\circ \tilde{X}]=\E[ (Y^*a)^{\mathsf{T}}\tilde{X}]=\lim_{k\in I} \E[(Y^*a)^\mathsf{T}X^k]\geq 0,
\]
so that $\tilde{X}\in \{X\in L^p(\R^n)\mid \E\sqb{Y^*\Lambda(X)}\geq 0 \}$, and this set is weakly closed, hence it is also strongly closed. The case $p=+\infty$ can be treated by Krein-\v{S}mulian theorem as above.

For (iii), let us first observe that $(L^p(\R^n_+))^{\#}=L^p(\R^n_{++})$ and $(L^p(\R_+))^{\#}=L^p(\R_{++})$. Now take $X,\bar{X}\in L^p(\R^n)$ with $X\leq_{L^p(\R^n_{++})}\bar{X}$. Hence, for almost every $\omega\in\Omega$, we have $X(\omega)\leq_{\R^n_{++}}\bar{X}(\omega)$. Since $\tilde{\Lambda}$ is regularly increasing, we have $\Lambda(X)(\omega) =\tilde{\Lambda}(X(\omega)) <\tilde{\Lambda}(\bar{X}(\omega))=\Lambda(\bar{X})(\omega)$ for almost every $\omega\in\Omega$. Therefore, $\Lambda(X)\leq_{L^p(\R_{++})}\Lambda(\bar{X})$. So $\Lambda$ is regularly increasing.
\end{proof}

\begin{proof}[Proof of Proposition \ref{propcom} ]
Let $Y^\ast \in L^q(\R_-)\setminus\{0\}$. Since we have $D$-concavity, finding the penalty function is a concave maximization problem. Moreover, since the strict sublevel set is nonempty, Slater's condition holds. 
Hence, we can use strong duality and obtain
\begin{align*}\alpha_{(Y^\ast \circ \Lambda)}(X^*,m)&=\sup_{X\in L^p(\mathbb{R}^n)}\cb{\mathbb{E}\sqb{-(X^{*})^\mathsf{T}X}\mid \E\sqb{-Y^\ast \Lambda(X) }\leq m}
\\&=\inf_{\lambda > 0}\sup_{X\in L^p(\R^n)}\of{\mathbb{E}\sqb{(X^{*})^\mathsf{T} X}-\lambda\mathbb{E}\sqb{Y^*\Lambda(X) }+\lambda m}
\\&=\inf_{\lambda > 0}\sup_{X\in L^p(\mathbb{R}^n)}\of{\mathbb{E}\sqb{(X^{*})^\mathsf{T}X-\lambda Y^*\Lambda(X)} +\lambda m}
\\&=\inf_{\lambda > 0}\of{\mathbb{E}\sqb{\sup_{x\in \R^n}\of{(X^{*})^\mathsf{T}x-\lambda Y^*\tilde{\Lambda}(x)}}+\lambda m},
\end{align*}
where the second equality is by strong duality (we can ignore the case $\lambda =0$ as it produces an objective value of $+\infty$) and the fourth equality is by \cite[Thm. 14.60]{Rockawets}. 

Note that for every $x^\ast\in \R^n$ and $y^\ast\in\R_-$, we have
\begin{equation}\label{lambdacalc}
\sup_{x\in \R^n}(x^{\mathsf{T}}x^\ast-\lambda y^\ast\tilde{\Lambda}(x))=
\begin{cases}
0& \text{ if } x^\ast=0, y^\ast=0,\\
\infty & \text{ if } x^\ast\neq 0, y^\ast=0,\\
-\lambda y^\ast\tilde{\Phi}\of{\frac{x^\ast}{\lambda y^\ast}}& \text{ if }y^\ast<0.
\end{cases}
\end{equation}
Therefore, $\alpha_{(Y^\ast \circ \Lambda)}(X^\ast,m)=+\infty$ if $Y^\ast\notin T_{X^\ast}$, and
\begin{equation}\label{eqbfr}
\alpha_{(Y^\ast \circ \Lambda)}(X^\ast,m)=
\inf_{\lambda > 0}\of{-\mathbb{E}\sqb{\lambda Y^\ast\Phi\of{\frac{X^\ast}{\lambda Y^\ast}}1_{\{Y^\ast<0\}}}+\lambda m}
\end{equation}
if $Y^\ast\in T_{X^\ast}$. Moreover, by \Cref{mainthm},
\[
\alpha_{\rho\circ\Lambda}(X^*,m)=\inf_{Y^*\in L^q(\R_-)\setminus\{0\}}\alpha_{(Y^\ast \circ \Lambda)}\of{X^*,\alpha_{\rho}\of{Y^*,m}}.
\]
By combining this equality with \eqref{eqbfr}, it follows that
\begin{align*}
\alpha_{\rho\circ\Lambda}(X^\ast,m) = \inf_{Y^\ast\in T_{X^\ast}}\inf_{\lambda > 0}\of{\E\sqb{-\lambda Y^\ast\Phi\of{\frac{X^\ast}{\lambda Y^\ast}}1_{\{Y^\ast<0\}}}+\lambda\alpha_{\rho}(Y^\ast,m)}.
\end{align*}
Then, since $T_{X^\ast}$ is a cone and $\alpha_{\rho}$ is positively homogeneous, we get 
\[
\alpha_{\rho\circ\Lambda}(X^\ast,m)
= \inf_{Y^\ast\in T_{X^\ast}}\of{-\mathbb{E}\sqb{Y^\ast\Phi\of{\frac{X^\ast}{Y^\ast}}1_{\{Y^\ast<0\}}}+\alpha_{\rho}(Y^\ast,m)},
\]
as desired.
\end{proof}

\begin{proof}[Proof of Proposition \ref{dualfn} ]
By \Cref{cor1} and \Cref{convexcase}, since $\dom (Y^\ast \circ \Lambda)=L^p(\R^n)$, we have
\begin{align}\beta_{\rho\circ\Lambda}(X^*,s)&=\sup_{Y^\ast \in L^q(\R_-)\setminus \{0\}}\beta_{\rho}\of{Y^\ast,\beta_{Y^\ast \circ \Lambda}(X^\ast,s)}\nonumber 
\\ 
&=\sup_{Y^\ast \in L^q(\R_+)\setminus \{0\}}\beta_\rho\of{Y^\ast, \sup_{\gamma \geq 0}\of{\gamma s-(Y^\ast \circ \Lambda)^\ast(\gamma X^\ast)}}\nonumber 
\\ 
&=\sup_{Y^\ast \in L^q(\R_-)\setminus \{0\}}\inf\cb{m\in \R\mid \alpha_{\rho}(Y^\ast,m)\geq \sup_{\gamma \geq 0}\of{\gamma s-(Y^\ast \circ \Lambda)^\ast(\gamma X^\ast)}} \nonumber 
\\&=\sup_{Y^\ast \in L^q(\R_-)\setminus \{0\}}\sup_{\gamma \geq 0} \beta_{\rho}\of{Y^\ast,\gamma s-(Y^\ast \circ \Lambda)^\ast(\gamma X^\ast)},\label{dualfneq}
\end{align}
where the last equality comes from \Cref{monlemma}. Let us calculate the second argument of $\beta_\rho$ for bounded case $\Phi(0)<+\infty$. For $\gamma=0$, by using \cite[Thm. 14.60]{Rockawets}, we have
\begin{align*}
-(Y^\ast \circ \Lambda)^\ast(0)&=-\sup_{Z\in L^p(\R^n)}\E\sqb{-Y^\ast \Lambda(Z)}
=-\E\sqb{\sup_{z\in \R^n}-Y^\ast \Lambda(z)}=\Phi(0)\E[Y^\ast].
\end{align*}
Here, the last equality follows by the following simple observation: for every $y^\ast\in\R_-$,
\[
\sup_{z\in \R^n} y^\ast \Lambda(z)=
\begin{cases}
0& \text{ if } y^\ast=0,\\
-y^\ast\Phi(0)& \text{ else}.
\end{cases}
\]
For $\gamma >0$, by \cite[Thm. 14.60]{Rockawets}, we get
\begin{align*}
(Y^\ast \circ \Lambda)^\ast(\gamma X^\ast)=\sup_{Z\in L^p(\R^n)}\of{\E\sqb{\gamma Z^{\mathsf{T}}X^\ast}-\E\sqb{Y^*\Lambda (Z)}}
=\E\sqb{\sup_{z\in \R^n}\of{\gamma z^{\mathsf{T}}X^\ast -Y^\ast \Lambda (z)}}.
\end{align*}
Using the calculation in \eqref{lambdacalc}, it follows that $(Y^\ast \circ \Lambda)^\ast(\gamma X^\ast)=+\infty$ if $Y^\ast\notin T_{X^\ast}$, and
\[(Y^\ast \circ \Lambda)^\ast(\gamma X^\ast)=
\E\sqb{-Y^\ast\Phi\of{\frac{\gamma X^\ast}{ Y^\ast}}1_{\{Y^\ast<0\}}}
\]
if $Y^\ast\in T_{X^\ast}$. Since $\beta_{\rho}$ is increasing in the second argument, we can ignore the case $Y^\ast\notin T_{X^\ast}$, since the second argument of $\beta_\rho$ will be $-\infty$ in \eqref{dualfneq}. By the positive homogeneity of $\alpha_{\rho}$, for $\gamma >0$, we have
\begin{align*}
\beta_{\rho}\of{Y^\ast,\gamma s+\E\sqb{Y^\ast\Phi\of{\frac{\gamma X^\ast}{ Y^\ast}}1_{\{Y^\ast<0\}}}}=\beta_{\rho}\of{\frac{Y^\ast}{\gamma}, s+\E\sqb{\frac{Y^\ast}{\gamma}\Phi\of{\frac{\gamma X^\ast}{ Y^\ast}}1_{\{Y^\ast<0\}}}}.
\end{align*}
By combining all the findings, we get
\begin{align*}
&\beta_{\rho\circ\Lambda}(X^*,s)=\sup_{Y^\ast \in L^q(\R_-)\setminus \{0\}}\sup_{\gamma \geq 0}\beta_{\rho}\of{Y^\ast,\gamma s-(Y^\ast \circ \Lambda)^\ast(\gamma X^\ast)}
\\&=\sup_{Y^\ast \in L^q(\R_-)\setminus \{0\}}\beta_{\rho}\of{Y^\ast,\Phi(0)\E[Y^\ast]} \vee
\sup_{\substack{Y^\ast \in T_{X^\ast},\\ \gamma > 0}}\beta_{\rho}\of{\frac{Y^\ast}{\gamma}, s+\E\sqb{\frac{Y^\ast}{\gamma}\Phi\of{\frac{\gamma X^\ast}{ Y^\ast}}1_{\{Y^\ast<0\}}}}
\\&=\sup_{Y^\ast \in L^q(\R_-)\setminus \{0\}}\beta_{\rho}\of{Y^\ast,\Phi(0)\E[Y^\ast]}\vee
\sup_{Y^\ast \in T_{X^\ast}}\beta_{\rho}\of{Y^\ast, s+\E\sqb{Y^\ast\Phi\of{\frac{ X^\ast}{ Y^\ast}}1_{\{Y^\ast<0\}}}},
\end{align*}
where the last equation comes from the fact that $T_{X^\ast}$ is a cone. Now we can pass to the probabilistic setting. For the left side, make the change-of-variables $Y^\ast=-\lambda \frac{d\Q}{d\P}$ where $\lambda>0$ and $\Q\in \mathcal{M}^q_1(\P)$. By using the positive homogeneity of $\alpha_{\rho}$, we have 
\begin{align*}
\beta_{\rho}\of{Y^\ast,\Phi(0)\E[Y^\ast]}&=\beta_{\rho}\of{-\lambda \frac{d\Q}{d\P} ,-\Phi(0)\E\sqb{\lambda\frac{d\Q}{d\P}}}\\
&=\beta_{\rho}\of{ -\frac{d\Q}{d\P} ,-\Phi(0)\E\sqb{\frac{d\Q}{d\P}}}
=\beta_{\rho}\of{ -\frac{d\Q}{d\P} ,-\Phi(0)},
\end{align*}
which gives
\[\sup_{Y^\ast \in L^q(\R_-)\setminus \{0\}}\beta_{\rho}\of{Y^\ast,\Phi(0)\E[Y^\ast]}=\sup_{\Q\in\mathcal{M}^q_1(\P)}\beta_{\rho}\of{ -\frac{d\Q}{d\P} ,-\Phi(0)}.\]
For the other part, we can make the change-of-variables $X^\ast=-w\cdot \frac{\mathbf{d\S}}{d\P}$ and $Y^\ast=-\lambda \frac{d\Q}{d\P}$ as before and get
\begin{align*}
&\sup_{Y^\ast \in T_{X^\ast}} \beta_{\rho}\of{Y^\ast, s+\E\sqb{Y^\ast\Phi\of{\frac{ X^\ast}{ Y^\ast}} 1_{\{Y^\ast<0\}}}}\\
&=\sup_{\substack{\Q\in \mathcal{M}_1^q(\P),\lambda >0\colon \\w_i\S_i\ll\Q\ \forall i}}\beta_{\rho }\of{-\frac{d\Q}{d\P},\frac{s}{\lambda}-\mathbb{E}_{\Q}\sqb{\Phi\of{\frac{w}{\lambda}\cdot\frac{ \mathbf{d\S}}{ d\Q}}}}.
\end{align*} 
Finally, we have 
\begin{align*}
&\beta_{\rho\circ\Lambda}\of{-w\cdot \frac{\mathbf{d}\S}{d\P},s}
\\&=\sup_{\Q\in\mathcal{M}^q_1(\P)}\beta_{\rho}\of{ -\frac{d\Q}{d\P} ,-\Phi(0)}\vee\sup_{\substack{\Q\in \mathcal{M}_1^q(\P),\lambda >0\colon\\w_i\S_i\ll\Q\ \forall i}}\beta_{\rho }\of{-\frac{d\Q}{d\P},\frac{s}{\lambda}-\mathbb{E}_{\Q}\sqb{\Phi\of{\frac{w}{\lambda}\cdot \frac{\mathbf{d}\S}{d\Q}}}}.
\end{align*}
For the unbounded case $\Phi(0)\negthinspace =\negthinspace \infty$, we can omit the first term above by the monotonicity of $\beta_{\rho}$.
\end{proof}

\begin{proof}[Proof of \Cref{dualsystemic}]
By \Cref{gconvex}, we have the following
\begin{equation*}
R(X)=\rho \circ \Lambda (X)=\sup_{X^\ast \in L^q(\R^n_-)\setminus \{0\}}\sup_{Y^\ast \in L^q(\R_-)\setminus \{0\}}\beta_{\rho}\of{Y^\ast, \E\sqb{X^{\mathsf{T}}X^\ast}-(Y^\ast \circ \Lambda)^\ast(X^\ast)}.
\end{equation*}
We calculate the second argument of $\beta_\rho$. By \cite[Thm. 14.60]{Rockawets}, we get
\begin{align*}
(Y^\ast \circ \Lambda)^\ast(X^\ast)=\sup_{Z\in L^p(\R^n)}\of{\E\sqb{Z^{\mathsf{T}}X^\ast}-\E\sqb{Y^*\Lambda (Z)}}
=\E\sqb{\sup_{z\in \R^n}\of{z^{\mathsf{T}}X^\ast -Y^\ast \tilde{\Lambda} (z)}}.
\end{align*}
By the calculation in \eqref{lambdacalc}, we have $(Y^\ast \circ \Lambda)^\ast(X^\ast)=+\infty$ if $Y\notin T_{X^\ast}$, and
\[(Y^\ast \circ \Lambda)^\ast(X^\ast)=
-\E\sqb{Y^\ast\Phi\of{\frac{X^\ast}{ Y^\ast}}1_{\{Y^\ast<0\}}}
\]
if $Y^\ast\in T_{X^\ast}$. Since $\beta_{\rho}$ is increasing in the second argument, we can ignore the case $Y^\ast\notin T_{X^\ast}$ since the second argument will be $-\infty$. Therefore, we have
\begin{equation*}
R(X)=\sup_{X^\ast \in L^q(\R^n_-)}\sup_{Y^\ast \in T_{X^\ast}}\beta_{\rho}\of{Y^\ast, \E\sqb{X^{\mathsf{T}}X^\ast}+\E\sqb{Y^\ast\Phi\of{\frac{X^\ast}{ Y^\ast}}1_{\{Y^\ast<0\}}}}.
\end{equation*}
We can make the change-of-variables $X^\ast=-w\cdot \frac{\mathbf{d\S}}{d\P}$ and $Y^\ast=-\lambda \frac{d\Q}{d\P}$ as before and we get
\begin{align*}
R(X)&=\sup_{X^\ast \in L^q(\R^n_-)}\sup_{Y^\ast \in T_{X^\ast}}\beta_{\rho}\of{Y^\ast, \E\sqb{X^{\mathsf{T}}X}+\E\sqb{Y^\ast\Phi\of{\frac{X^\ast}{ Y^\ast}}1_{\{Y^\ast<0\}}}}
\\&=\sup_{\substack{w \in \mathbb{R}^n_+\setminus\{0\},\S\in \mathcal{M}_n^q(\P),\\ \Q \in \mathcal{M}^q_1(\P)\colon w_i\S_i\ll\Q\ \forall i}}\beta_{\rho }\of{-\frac{d\Q}{d\P},-\mathbb{E}_{\Q}\sqb{\Phi\of{\frac{w\cdot \mathbf{d\S}}{d\Q}}}-w^{\mathsf{T}}\E_{\S}\sqb{X}},
\end{align*}
after using the positive homogeneity of $\alpha_{\rho}$ and writing $w$ instead of $\frac{w}{\lambda}$.
\end{proof}

\begin{proof}[Proof of Proposition \ref{propeisnoe1}]
Since we have concavity, finding the penalty function is a concave maximization problem. Thanks to Slater's condition holds, we can use strong duality and obtain
\begin{align*}
\alpha_{(Y^\ast \circ \Lambda)}\of{X^\ast,m}
&=\sup_{X\in L^p(\R^n_+)}\cb{\E\sqb{X^{\mathsf{T}}X^{*}}\mid \E\sqb{Y^\ast\Lambda(X)}\leq m}\\
&=\inf_{\lambda\geq 0}\sup_{X\in L^p(\R^n_+)} \of{\E\sqb{X^{\mathsf{T}}X^{\ast}-\lambda Y^\ast\Lambda(X)}+\lambda m}\\
&=\inf_{\lambda\geq 0}\E\sqb{\sup_{x\in \R^n_+}\of{x^{\mathsf{T}}X^\ast-\lambda Y^\ast\tilde{\Lambda}(x)+\lambda m}},
\end{align*}
where last equality is by \cite[Thm. 14.60]{Rockawets}. For $\lambda=0$, by using the fact that $X^\ast\in L^q(\R^n_-)$, we reach
\[
\sup_{X\in L^p(\R^n_+)} \of{\E\sqb{X^{\mathsf{T}}X^\ast-\lambda Y^\ast\Lambda(X)}+\lambda m}=\sup_{X\in L^p(\R^n_+)} \E\sqb{X^{\mathsf{T}}X^\ast}=0.
\]
On the other hand, by the calculation in \eqref{lambdacalc}, we have
\[
\alpha_{(Y^\ast \circ \Lambda)}\of{X^\ast,m}=0\wedge \inf_{\lambda>0}\of{\lambda m-\E\sqb{1_{\{Y^\ast<0\}}\lambda Y^\ast\Phi\of{\frac{X^\ast}{\lambda Y^\ast}}}},
\]
and by \Cref{convexthm}, we obtain
\begin{align*}
\alpha_{\rho \circ \Lambda}\of{X^*,m}&=\inf_{Y^*\in L^q(\R_-)\setminus\{0\}} \alpha_{(-h^{\Lambda}_{Y^\ast})}\of{X^*,\alpha_{\rho}\of{Y^*,m}}
\\&=\inf_{Y^*\in L^q(\R_-)\setminus\{0\}}0 \wedge \inf_{\lambda>0}\of{\lambda \alpha_{\rho}\of{Y^*,m}-\E\sqb{1_{\{Y^*<0\}}\lambda Y^*\Phi\of{\frac{X^*}{\lambda Y^*}}}}
\\&=0 \wedge \inf_{Y^*\in L^q(\R_-)\setminus\{0\}}\of{\alpha_{\rho}\of{Y^*,m}-\E\sqb{1_{\{Y^*<0\}} Y^*\Phi\of{\frac{X^*}{ Y^*}}}},
\end{align*}
where last line follows as $\alpha$ is positively homogeneous in its first component and $L^q(\R_-)$ is a cone.

Next, let us fix some arbitrary $n\in\N$ and take
\[
Y^\ast_n \coloneqq \of{1-\frac{1}{n}}Y^\ast 1_{\cb{Y^\ast<0}}-\frac{1}{n}1_{\cb{Y^\ast=0}}\in L^q(\R_{--}).
\]
Then, we have
\begin{align*}
&\inf_{\bar{Y}^*\in L^q(\R_{--})}\of{\alpha_{\rho}\of{\bar{Y}^*,m}-\E\sqb{1_{\{\bar{Y}^*<0\}} \bar{Y}^*\Phi\of{\frac{X^*}{ \bar{Y}^*}}}}
\\&\leq \alpha_{\rho}\of{Y_n^*,m}-\E\sqb{1_{\{Y^*_n<0\}} Y^*_n\Phi\of{\frac{X^*}{ Y^*_n}}} \\
&=\sup_{Y\in S^{\rho}_m}\E\sqb{YY^*_n}+\E\sqb{1_{\{Y^*_n<0\}}\sup_{x\in \R^n_+}\of{X^{*T}x-Y^*_n\tilde{\Lambda}(x)}}
\\&\leq \of{1-\frac{1}{n}}\alpha_{\rho}\of{Y^*1_{\cb{Y^*<0}},m}+\frac{1}{n}\alpha_{\rho}\of{-1_{\cb{Y^*=0}},m}+\of{1-\frac{1}{n}}\E\sqb{-1_{\{Y^*<0\}} Y^*\Phi\of{\frac{X^*}{ Y^*}}}
\\&\quad +\frac{1}{n}\E\sqb{1_{\{1_{\cb{Y^*=0}}>0\}} 1_{\cb{Y^*=0}}\Phi\of{\frac{X^*}{ -1_{\cb{Y^*=0}}}}},
\end{align*}
where the last inequality comes from the fact that supremum of affine functions is convex and indicator function of a convex set is a convex function. These inequalities are valid for every $n\in \N$, hence by sending $n$ to $\infty$, we get 
\begin{align*}
&\inf_{\bar{Y}^\ast\in L^q(\R_{--})}\of{\alpha_{\rho}\of{\bar{Y}^\ast,m}-\E\sqb{1_{\{\bar{Y}^\ast<0\}} \bar{Y}^\ast\Phi\of{\frac{X^\ast}{ \bar{Y}^\ast}}}}
\\&\leq \alpha_{\rho}\of{Y^\ast 1_{\cb{Y^\ast<0}},m}-\E\sqb{1_{\{Y^\ast<0\}} Y^\ast\Phi\of{\frac{X^\ast}{ Y^\ast}}}
=\alpha_{\rho}\of{Y^\ast,m}-\E\sqb{1_{\{Y^\ast<0\}} Y^\ast\Phi\of{\frac{X^\ast}{ Y^\ast}}}, 
\end{align*}
where last equality is trivial since it is the set where $Y^*=0$ and does not affect the expectation. Since this inequality true for every $Y^*\in L^q(\R_-)\setminus \{0\}$, by taking infimum we will have the following
\begin{align*}
&\inf_{Y^\ast\in L^q(\R_{--})}\of{\alpha_{\rho}\of{Y^*,m}-\E\sqb{1_{\{Y^*<0\}} Y^*\Phi\of{\frac{X^*}{ Y^*}}}}
\\&\leq \inf_{Y^*\in L^q(\R_-)\setminus \{0\}}\of{\alpha_{\rho}\of{Y^*,m}-\E\sqb{1_{\{Y^*<0\}} Y^*\Phi\of{\frac{X^*}{ Y^*}}}}.
\end{align*}
Also since $L^q(\R_{--}) \subseteq L^q(\R_-)\setminus \{0\}$, the reverse inequality holds as well, hence we obtain
\begin{align}\label{ciksequation}
&\inf_{Y^\ast\in L^q(\R_{--})}\of{\alpha_{\rho}\of{Y^\ast,m}-\E\sqb{1_{\{Y^\ast<0\}} Y^\ast\Phi\of{\frac{X^\ast}{ Y^\ast}}}}\nonumber 
\\&= \inf_{Y^\ast\in L^q(\R_-)\setminus \{0\}}\of{\alpha_{\rho}\of{Y^\ast,m}-\E\sqb{1_{\{Y^\ast<0\}} Y^\ast\Phi\of{\frac{X^\ast}{ Y^\ast}}}},
\end{align}
as desired.
\end{proof}

\begin{proof}[Proof of \Cref{propeisnoe3}]
By \Cref{convexconvexcor} we have
\begin{equation*}
R(X)=\rho \circ \Lambda (X)=\sup_{X^\ast \in L^q(\R^n_-)\setminus \{0\}}\sup_{Y^\ast \in L^q(\R_-)\setminus \{0\}}\beta_{\rho}\of{Y^\ast, \E\sqb{X^{\mathsf{T}}X^\ast}-(Y^\ast \circ \Lambda)^\ast(X^\ast)}.
\end{equation*}
We will calculate the second argument. By using \cite[Thm. 14.60]{Rockawets}, we get
\begin{align*}
(Y^\ast \circ \Lambda)^\ast(X^\ast)=\sup_{Z\in L^p(\R^n_+)}\of{\E\sqb{Z^{\mathsf{T}}X^\ast}-\E\sqb{Y^*\Lambda (Z)}}
=\E\sqb{\sup_{z\in \R^n_+}\of{z^{\mathsf{T}}X^\ast -Y^\ast \tilde{\Lambda} (z)}}.
\end{align*}
By \eqref{lambdacalc}, we have
\[
(Y^\ast \circ \Lambda)^\ast(X^\ast)=-\E\sqb{1_{\{Y^\ast <0\}}Y^\ast\Phi\of{\frac{X^\ast}{Y^\ast}}}.
\]
Now, let us complete the proof by using \Cref{monlemma} as follows:
\begin{align*}
&\sup_{Y^\ast \in L^q(\R_-)\setminus \{0\}}\beta_{\rho}\of{Y^\ast, \E\sqb{X^{\mathsf{T}}X^\ast}+\E\sqb{1_{\{Y^\ast <0\}}Y^\ast\Phi\of{\frac{X^\ast}{Y^\ast}}}}
\\&= \sup_{Y^\ast \in L^q(\R_-)\setminus \{0\}} \inf \cb{m\in \R\mid  \alpha_{\rho}(Y^\ast,m)\geq \E\sqb{X^{\mathsf{T}}X^\ast}+\E\sqb{1_{\{Y^\ast <0\}}Y^\ast\Phi\of{\frac{X^\ast}{Y^\ast}}}  }
\\&= \sup_{Y^\ast \in L^q(\R_-)\setminus \{0\}} \inf \cb{m\in \R\mid  \alpha_{\rho}(Y^\ast,m)-\E\sqb{1_{\{Y^\ast <0\}}Y^\ast\Phi\of{\frac{X^\ast}{Y^\ast}}}\geq \E\sqb{X^{\mathsf{T}}X^\ast}  }
\\&=\inf \cb{m\in \R\mid \forall Y^\ast \in L^q(\R_-)\setminus \{0\}\colon \alpha_{\rho}(Y^\ast,m)-\E\sqb{1_{\{Y^\ast <0\}}Y^\ast\Phi\of{\frac{X^\ast}{Y^\ast}}}\geq \E\sqb{X^{\mathsf{T}}X^\ast}  }
\\&=\inf \cb{m\in \R\mid  \inf_{Y^\ast \in L^q(\R_-)\setminus \{0\}}\of{  \alpha_{\rho}(Y^\ast,m)-\E\sqb{1_{\{Y^\ast <0\}}Y^\ast\Phi\of{\frac{X^\ast}{Y^\ast}}}}\geq \E\sqb{X^{\mathsf{T}}X^\ast}  }
\\&=\inf \cb{m\in \R\mid  \inf_{Y^\ast\in L^q(\R_{--})}\of{\alpha_{\rho}\of{Y^\ast,m}-\E\sqb{1_{\{Y^\ast<0\}} Y^\ast\Phi\of{\frac{X^\ast}{ Y^\ast}}}}\geq \E\sqb{X^{\mathsf{T}}X^\ast}  }
\\&=\sup_{Y^\ast\in L^q(\R_{--})}\inf \cb{m\in \R\mid \alpha_{\rho}(Y^\ast,m)\geq \E\sqb{X^{\mathsf{T}}X^\ast}+\E\sqb{1_{\{Y^\ast <0\}}Y^\ast\Phi\of{\frac{X^\ast}{Y^\ast}}}  }
\\&=\sup_{Y^\ast\in L^q(\R_{--}) }\beta_{\rho}\of{Y^\ast, \E\sqb{X^{\mathsf{T}}X^\ast}+\E\sqb{1_{\{Y^\ast <0\}}Y^\ast\Phi\of{\frac{X^\ast}{Y^\ast}}}}.
\end{align*}
Here, we use \eqref{ciksequation} in the fifth equality and \Cref{monlemma} in the sixth equality.
\end{proof}
\section*{Acknowledgments}

This work is partially supported by the 3501 program of T\"{U}B{\.I}TAK (Scientific \& Technological Research Council of Turkey), Project No. 117F438. The authors thank \"{O}zlem \c{C}avu\c{s} and Elisa Mastrogiacomo for their valuable feedback on a draft version of the paper.

\bibliographystyle{siamplain.bst}
\bibliography{references.bib}
\end{document}